\documentclass[traditabstract]{aa}
\usepackage{graphicx}
\usepackage{txfonts}
\usepackage[breaklinks,colorlinks,citecolor=blue]{hyperref}
\usepackage{color}
\usepackage{fixltx2e}
\usepackage{natbib,twoopt}
\usepackage{url}
\usepackage{multirow}
\usepackage{epsf}
\usepackage{epsfig}
\usepackage{longtable}
\usepackage{float}
\usepackage{subfig}
\usepackage{caption}
\usepackage[dvipsnames]{xcolor}

\definecolor{mypink1}{RGB}{219, 48, 122}

\usepackage{ifthen}
\usepackage[T1]{fontenc}
\usepackage{lmodern}
\usepackage{ifxetex,ifluatex}
\usepackage{latexsym}
\usepackage{pdfpages}
\usepackage{dblfloatfix}
\usepackage{morefloats}
\usepackage{caption}
\usepackage{lscape}
\usepackage{mathtools}

\bibpunct{(}{)}{;}{a}{}{,} 
\makeatletter
\newcommandtwoopt{\citeads}[3][][]{\href{http://adsabs.harvard.edu/abs/#3}%
{\def\hyper@linkstart##1##2{}%
\let\hyper@linkend\@empty\citealp[#1][#2]{#3}}}
\newcommandtwoopt{\citepads}[3][][]{\href{http://adsabs.harvard.edu/abs/#3}%
{\def\hyper@linkstart##1##2{}%
\let\hyper@linkend\@empty\citep[#1][#2]{#3}}}
\newcommandtwoopt{\citetads}[3][][]{\href{http://adsabs.harvard.edu/abs/#3}%
{\def\hyper@linkstart##1##2{}%
\let\hyper@linkend\@empty\citet[#1][#2]{#3}}}
\newcommandtwoopt{\citeyearads}[3][][]%
{\href{http://adsabs.harvard.edu/abs/#3}
{\def\hyper@linkstart##1##2{}%
\let\hyper@linkend\@empty\citeyear[#1][#2]{#3}}}
\makeatother

\begin{document}

   \title{Search and analysis of giant radio galaxies with associated nuclei (SAGAN) - I}
   \subtitle{New sample $\text{and}$ multi-wavelength studies}
   \titlerunning{SAGAN-I}
%
%

\author{P. Dabhade\inst{1,2}\thanks{E-mail: pratik@strw.leidenuniv.nl} 
\and M. Mahato\inst{2}
\and J. Bagchi\inst{2}
\and D. J. Saikia\inst{2}
\and F. Combes\inst{3,4}
\and S. Sankhyayan\inst{2,5,6} 
\and H. J. A. R\"{o}ttgering\inst{1}
\and L. C. Ho\inst{7,8}
\and M. Gaikwad\inst{9}
\and S. Raychaudhury\inst{2,10}
\and B. Vaidya\inst{11}
\and B. Guiderdoni\inst{12}
}
  
\authorrunning{Dabhade et al}
\institute{$^{1}$Leiden Observatory, Leiden University, P.O. Box 9513, NL-2300 RA, Leiden, The Netherlands \\
$^{2}$Inter-University Centre for Astronomy and Astrophysics (IUCAA), Pune 411007, India\\ 
$^{3}$Sorbonne Universit\'e, Observatoire de Paris, Universit\'e PSL, CNRS, LERMA, 75014 Paris, France\\
$^{4}$Coll\`ege de France, 11 Place Marcelin Berthelot, 75231 Paris, France\\
$^{5}$Indian Institute of Science Education and Research (IISER), Dr. Homi Bhabha Road, Pashan, Pune 411008, India \\
$^{6}$National Centre for Radio Astrophysics, TIFR, Post Bag 3, Ganeshkhind, Pune - 411007, India \\
$^{7}$Kavli Institute for Astronomy and Astrophysics, Peking University, Beijing 100871, People's Republic of China\\
$^{8}$Department of Astronomy, School of Physics, Peking University, Beijing 100871, People's Republic of China\\
$^{9}$ Max-Planck-Institut f\"ur Radioastronomie, Auf dem Hugel 69, 53121 Bonn, Germany \\
$^{10}$Department of Physics, Presidency University, 86/1 College Street, Kolkata 700073, India \\
$^{11}$ Discipline of Astronomy, Astrophysics and Space Engineering, Indian Institute of Technology Indore, 453552, India \\
$^{12}$ Univ. Lyon, ENS de Lyon, CNRS, Centre de Recherche Astrophysique de Lyon, 69230 Saint-Genis-Laval, France}

 \date{\today} 
 
  \abstract
{We present the first results of a project called SAGAN, which is dedicated solely to the studies of relatively rare megaparsec-scale radio galaxies in the Universe, called giant radio galaxies (GRGs). We 
have identified 162 new GRGs primarily from the NRAO VLA Sky Survey (NVSS) with sizes ranging from $\sim$\,0.71 Mpc to $\sim$\,2.82 Mpc in the redshift range of $\sim$\,0.03 - 0.95, of which 23 are 
hosted by quasars (giant radio quasars, GRQs). As part of the project SAGAN, we have created a database of all known GRGs, the GRG catalogue, from the literature (including our new sample); it includes 820 sources. 
For the first time, we present the multi-wavelength properties of the largest sample of GRGs. This provides new insights into their nature.
   
Our results establish that the distributions of the radio spectral index and the black hole mass of GRGs do not differ from the corresponding distributions of normal-sized radio galaxies (RGs). However, GRGs have a lower Eddington ratio than RGs. Using the mid-infrared data, we classified GRGs in terms of their accretion mode: either a high-power radiatively efficient high-excitation state, or a radiatively inefficient low-excitation state. This enabled us to compare key physical properties of their active galactic nuclei, such as the black hole mass, spin, Eddington ratio, jet kinetic power, total radio power, magnetic field, and size. We find that GRGs in high-excitation state statistically have larger sizes, stronger radio power, jet kinetic power, and higher Eddington ratio than those in low-excitation state. Our analysis reveals a strong correlation between the black hole Eddington ratio and the scaled jet kinetic power, which suggests a disc-jet coupling.

Our environmental study reveals that $\sim$\,10\% of all GRGs may reside at the centres of galaxy clusters, in a denser galactic environment, while the majority appears to reside in a sparse environment. The probability of finding the brightest cluster galaxy (BCG) as a GRG is quite low and even lower for high-mass clusters. We present new results for GRGs that range from black hole mass to large-scale environment properties. We discuss their formation and growth scenarios, highlighting the key physical factors that cause them to reach their gigantic size.}


\maketitle

\section{Introduction} \label{sec:intro}
 In the 1950s, it was revealed that some galaxies dominantly emit at radio wavelengths \citep{JD53,BW54} through synchrotron radiation \citep{Shklovskii55,Burbidge56}. These galaxies later 
came to be known as radio galaxies (RGs), whose radio emission often extends well beyond the physical extent of the galaxies as seen at optical wavelengths. Thereafter, it was realised by theoretical 
efforts \citep{Salpeter64,bell69,bardeen70} that a supermassive black hole (10$^{6}$ - 10$^{10}$ \rm M$_{\odot}$) residing at the centre of host galaxy must be responsible for powering \citep{rees71} 
the radio galaxy through twin collimated and relativistic jets \citep{Blanford_Rees74,Scheuer74}. The creation of the relativistic radio jets is not completely understood and is currently under 
investigation. Astrophysical models (\citealt{bz77}; hereafter B-Z, \citealt{bp82,meier01}) show that they are created by mass-accreting black holes with a certain geometry of the magnetic field that threads the accretion disc.

The B-Z model describes the process that is thought to be responsible for powering jets in galactic microquasars and gamma-ray bursts in addition to radio galaxies and quasars.
It has been established in many studies over the years that supermassive black holes (SMBHs) reside at the centres of almost all massive galaxies \citep{soltan82,rees84,Begelman84,Magorrian98,Kormendy_2013}, and their active phase is triggered only under certain circumstances. These active forms of SMBHs are known as the active galactic nuclei (AGNs), whose signatures can be observed at almost all wavelengths, ranging from radio to gamma-rays. 

Those AGNs that predominantly emit at radio wavelengths are called radio-loud AGNs (RLAGNs). A smaller fraction and more powerful class of AGNs are quasars, which are among the most energetic and brightest objects known in the Universe. When quasar AGNs emit radiation at radio wavelengths, they are called radio-loud quasars (RQs).
In high-luminosity RGs or RQs, the jets tend to terminate in high-brightness regions, called hotspots, at the outer edges of the radio lobes, which are filled with relativistic non-thermal plasma. This class of RGs is the  Fanaroff-Riley II (FR-II) class \citep{FR74}. FR-IIs are edge-brightened RGs, and the Fanaroff-Riley I (FR-I) class RGs are less powerful than FR-II. The FR-I structure is mainly edge-darkened and lobe brightness peaks within the inner half of their extent; they have no hotspot.
The largest angular size is usually measured between the peaks in the hotspots for FR-II sources and the outermost contours at the 3$\sigma$ level for FR-I sources. The projected linear size of RGs or RQs extends from less than a few kilo parsecs (pc) to several megaparsecs (Mpc).

In the past six decades, thousands of RGs have been found and catalogued, but only a few hundred RGs have been discovered so far to exhibit Mpc-scale sizes. Since their discovery in the 1970s by \citet{willis74}, this relatively rare sub-class of RGs has been referred to by several names, such as `giant radio sources' (GRSs), `large radio galaxies' (LRGs), and `giant radio galaxies' (GRGs). In order to avoid confusion and to maintain uniformity, we refer to this giant sub-class of RGs as `giant radio galaxies' (GRGs), as previously adopted in several works \citep{Schoenmakers2001,D17,Ursini18xgrg,PDLOTSS}. 
 
 Since the discovery of GRGs in the 1970s to early-2000s, the Hubble constant (H$_0$) that is used to derive the physical properties of the GRGs had a range of values between 50 to 100 km s$^{-1}$ Mpc$^{-1}$ based on available measurements at that time. This led to over- or under-estimating the sizes of these sources and eventually to inaccurate statistics of their population.
 With the advent of precision cosmology derived from the cosmic microwave background radiation observed with the Wilkinson Microwave Anisotropy Probe (WMAP; \citealt{wmap13}) and Planck mission 
\citep{2016A&A...594A..13P}, the value of H$\rm _0$ was set to $\sim$\,68 km $\rm s^{-1}$ Mpc$^{-1}$. The first GRGs discovered by \citet{willis74} were 3C236 and DA240, both of which are more than 2 
Mpc in size, and hence they did not set a lower size limit for RGs to be classified as GRGs. Later studies \citep{Ishwar-saikia00,Schoenmakers2001} adopted a lower limit of 1 Mpc, assuming H$_0$ to 50 km s$^{-1}$ Mpc$^{-1}$. Recent studies \citep{D17,Kuzmicz2018,Ursini18xgrg,PDLOTSS} have adopted 700 kpc as the lower size limit of GRGs with the updated H$\rm_0$ value obtained from observations.
 
In the past six decades, owing to radio surveys such as the third Cambridge radio survey (3CR; \citealt{3cr,3crr}), the Bologna Survey (B2; \citealt{b2}), the Faint Images of the Radio Sky at Twenty-Centimeters survey (FIRST; \citealt{beckerfirst95}), the NRAO VLA Sky Survey (NVSS; \citealt{nvss}), the Westerbork Northern Sky Survey (WENSS; \citealt{wenss97}), the Sydney University Molonglo Sky Survey (SUMSS; \citealt{sumss99}), the TIFR GMRT Sky Survey (TGSS; \citealt{tgss_intema}), and the LOFAR Two-metre Sky Survey (LoTSS; \citealt{lotssshimwell}), millions of RGs  have been found, and a considerable fraction has been studied in detail. 
However, only a few hundred of these RGs have been identified as giants or GRGs. The relatively small number of GRGs might partly be due to difficulties in identifying them. Possible factors include the detection of low surface brightness components, ensuring that two radio components are indeed related, and finding confirmed optical counterparts. However, when  we take a complete radio sample such as the 3CRR sample \citep{3crr}, the median size\footnote{\url{https://3crr.extragalactic.info/}} of the  RGs or RQs is $\sim$\,350 kpc, and only $\sim$\,7\% of the sample are GRGs.

Over a course of about 45 years, about 40 research papers 
\citep{willis74,Bridle1976,Waggett77,3crr,kronberg1986,Bruyn1989,jones89,ekers89,lacy93,lawgreen95,cotter96,7McCarthy,ravi96,Ishwara1999,lara2001,Machalski_2001,Schoenmakers2001,sadler2002,letawe2004,
Saripalli2005,saikia2006,Machalski_2007,Huynh2007,Machalski_2008,koziel11,hotaspeca,Solovyov14,molina14,bagchi14,Amirkhanyan2015,Tamhane2015,Amirkhanyan2016,D17,clarke2017,kapinska17,prescott18,
binny18,Kuzmicz2018,koziel19,PDLOTSS} have reported about 662 GRGs spread throughout the sky. This estimate is based on a database of GRGs that is compiled by us, for which we adopted 700 kpc as the lower limit of GRG 
size. We used concordant cosmological parameters from Planck (H$\rm _0$ = 67.8 km $\rm s^{-1}$ Mpc$^{-1}$, $\Omega\rm _m$ = 0.308, $\Omega\rm _{\Lambda}$ = 0.692; \citealt{2016A&A...594A..13P}). 

Some of the open questions related to GRGs are the processes through which they grow to Mpc-scale sizes, how rare they are with respected normal radio galaxies, do they host most powerful SMBHs, nature of their accretion states, and how fast their SMBHs are spinning. Owing to their enormous sizes, it essential to investigate whether they grow only in sparser environments, and their contribution to large-scale processes.

Giant radio galaxies are the largest single objects in the Universe. They pose interesting questions related to the formation and evolution of such large structures through relativistic jets that originate in SMBHs. These large RGs must be supplied with energy either continuously or quasi-continuously over timescales of at least 10$^{7}$ - 10$^{8}$ years when we consider that the hotspots (terminal shocks of the jet) move with sub-relativistic speeds \citep{best95,scheuer95,askalongair00} of $\sim$\, 0.1c, although highly asymmetric sources such as the apparently one-sided GRGs may require higher velocities \citep{saikia90}. 
It has been suggested that their evolution to such large sizes is due to their occurrence in low-density environments \citep{mack98,pirya,malarecki15,Saripalli15}. \citet{pirya} found that GRGs are usually located in low galaxy density environments by counting galaxies in the vicinity of GRGs. However, the structural asymmetries in their sample of giants, in which the nearer component is often brighter, suggest asymmetries in the intergalactic medium that are not apparent from  examining the galaxy distribution. Only in the case of 3C326 did they find the closer brighter arm to be possibly interacting with a group of galaxies that is part of a larger filamentary structure. \citet{Machalski_2008} reported the discovery of the largest GRG J1420-0545 and suggested that the source is located in a very low-density environment and has a high expansion speed along the jet axis. However, a few studies, such as  \citet{Komberg09} and \citet{D17,PDLOTSS}, have shown dozens of GRGs residing in dense cluster environments. A few other studies \citep{ravi96,Saripalli2005,bruni19gpsgrg,bruni20} have also suggested that GRGs may attain their gigantic size due to the restarted AGN activity. Different explanations may be applicable to different subsets of GRGs, which detailed studies of large samples and comparisons with theoretical models and simulations will help clarify (e.g. \citet{Qjet_Hardcastle} and references therein).

It has also been suggested that GRGs contain exceptionally powerful central engines that are powered by massive black holes, which are responsible for their gigantic sizes \citep{gk89}. Based on the current understanding, the environment alone therefore cannot be the only determining factor for the giant size of GRGs, but a combination of AGN power (including jet power and accretion states), environmental factors, and the longevity of AGN activity (duty cycle) might be playing an equally important role.
 
Moreover, multi-wavelength studies of only a small fraction of GRGs have been carried out so far to specifically address the important questions related to their unusual nature. This has restricted a comprehensive statistical analysis of the properties of GRGs to understand their true nature.

In order to address these questions in a systematic study of a large number of GRGs, we have initiated a project that is completely dedicated to the study of GRGs, We describe this in the following parts of this paper along with our first results.
For the first time, we have been able to obtain a good understanding of the GRG astrophysics here, spanning an enormous range ($\sim 10^{11}$) of physical scales from $\sim 10^{-5}$ pc in the jet launch zone that lies in the vicinity of the black hole to $\sim$\,1 Mpc, where the jet termination point is located.

The paper is organised as follows: In Sec.\ \ref{sec:sagan} we present an overview and the goals of project SAGAN. In Sec.\ \ref{sec:nsamp} we describe the  search criteria and  method for  identifying the new  GRG sample from the NVSS survey, followed by a discussion of  the newly created GRG database in Sec.\ \ref{sec:grgcat}. In Sec.\ \ref{sec:analysis} we describe the analysis methods we employed on multi-wavelength data of GRGs to estimate their various properties. Next, we present the results of the analysis along with a discussion and their implications in Sec.\ \ref{SGS}, which is further sub-divided into several subsections, each dedicated to a property of GRGs. We conclude our study of GRGs in project SAGAN and outline its future prospects in Sec.\ \ref{sec:sum}. Lastly, in Appendix\ \ref{sec:aptab} we present three main tables consisting of the properties of our new GRG sample, and in Appendix\ \ref{sec:apfig} we show the multi-frequency radio maps of the GRGs in our new sample.

Throughout this paper, the flat $\Lambda$CDM cosmological model is adopted based on the Planck results (H$\rm _0$ = 67.8 km s$^{-1}$ Mpc$^{-1}$, $\Omega\rm _m$ = 0.308,  and $\Omega\rm _{\Lambda}$ = 0.692 \citealt{2016A&A...594A..13P}). All the images are presented in a J2000 coordinate system. We use the convention $S_{\nu}\propto \nu^{-\rm \alpha}$, where S$_{\nu}$ is the flux density at frequency $\nu,$ and $\rm \alpha$ is the spectral index.


\section{Project SAGAN}\label{sec:sagan}
To understand the physics of these extreme cosmic radio sources much better, and specifically, to  address the key questions that remain, we have initiated a project called SAGAN\footnote{\url{https://sites.google.com/site/anantasakyatta/sagan}} (Search and analysis of GRGs with associated nuclei), whose pilot study results were presented in the previous paper \citep{D17}. Some of the main goals for this project are listed below.

\begin{enumerate}
\item We aim to create a complete and uniform database of GRGs from the literature, which spans five decades starting from 1970s, using a single cosmological model with H$\rm _0$ = 67.8 km s$^{-1}$ Mpc$^{-1}$, $\Omega\rm _m$ = 0.308, and $\Omega\rm _{\Lambda}$ = 0.692 (flat $\Lambda$CDM). 

\item  We search for more GRGs from existing radio and optical or infrared survey data.
\item Using the newly created large database of GRGs, we carry out  multi-wavelength studies of the host AGNs of the GRGs. We intend to focus on some key  physical properties such as the accretion rate ($\rm \Dot{m}$) of the black hole, excitation type, black hole mass (M$_{\rm BH}$), Eddington ratio, spin, host galaxy star formation rate (SFR), and  high-energy gamma-ray emission from jets.
\item We explore effects of the environment on the morphology and growth of the GRGs.
\item We use  magneto-hydrodynamical (MHD) simulations to investigate  the  jet physics and the  conditions required for the collimation and stability of relativistic jets, propagating  to megaparsec or larger physical distances from the host AGN.

\end{enumerate}

Broadly, the goal is to understand birth, growth, and evolution of GRGs and their possible contribution to other processes in the Universe.

In this first paper, we present the results of our search for GRGs from the NVSS along with GRGs from other published works and investigate their multi-wavelength properties (in radio, optical, and mid-infrared bands). We not only report a larger sample of 162 hitherto unidentified GRGs, but also shed light for the first time on their AGN and host galaxy physical properties. 
\subsection{New sample of GRGs from NVSS}\label{sec:nsamp}
The NVSS provides radio maps ($\delta > -40^{\circ}$, 82\% of the sky) at 1400 MHz with a modest resolution of 45$\arcsec$ and has root mean square (rms) brightness fluctuations of $\sim$\,0.45 mJy beam$^{-1}$. The NVSS was released more than 20 years ago, and still it continues to be a source of many interesting discoveries \citep[e.g. for GRGs;][]{solovyov11,Amirkhanyan2016,proctorGRS,D17}.

\citet{proctorGRS} produced a catalogue of 1616  possible  giant radio sources (GRSs) from automated 
pattern-recognition  techniques using NVSS data. This catalogue of 1616 sources represents the 
radio  objects that are candidates for GRGs with projected angular size $\geq$ 4\arcmin. Therefore this catalogue serves as a useful database for finding new GRGs.

Following-up on our pilot study, \citet{D17}, we further carried out our independent 
manual visual search for GRGs from the NVSS, and the results of the search were combined with the fraction of GRGs we confirmed from the \citet{proctorGRS} sample.
In order to confirm potential GRGs from Proctor's sample, we used the following radio surveys, which complement each other, to understand the radio morphology of the sources:
\begin{itemize}
    \item NVSS: It has a high sensitivity for large-scale diffuse emission and is one of the best all-sky radio surveys to date.
    \item FIRST: This survey is at 1400 MHz with a resolution of $\sim$\,5$\arcsec$ and 0.15 mJy beam$^{-1}$ rms. Its high-resolution maps provide vital information about the radio cores and hotspots of the sources.
    \item TGSS: This is a low-frequency radio survey at 150 MHz with a resolution of $\sim$\,25$\arcsec$ and an rms of $\sim$\,3.5 mJy beam$^{-1}$, covering the entire NVSS footprint. It is sensitive to diffuse low radio frequency emission and particularly good at detecting  sources with very steep spectra.
    \item VLASS\footnote{\url{https://archive-new.nrao.edu/vlass/HiPS/VLASS_Epoch1/Quicklook/}}: The Very Large Array Sky Survey \citep{vlass}. This is the most recent all-sky radio survey at 3000 MHz with a resolution of $\sim$\,2.5$\arcsec$ and rms of $\sim$\,100 $\mu$Jy and covers the same footprint as the NVSS. This survey is deeper, has a better resolution, and covers a greater sky area than the FIRST. It is very useful for identifying compact structures such as the radio cores of the sources.
\end{itemize}

After the overall morphology of the sources was determined using the available radio surveys, optical and mid-IR data from the Sloan Digital Sky Survey (SDSS; \citealt{sdss00,sdssdr14}), the Panoramic Survey Telescope and Rapid Response System (Pan-STARRS; \citealt{kaiser02,kaiser10,chambers16}), and the Wide-field Infrared Survey Explorer (\textit{WISE}; \citealt{wright10-wise}), respectively, were used to identify the host galaxies of the candidate GRGs.

The following steps and criteria were used to create the final sample of confirmed GRGs from the  \citet{proctorGRS} GRS catalogue:

\begin{enumerate}
\item Optical/mid-IR and radio maps were overlaid to identify the host galaxy/AGN that coincides with the radio core. Sources that did not have a radio core-host galaxy association were rejected.

\item We selected only those sources via thorough manual inspection whose various components (core, jets, or lobes) were sufficiently well resolved, with no ambiguity in their radio morphology.

\item All the sources selected by the above steps were checked for redshift ($z$) information of the host galaxy (photometric or spectroscopic) from publicly available optical surveys and databases.

\item The angular sizes of the sources were computed using NVSS radio maps for uniformity, and to ensure that there is no flux or structure loss, which the other higher resolution radio surveys (FIRST, TGSS, and VLASS) are prone to. We measured the largest angular separation of the two components (lobes, hotspots, tails, and jets) of the sources after considering only the parts of the sources that are seen above 3$\sigma$. In this way, the angular sizes of all the sources were revised, and the angular extent of some sources was $<$ 4\arcmin, which is the lower limit of the \citet{proctorGRS} sample.

\item Lastly, we made use of redshift and angular size information to compute the projected linear size of the sources, and only those greater than 700 kpc were considered for our GRG sample (SAGAN GRG sample, or SGS henceforth).

\end{enumerate}

\begin{table}[h]
\caption{Short summary of the classification of sources from \citet{proctorGRS}.}
\begin{center}
\begin{tabular}{lcll}
\hline
Classification & No. of objects \\ 
\hline
Ambiguous morphology &  156 \\
Independent sources & 266 \\ 
New GRGs        & 151 \\
Known GRGs &    165 \\ 
Narrow-angle tailed RGs & 24 \\ 
No core & 143 \\ 
No host & 20\\
No redshift & 311\\
RGs     & 311 \\ 
Supernova remnant       & 6 \\ 
Spiral/disc galaxies & 32 \\ 
Wide-angle tailed RGs   & 31 \\
\hline
Total & 1616 \\
\hline
\end{tabular}
\end{center}
\label{tab:sumtab}
\end{table}

 \begin{figure*}
 \centering
 \includegraphics[scale=0.84]{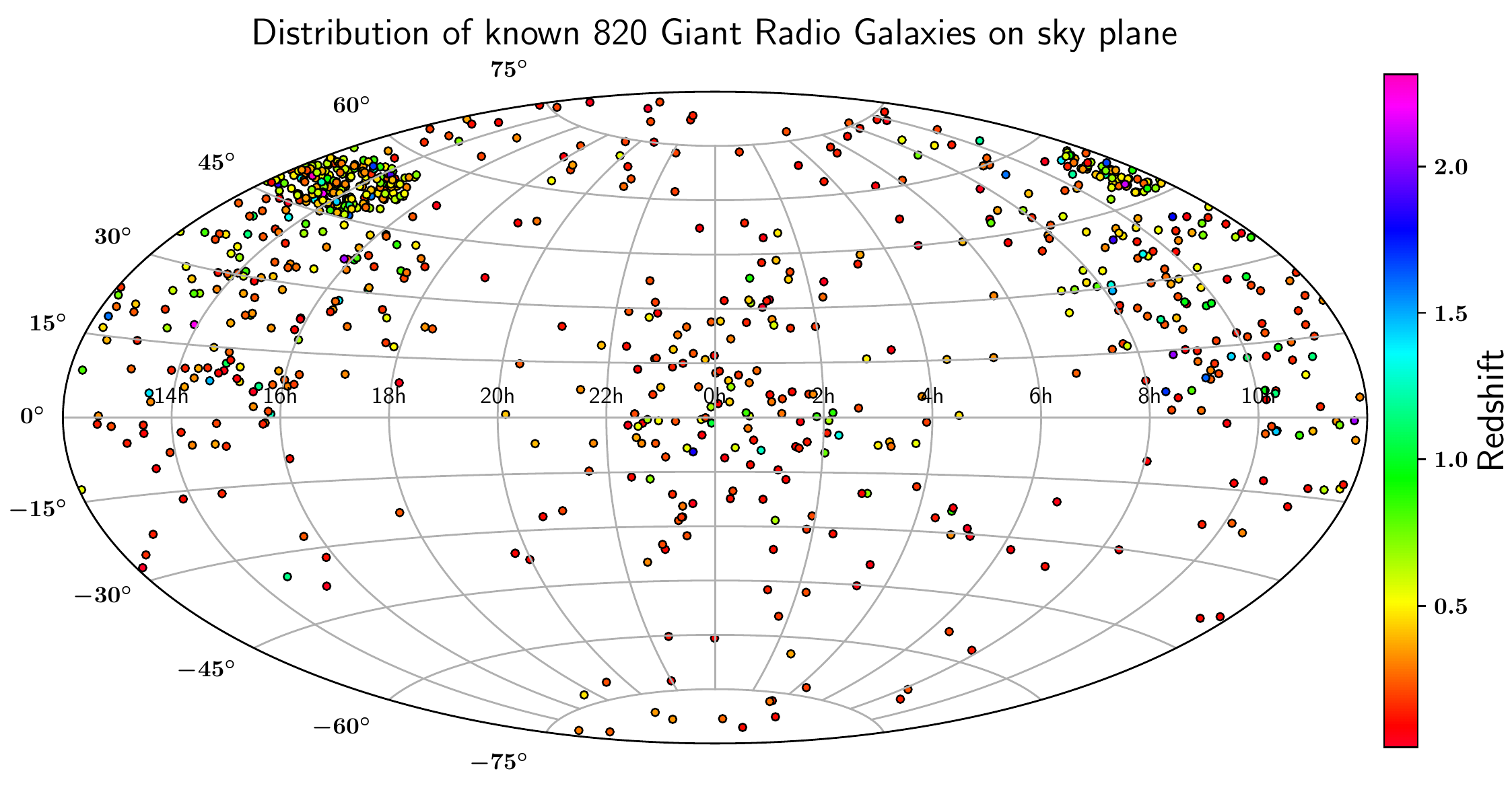} 
 \caption{Sky distribution of all the known GRGs from 1974 to 2020 together with our SAGAN GRG sample in an Aitoff projection. The total number of GRGs plotted here is 820 (LoTSS: 225 and SAGAN: 162, and  all others from literature: 433). The large clustering of GRGs in the northern region of the plot (right ascension 10h45m to 15h30m and declination $45^\circ 00\arcmin$ to $57^\circ 00\arcmin$) is the result of finding the large sample of GRGs (225) in the LoTSS by us \citep{PDLOTSS}. The colour of the points on the plot corresponds to their redshift, which is indicated in the vertical colour bar at the right side of the plot. We do not use all the 820 GRGs for our analysis in this paper, but only those (762 GRGs) with a redshift lower than 1.}
 \label{fig:skydistribution}
 \end{figure*}

 \begin{figure}
 \centering
 \includegraphics[scale=0.31]{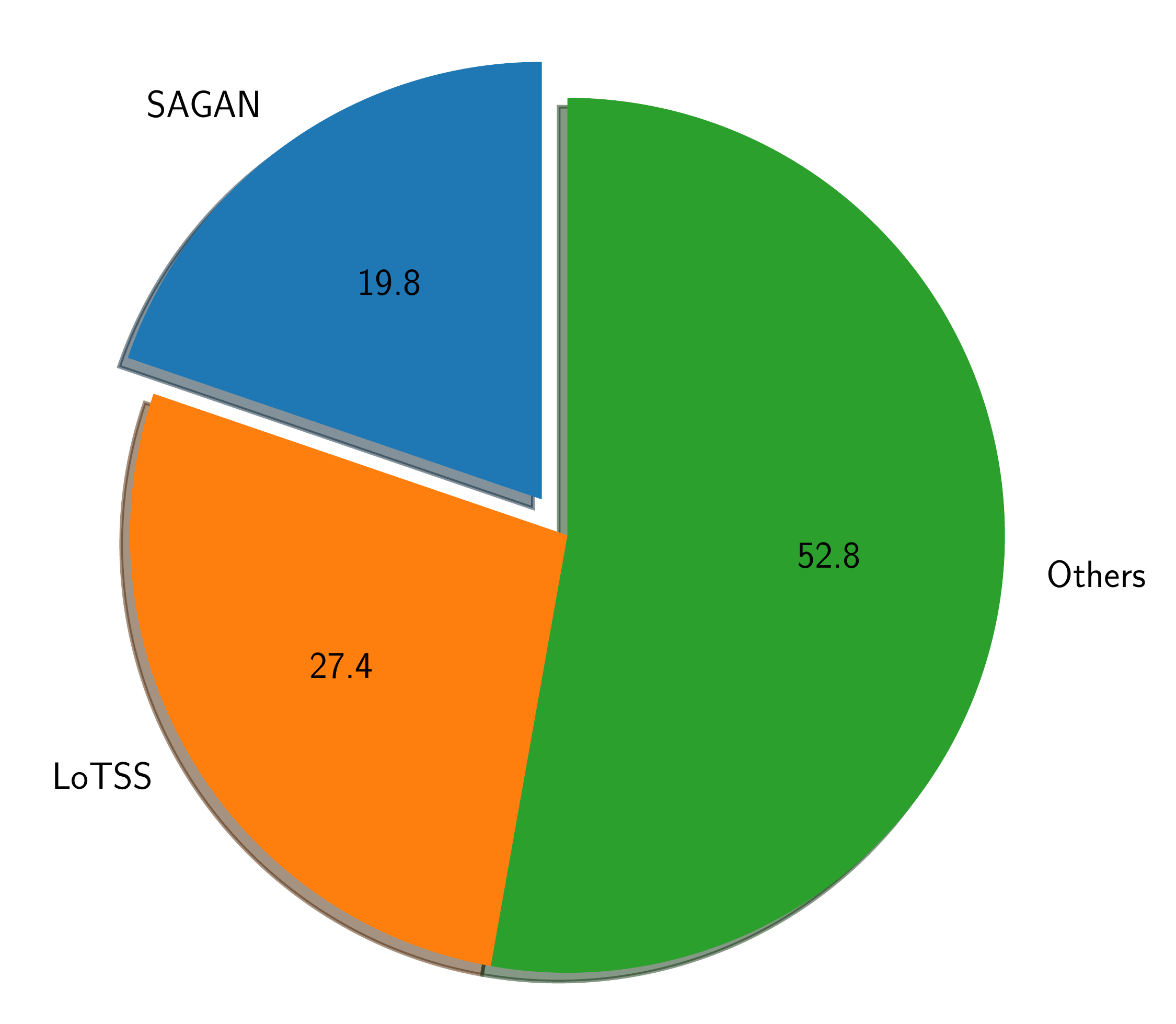} 
 \caption{Pie diagram representing the contribution of SGS (blue: $\sim$\,20\%) and LoTSS-GRGs (orange: $\sim$\,27\%) to the total GRG population. The GRGs reported in the literature until March 2020 are plotted in green. We show all the known (820) sources without any filters.}
 \label{fig:piechart}
 \end{figure}

These steps resulted in identifying 151 new GRGs from the \citet{proctorGRS} sample. We also classified the remaining sources into different categories, which can be useful to the scientific community for future work. The classification was made based on the availability of radio and optical data, which are given in Table~\ref{tab:sumtab}.
Many sources from our independent manual search were in common with \citet{proctorGRS} sample, and a total of 11 GRGs were found to be unique (not in the \citealt{proctorGRS} sample). After combining the two, we therefore report our final sample (SGS) of 162 GRGs as seen in Table~\ref{tab:maintab}. The sample is discussed in more detail in Sec.\ \ref{SGS}.

The basic information of the SGS, that is, right ascension (RA) and declination (Dec) of host galaxies in optical, AGN type (galaxy or quasar), redshift, angular size (arcminute), physical size or projected linear size (Mpc), flux density (mJy), and radio powers (W Hz$^{-1}$) at 1400 MHz and 150 MHz, and spectral index with error estimates are presented in the Table~\ref{tab:maintab}.
About 37.7\% of the SGS have photometric redshifts, and the remainder has spectroscopic redshift information. When we consider the lower end of the uncertainties of the photometric redshifts obtained from the SDSS, only three sources in our SGS, SAGANJ010052.6+061641.01, SAGANJ075931.84+082534.59, and SAGANJ125204.82-222645.70, are below the cutoff of 0.7 Mpc. The lower limits of their projected linear sizes are 0.68, 0.67, and 0.69 Mpc,
respectively.

\subsection{The GRG catalogue}\label{sec:grgcat}
In order to explore and study the trends of GRG properties using a statistically significant sample, we combined our SGS with all other known GRGs from the literature (as of April 2020) given in Sec.\ \ref{sec:intro}, and we refer it as the \texttt{GRG catalogue} throughout this paper. The total number of GRGs in the \texttt{GRG catalogue}, that is, the total number of GRGs known to date, is 820, and it is a unique complete compendium of known GRGs to date. 
Fig.\, \ref{fig:skydistribution} shows the distribution of all the known GRGs (including the GRG sample of this paper) on the sky. The high concentration of GRGs  in the northern region of the plot (right ascension 10h45m to 15h30m and declination $45^\circ 00\arcmin$ to $57^\circ 00\arcmin$) is primarily due to the recent discovery of a large sample of new GRGs (225) from the LoTSS by us \citep{PDLOTSS}, which has contributed about 30\% to the known GRG population as shown in Fig.\, \ref{fig:piechart}. Our reporting sample from this paper, called the SGS, has contributed an additional $\sim$\,20\% to the overall known population of GRGs. This means that we contributed about 50\% of all known GRGs to date.

We point out our analysis here is restricted to 762 out of 820 GRGs ($\sim$\,93\% of the \texttt{GRG catalogue}), and 58 GRGs with $z>1$ were not considered to avoid any kind of bias. Beyond a redshift of 1, we are limited because no optical data are available and because the radio source properties (luminosity and size) evolve fast in the early cosmic epoch of $z>1$. SGS, which is now part of the \texttt{GRG catalogue,} has all sources with $z<1$.


\section{Analysis}\label{sec:analysis}

\subsection{Size}
The projected linear size of the sources is taken as the end-to-end distance between the two hotspots (peak fluxes) in case of FR-II sources, and for FR-Is, it is the distance between the maximum extents defined by the outer lobes. For the measurement of angular sizes, only the NVSS maps are considered for uniformity. The projected linear sizes of sources are estimated using the following formula, and they are tabulated in Table\ \ref{tab:maintab} (Col. 8):
\begin{equation}
  \rm  D = \frac{\theta \times D_{c}}{(1+z)} \times \frac{\pi}{10800}
,\end{equation}
where $\theta$ is the angular extent of the GRG in the sky in units of arcminutes, D$_{c}$ is the comoving distance in Mpc, $z$ is the redshift of the GRG host galaxy, and D is the projected linear size of the GRG in Mpc.

\subsection{Flux density and radio power}
The integrated flux density of GRGs was estimated using Common Astronomy Software Applications ({\tt CASA}) \citep{casa} with the task {\tt CASA-VIEWER} by manually selecting the regions of emission associated with each GRG in NVSS\footnote{\url{https://www.cv.nrao.edu/nvss/postage.shtml}} and TGSS\footnote{\url{https://vo.astron.nl/tgssadr/q_fits/cutout/form}} maps. The TGSS maps were convolved to the resolution of NVSS. We used the scheme of \citet{klein03} for measuring flux density errors for each source, where we adopted 3$\%$ and 20$\%$ flux calibration errors, as mentioned in the literature for the NVSS and the TGSS, respectively.

The radio powers of GRGs were calculated using Eq. 2 and are given in Table.\ \ref{tab:maintab} (Cols. 10 and 12): 
\begin{equation}
\rm P_{\nu}  = 4\pi D_L^{2}S_{\nu} (1+z)^{\alpha - 1}
 \label{power}
,\end{equation}
where $\rm D_{\rm L}$ is the luminosity distance, $\rm S_ {\rm\nu}$ is the measured radio flux density at frequency $\rm \nu$,  $(1+z)^{\rm \alpha - 1}$ is the standard k-correction term, and $\rm \alpha$ is the radio spectral index (as described in Sec.\ \ref{sec:SI}). 

\subsection{Jet kinetic  power}\label{sec:jp}
AGN jets, made of relativistic charged particles and magnetic fields, emanate from the central engine and pierce the interstellar medium.
Low frequency radio observations enable us to estimate the jet kinetic power, which helps us to study other characteristics of the radio-loud SMBH system as discussed in later sections.
High radio frequencies ($\sim$\,1 GHz) are ideal for observing nuclear jet components owing to their flatter spectral nature. Because these components have high velocities, relativistic effects such as the Doppler enhancement effects are prominent. Therefore, lower radio frequencies are more suitable for probing the jet kinetic power because the contribution from Doppler enhancement is negligible.
We used the following relation from the simulation-based analytical model of \citet{Qjet_Hardcastle} to estimate the jet kinetic power:
\begin{equation}
\rm  L_{150} = 3 \times 10^{27}\frac{Q_{\rm Jet}}{10^{38} \ W} W \ Hz^{-1}
   \label{jet_kp2}
,\end{equation} 
 where $\rm L_{150}$ is the radio luminosity at 150 MHz, at which Doppler boosting is negligible, and $\rm Q_{Jet}$ is the jet kinetic power. There is a scatter of 0.4 dex (rms) about the relation for the sources with $z$ < 0.5.
However, significant scatter is observed in the relationship at higher redshifts, with a major contributor being the increased inverse-Compton scattering. Other contributors include the environment and evolutionary state of the sources \citep{Qjet_Hardcastle}. In the tracks of the P-D diagram for sources of different jet powers in the models presented by \citet{Qjet_Hardcastle}, the luminosity of the source appears to decrease with size during the giant phase for the different jet powers. Their modelling shows that it is possible to obtain more accurate results by taking environment and age into account. The relation is consistent with the regression line of \citet{ineson17} and earlier work by \citet{Willott1999}.
 
 Sources in the \texttt{GRG catalogue} taken from the \cite{PDLOTSS} GRG sample have flux density measurements at 144 MHz from the LoTSS, which is used to derive the 150 MHz radio luminosity. The TGSS was used for the remaining sources in the \texttt{GRG catalogue} to obtain the 150 MHz radio luminosity. This was mainly done for our newly found SGS, and only sources with full structure detection in TGSS were considered for the same. The results are presented in Col. 8 of Table \ref{tab:table_2}.
       
\subsection{Spectral index ($\rm \alpha$)}\label{sec:SI}
For a radio source, its spectral index ($\rm \alpha$) represents the energy distribution of the relativistic electrons \citep{scheuer68}, and its measurement should therefore ideally involve covering wide frequency range. Studies have shown that $\rm \alpha$ correlates with radio power and redshift.
 For synchrotron radiation, unless it is affected by radiative losses and optical depth effects, it is well known that the radio flux density varies with frequency as $\rm S_{\nu}$ $\propto$ $\nu^{-\rm \alpha}$ , resulting in a two-point spectral index measurement, given as follows:
\begin{equation}
\rm \alpha = \frac{\ln S_{\nu_{1}} - \ln S_{\nu_{2}}}{\ln\nu_{2} - \ln\nu_{1} }
.\end{equation}

The integrated spectral index between 150 and 1400 MHz ($\rm \alpha^{\rm 1400}_{\rm 150}$) for the GRGs was computed using the TGSS and the NVSS radio maps (Table.\ \ref{tab:maintab}, Col. 13). We did not include GRGs with incomplete structure detection in the TGSS for the $\rm \alpha^{\rm 1400}_{\rm 150}$ studies. For these sources, the $\rm \alpha^{\rm 1400}_{\rm 150}$ is assumed to be 0.75 to determine the radio power at 1400 MHz ($\rm P_{1400}$). See Sec.\ \ref{sec:SISGS} and Sec.\ \ref{sec:SIdisc} for a more detailed discussion.

For sources SAGANJ090111.78+294338.00 and SAGANJ091942.21+260923.97, TGSS data are lacking because they fall in a sky area that is not covered in TGSS-ADR-1. Therefore, no spectral index measurements were possible for them.

\subsection{Absolute r-band magnitude}\label{sec:analysis-rband}
Using the SDSS, we obtained the apparent r-band magnitudes (m$_{\rm r}$) of host galaxies of GRGs in the SGS as well as for all the other objects in the \texttt{GRG catalogue}. The absolute r-band magnitudes (M$_{\rm r}$) of galaxies were computed after applying the k-correction on extinction-corrected r-band apparent magnitudes (m$_{\rm r}$) of SDSS. The k-correction was computed using the \texttt{K-CORRECT} v4.3 software \citep{kcorr} for a rest frame at $z$ = 0.
Column 5 of Table~\ref{tab:maintab} shows m$_{\rm r}$ of the sources from the SGS.

\subsection{Black hole mass}
We estimated the black hole masses associated with the AGNs in the host galaxies of GRGs using the  M$_{\rm BH}$-$\sigma$  relation.
The M$_{\rm BH}$-$\sigma$  relation is based on a strong correlation between the central galactic 
black hole mass (M$_{\rm BH}$) and the effective stellar velocity dispersion ($\sigma$) in the 
galactic bulge \citep{Ferrarese2000, Gebhardt2000} given by
\begin{equation}
\rm \log \left(\frac{M_{BH}}{M_{\odot}}\right) = \alpha + \beta \log  \left(\frac{\sigma}{200\ {\rm km~s}^{-1}}\right)
,\end{equation}
where $\rm \alpha = -0.510 \pm 0.049$ and $\rm \beta = 4.377 \pm 0.290$ \citep{Kormendy_2013}. 
Estimates for $\sigma$ (Col. 4 of Table~\ref{tab:table_2}) were available for only 46 host galaxies of GRGs from the SGS in the SDSS, and hence the M$_{\rm BH}$ of  GRGs (Table.\ \ref{tab:table_2}, Col. 
5) could be computed with this method. 

\subsection{Eddington ratio}
The dimensionless Eddington ratio ($\lambdaup_{\rm Edd}$) is the ratio of the AGN bolometric luminosity to the maximum Eddington luminosity, which in turn is the estimate of the accretion rate of the SMBH in terms of the Eddington accretion rate. The Eddington ratio is expressed by the following relation:
\begin{equation}
 \rm \lambdaup \equiv \frac{L_{\rm bol}}{L_{\rm Edd}}  
,\end{equation}
where L$_{\rm bol}$ represents the  bolometric luminosity, and the Eddington luminosity is L$_{\rm Edd}$. The L$_{\rm bol}$ is calculated from the luminosity of the [OIII] emission line using the relation
L$_{\rm bol}$ = 3500 $\times$ L$_{\rm [OIII]}$ \citep{Heckman2004}. The values of $\lambdaup_{\rm Edd}$ are tabulated in Col. 7 of Table.\ \ref{tab:table_2}.
The Eddington luminosity (also known as the Eddington limit) is derived from the black hole mass, and it is the maximum luminosity that an object could have when the force of radiation acting outwards and the gravitational force acting inwards are balanced. The equation of the Eddington luminosity for pure ionised hydrogen plasma is
 L$_{\rm Edd}$ = 1.3 $\times$ 10$^{38}$ $\times$ ($\rm \frac{M_{BH}}{\rm M_{\odot}}$) erg s$^{-1}$.

\subsection{Black hole spin} \label{sec:spininfo}
Spin and angular momentum ($a$ and $J$) are fundamental properties of black holes, together with the mass. This can help us reconstruct the history of mergers and accretion activity \citep{Hughes_2003,Volonteri_2007,King,Daly} that have occurred in the central engine in the past billion years and thus paves the way to understanding the energetic astrophysical jets. In the B-Z model, the relativistic jet is the outcome of the combined effect of rotation (frame dragging) and the accumulated magnetic field near the black hole \citep{bz77}, which is fed matter by the rotating accretion disc that surrounds it. In the alternative model of \citet{bp82}, the jet power can come from the rotation of the accretion disc with the help of magnetic threading, without invoking a spinning black hole. 
However, in both the processes, the intensity and geometry of the poloidal component of the magnetic field near the black hole horizon strongly affect the Poynting flux of the emergent jet \citep{BeckwithHawley08}.

According to the B-Z model \citep{bz77,blandford90}, the relationship between the jet power (Q$_{\rm Jet}$),  the black hole mass (M$_{\rm BH}$), the black hole dimensionless spin ($a$ = Jc/(GM$^{2}$)), and the poloidal magnetic field (B) threading the accretion disc and ergosphere takes the following form:
\begin{equation}\label{eq:bhspin}
\rm Q_{Jet} \propto B^2 M_{BH}^2 \textit{a}^{2}
,\end{equation}
where  Q$_{\rm Jet}$ is in units of 10$^{44}$ erg s$^{-1}$, B  is
in units of 10$^4$ G, M$_{\rm BH}$ is in units of 10$^8$  M$_{\odot}$ , and $a$ is the dimensionless spin parameter ($a$=0 refers to a non-rotating black hole, and $a$=1 is a maximally spinning black hole). The spin ($a$) can be quantified using the above relation when the other contributing parameters are known or fixed. The constant of proportionality is taken to be $\sim$\,$\sqrt{0.5}$ , as in the B-Z model. Owing to the challenges of estimating the magnetic field in the vicinity of the black hole to compute the spin, we considered the Eddington magnetic field  strength (B$_{\rm Edd}$) \citep{Beskkin,Daly}, which is 
 \begin{equation}\label{eq:bedd}
\rm  B \sim B_{Edd} \approx 6 \times 10^{4}\left(\frac{M_{BH}}{10^{8}M_{\odot}} \right)^{-1/2} Gauss
 .\end{equation}
 This is the upper limit of the magnetic field strength close to the central engine, and it is based on the assumption that the magnetic field energy density balances the total energy density of the accreting plasma that has a radiation field of Eddington luminosity. 
These estimates are presented in Table\ \ref{tab:spin}.

The X-ray reflection is the most robust and effective technique employed to date to estimate the spin of black holes \citep{reynoldspinnat19}. However, owing to the weakness of the signal and because the data for the objects need to be rich, it has been convincingly performed for only $\sim$\,20 sources so far. For radio galaxies, which are jetted sources and mostly show weak X-ray reflection signatures, it is possible to estimate the spin assuming the B-Z mechanism \citep{Daly,Mikhailovspin}. In this radio-driven method, the spin ($a$) of the black hole can be estimated when we have estimates of M$_{\rm BH}$ and Q$_{\rm Jet}$ and adopt $\rm B_{\rm Edd}$ as B. This method provides an indirect estimate of the spin of the black hole.

\subsection{\textit{WISE} mid-IR properties}\label{sec:wisean}
We used the \textit{WISE} survey to study the hosts of GRGs at mid-IR wavelengths. \textit{WISE}, which is a space-based telescope, carried out an all-sky survey in four mid-IR bands [W1 (3.4$\mu$m), W2 (4.6$\mu$m), W3 (12$\mu$m), and W4 (22$\mu$m)] with an angular resolution of 6.1$\arcsec$, 6.4$\arcsec$, 6.5$\arcsec$  , and 12$\arcsec$ , respectively. 

Using the mid-IR colours, we obtained the properties of possible dust-obscured AGN, and estimated its radiative efficiency. The mid-IR information of host galaxies of GRGs is very useful in gauging the radiative efficiency because the optical-UV radiation from the  accretion disc of the AGN is absorbed by the surrounding dusty torus (if present) and is re-radiated at mid-IR wavelengths. Moreover, it has been shown in the literature that \textit{WISE} mid-IR colours can effectively distinguish AGNs from star-forming and passive galaxies, and within the AGN subset itself, high-excitation radio galaxies (HERGs) and low-excitation radio galaxies (LERGs) stand out in the mid-IR colour-colour and mid-IR-radio plots \citep{stern12,gurkan14}. In the absence of any dedicated multi-wavelength survey of host galaxies of GRGs, \textit{WISE} data are therefore ideal for exploring GRG properties.

The {\tt WISE All-Sky Source Catalogue}  was used to obtain magnitudes of host galaxies of GRGs in the four mid-IR bands. After applying photometric quality cuts of 3$\sigma$, we obtained reliable mid-IR magnitudes for sources in the \texttt{GRG catalogue}. Upper limits of the magnitudes in the relevant bands were estimated for sources that did not have a 3$\sigma$ detection as derived with the method prescribed in the \textit{WISE} documentation.

\citet{gurkan14} have demonstrated that data from \textit{WISE} can be used to distinguish accretion modes in RLAGNs. To achieve this, \citet{gurkan14} used four complete radio samples with emission line-based classification of RGs into low-excitation radio galaxies (LERGs), high- excitation radio galaxies (HERGs), quasars, star-forming galaxies (SFGs), narrow line radio galaxies (NLRGs), and ultra-luminous infrared radio galaxies (ULIRGs). These were further placed in the WISE colour-colour plots, where these sub-classes occupied different regions and thereby enabled distinction. Furthermore, other works \citep{Sullivan15,mingo16,Whittam18} have also confirmed this finding with other radio samples. 
Therefore we employed the scheme from \citet{mingo16}, which is based on the earlier work of \citet{wright10-wise}, \citet{lake12-wise}, and \citet{gurkan14}, to classify the host AGN and host galaxies of GRGs into LERGs, HERGs, quasars, SFGs, and ULIRGs.

\begin{figure*}
\centering
\includegraphics[scale=0.6]{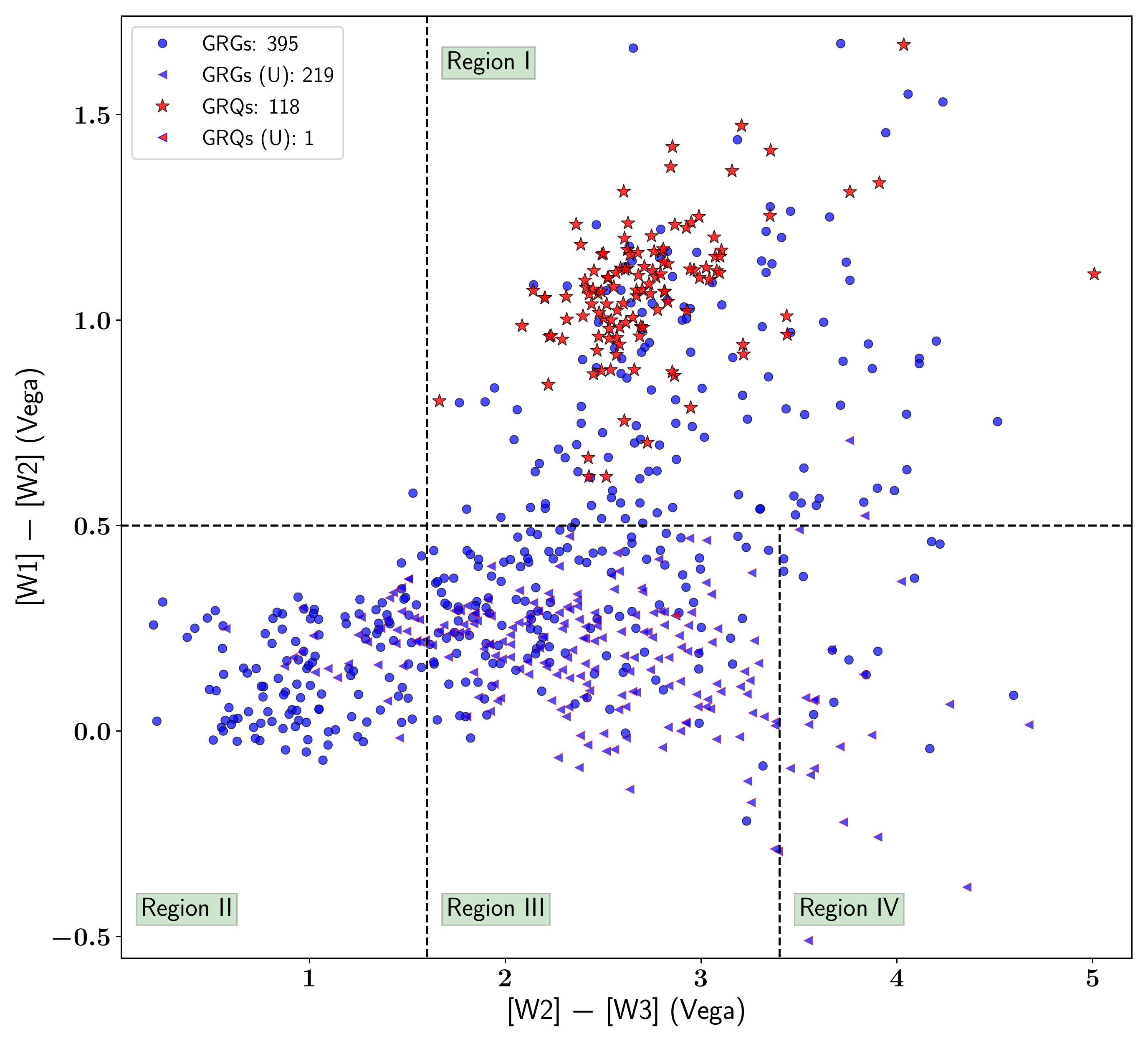} 
\caption{Position of GRGs and GRQs (733 sources from \texttt{GRG catalogue}) in the mid-IR colour-colour plot using \textit{WISE} mid-IR magnitudes (W1, W2, W3, and W4 have 3.4, 4.6, 12, and 22 $ \rm \mu m $ Vega magnitudes, respectively). Objects from the \texttt{GRG catalogue} are included.  Region I: [W1]$-$[W2] $\geq 0.5$ and [W2]$-$[W3] < 5.1 is mostly occupied by HERGs and quasars. Region II: Objects that have [W1]$-$[W2] < 0.5 and 0 < [W2]$-$[W3] < 1.6 are basically LERGs. Region III: Star-forming galaxies and LERGs lie mostly in this region ([W1]$-$[W2] < 0.5 and 1.6 $\leq$ [W2]$-$[W3] < 3.4 ). Region IV: ULIRGs lie in the region of [W1]$-$[W2] $<$ 0.5  and [W2]$-$[W3] $\geqslant$ 3.4. All the sources have z $<$ 1. The  triangle in the plot and the U in the legend indicate the upper limits of the W3 magnitudes. }
\label{fig:WISE}
\end{figure*}

Fig.\ \ref{fig:WISE} is a colour-colour plot using four mid-IR bands [W1 (3.4$\mu$m), W2 (4.6$\mu$m), W3 (12$\mu$m), and W4 (22$\mu$m)] and shows not only the distinction between LERGs and HERGs, but also effectively distinguishes  AGNs from star-forming and passive galaxies. Fig.\ \ref{fig:WISE} includes 733 sources from the \texttt{GRG catalogue} based on the availability of data and detection in \textit{WISE} database. It has been separated into four regions, which signify the following:
\begin{enumerate}
\item Region I: HERGs and quasars (W1 $-$ W2 $\geqslant$ 0.5 , W2 $-$ W3 $<$ 5.1). This region consists of 250 GRGs, 103 of which are hosted by quasars. 
\item Region II: LERGs (W1 $-$ W2 $<$ 0.5 , 0 $<$ W2 $-$ W3 $<$ 1.6). This region hosts 153 sources.
\item Region III: LERGs and star-forming galaxies (W1 $-$ W2 $<$ 0.5 , 1.6  $\leqslant$ W2 $-$ W3 $<$ 3.4)- 297 giants lie in this region.
\item Region IV: ULIRGs (W1 $-$ W2 $<$ 0.5 , W2 $-$ W3 $\geqslant$ 3.4). Only 33 sources lie in this region.
\end{enumerate}

Fig.\ \ref{fig:WISE}  shows that host galaxies of GRGs reveal a variety of AGN excitation types similar to the RG types. This also indicates that the GRG host galaxy or AGN do not preferentially show any specific AGN excitation type. The quasars in this plot are not identified with this method, but were previously classified from the SDSS and other available literature data. Similar results were presented in \citet{D17}, but with a much smaller sample. We here placed almost the entire \texttt{GRG catalogue} on the \textit{WISE} colour-colour plot for AGN diagnostic and classified the known GRG population into their excitation types. We focused on the low- and high-excitation part of the classification, and efforts were made to ensure a clean classification. For a sub-sample (based on the  availability of data), we also compared our LERG and HERG classification of GRGs from \textit{WISE} with the standard classification method based on emission line ratios and found them to be consistent with each other.

We refer to the GRGs with low- and high-excitation types as LEGRG and HEGRG, respectively. Because regions III and IV in Fig.\ \ref{fig:WISE} both have a mixed population of LERGs as well as star-forming galaxies and ULIRGs, respectively (thereby confusing the classification), we did not consider objects in these regions for our analysis, and considered only sources in region II to be LEGRGs to make a clean sample. Similarly, we excluded from region I known quasars and created a HEGRG sample for our analysis. These two criteria reduce the number of objects available for our further analysis.


\section{SAGAN GRG sample: Results} \label{SGS}
The classification of the sources in the SGS is shown in Table\ \ref{tab:sumtable} below. Out of 162 GRGs, 23 sources are found to be hosted by galaxies with quasars as their AGN (we refer to them as GRQs). The quasar nature of these 23 GRQs is identified using the 
spectroscopic data from the SDSS, \citet{parisdr14}, and other available literature data. 
All the GRGs have been detected in the redshift range of $\sim$\, 0.03 - 0.95, with projected linear sizes varying from $\sim$\,0.71 - 2.82 Mpc. Three GRGs in our sample have projected linear sizes $\geq$ 2 Mpc.

\begin{table}[h]
\caption{Summary of classified sources.}
\centering
\begin{tabular}{cc}
\hline
\hline
Types & No. \\ 
\hline
GRQ & 23 \\ 
BCG & 18 \\ 
FR I & 8 \\ 
FR II & 149 \\ 
HyMoRS & 4 \\ 
DDRG & 1 \\ \hline
\end{tabular}
\label{tab:sumtable}
\end{table}

\subsection{Spectral index distribution of the SGS}\label{sec:SISGS}
We were able to estimate $\rm \alpha^{\rm 1400}_{\rm 150}$ for a total of 123 from the SGS, 18 of which are GRQs (see Col. 13 of Table \ref{tab:maintab}). For the remaining 39 sources in the SGS, TGSS had only partial or no detection, and we therefore assumed the spectral index to be 0.75 to calculate P$_{\rm 1400}$. Therefore the $\rm \alpha^{\rm 1400}_{\rm 150}$ data for the remaining 39 sources are not included for further studies in this paper.
The median value of $\rm \alpha^{\rm 1400}_{\rm 150}$  for the GRGs in our sample (0.69 $ \pm$ 0.02) is similar to that of GRQs (0.69 $\pm$ 0.04).
\citet{PDLOTSS} also reported that $\rm \alpha^{\rm 1400}_{\rm 150}$ of GRGs and GRQs was similar, with almost similar sample size. However, because their sample was chosen from a low-frequency survey such as the LoTSS, they found it to be slighter steeper than our SGS. 

\subsection{PD$z\Large\alpha$ parameters of GRGs}
We investigated the correlation between various properties of sources in our sample, such as radio power (P), projected linear size (D), redshift ($z$), and spectral index ($\rm \alpha$) (Fig.\,\ref{PDZA}). The inferences are presented below.

\begin{enumerate}
\item Radio power (P) versus redshift ($z$): Fig.\ \ref{PDZA} (a) represents the distribution of GRGs in the P-$z$ plane, with the radio power spanning over three orders of magnitude up to a redshift range of $\sim$\,0.95. Most of the sources are  within the redshift range of 0.1 to 0.5, and the number of radio sources decreases with increasing redshift beyond $z$ = 0.5.

 The non-availability of sources in the lower right quadrant is most likely due to the non-detection of low powered sources at high redshift because of the sensitivity limit of the survey, known as the Malmquist bias. The GRQs occupy the high radio luminosity and high-redshift regime because more optical data are available than for GRGs. The weakest source in our sample is SAGANJ090640.80+142522.97, with a flux density of $\sim$\,24 mJy at 1400 MHz. The dashed line represents the minimum luminosity at different  redshifts, corresponding to a minimum flux density of $\sim$\,24mJy assuming a spectral index of 0.75. This is the NVSS limit for detecting objects with low surface brightness, and hence the absence of any source below this line. Recently, \citet{PDLOTSS} discovered a large sample of new GRGs, and significant fractions of them were of low luminosity. When we place them on Fig.\ \ref{PDZA} (a), they tend to lie below the sensitivity line of NVSS, implying higher sensitivity of LoTSS.

\begin{figure*}
\includegraphics[scale=0.58]{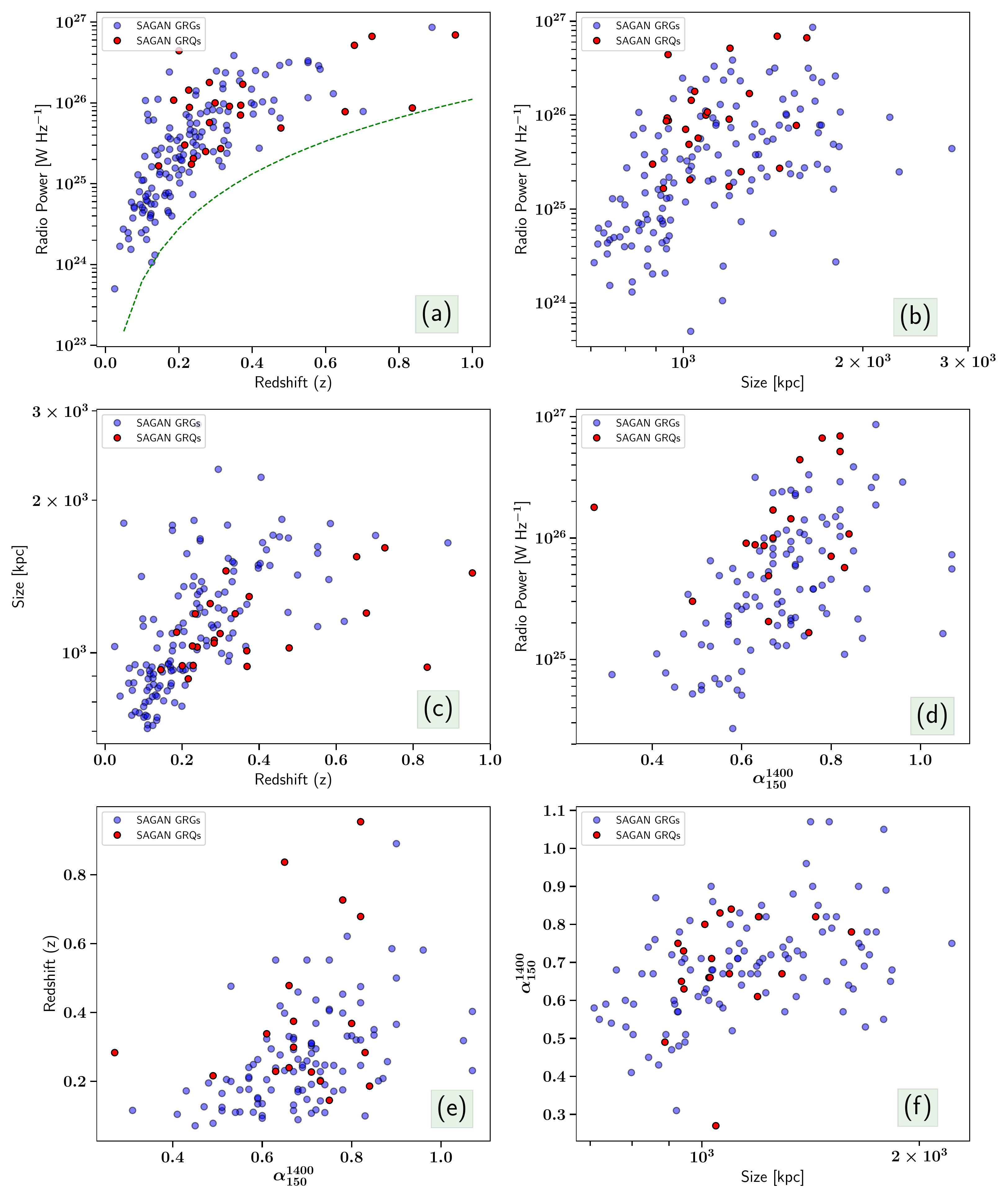}
\caption{Correlations of radio power [W Hz$^{-1}$] at 1400 MHz , size [kpc], redshift, and spectral index for GRGs and GRQs. Red circles indicate GRQs, and blue circles indicate GRGs. The dashed green line in panel (a) represents the minimum radio luminosity corresponding to a flux density $\sim$\,24mJy of source SAGANJ090640.80+142522.97 at different redshifts, assuming a spectral index of 0.75.}
\label{PDZA}
\end{figure*}

\item Radio power (P) versus projected linear size (D): The projected linear sizes of the radio sources are plotted against their radio power, measured at 1400 MHz, as shown in Fig.\ \ref{PDZA} (b). This plot is the radio astronomer's equivalent of the traditional Hertzsprung-Russell diagram and is commonly known as the P-D diagram. We can draw the following conclusions from this diagram: 
\begin{itemize}
     \item Despite the systematic search for giants, very few sources are found to be of extremely large size and very low radio power ($\leq$ 10$^{24}$ W Hz$^{-1}$ at 1400 MHz).
    \item The upper right region of PD diagram, that is, the region of sources with high radio power and large projected linear size, is devoid of any source. This is indicative of an increase in radiative losses as the sources grow, which decreases the surface brightness and renders them inaccessible to the surveying telescope because of its sensitivity limit. 
    \item There is a sudden drop in the number of giants with a projected linear size beyond 2 Mpc. The projected linear sizes of only three sources, SAGANJ064408.04+104341.40, SAGANJ225934.13+082040.78, and SAGANJ231622.32+224650.28 in our sample, exceed 2 Mpc. All of them have low redshifts;  the highest is  0.405 for source SAGANJ225934.13+082040.78.
    The projected linear size of only 66 of $\sim$\,762 known GRGs from the \texttt{GRG catalogue} ($z < 1$) is greater than 2 Mpc. Among them, four sources have extraordinary large projected linear sizes between 3 to 4 Mpc, while another four sources have projected linear sizes $\geq$ 4 Mpc, and the largest GRG known to date spans up to 5.2 Mpc at a redshift of 0.3067 \citep{Machalski_2008}. About 50$\%$ of the sources are at low redshifts (z $\leq$ 0.4), and the high-redshift objects are mostly dominated by quasars. This might be attributed to the sensitivity limit of radio surveys. Another possibility might be the limited lifetime  of radio sources \citep{Schoenmakers2001}. A large fraction of sources may have switched off before they reach 2 Mpc and beyond. 
\end{itemize}
    
\item Projected linear size (D) versus redshift (z): Fig.\ \ref{PDZA} (c) shows the positive correlation between projected linear sizes of our sources with redshift (z$<1$), which is expected because luminosity strongly correlates with these two parameters. This may be an outcome of the selection bias of the sources considered in this plot. A negative correlation between projected linear size and redshift is expected from the systematic increase in density of the environment at earlier epochs. 

\item Radio power (P) versus spectral index ($\rm \alpha$): 

 Fig.\ \ref{PDZA} (d) shows that the spectral index increases with radio power. The correlation is significant for our sources, which are selected at high frequency (1400 MHz), similar to the results of \citet{Laing1980} for extended sources. It is also consistent with results of \citet{Blundell}, who reported that the spectra of hotspots in powerful radio sources are steeper than those in weaker radio sources. Sources with high Q$\rm _{Jet}$ form powerful hotspots with enhanced magnetic fields \citep{Blundell}, which in turn leads to the rapid synchrotron cooling of relativistic electrons (cooling time $\rm \tau$ $\propto$  $\rm 1/B^{2}$). This eventually results in an increase in synchrotron losses, and thus electrons with a steeper energy distribution are injected into the lobes. 

\item Redshift ($z$) versus spectral index ($\rm \alpha$):  
Fig.\ \ref{PDZA} (e) shows that at higher $z,$ we obtain relatively steeper $\rm \alpha$, although there is a large scatter and sources with steep spectra are also seen at low redshifts. 

In a flux-density limited sample where luminosity and redshift are strongly correlated, it is difficult to determine whether the primary dependence is on luminosity or redshift (e.g. \citealt{Blundell}). 
 In addition to the effects discussed above, other suggested explanations for the $z$-$\alpha$ correlation are as follows:
 \begin{itemize}
 \item At higher redshifts, the circum-galactic medium is denser, which slows the hotspots down. The hotspots along with the lobes therefore remain in the high-pressure medium for a prolonged duration. A reduction in the velocity will result in the production of electrons with steeper energy distribution \citep{Kirk,Mangalam,Athreya,Klamer,GK}.
  
 \item In addition to higher synchrotron radiative losses due to a higher magnetic field in the more luminous sources, an important factor at high redshifts is inverse-Compton losses. As the redshift increases, there is a rapid increment in the energy density of the cosmic microwave background radiation (CMBR energy density $\propto (1+z)^4$). When the high-energy electrons of plasma interact with the CMBR photons, energy loss due to inverse Compton radiation increases, and thus the spectra steepen \citep{Krolik,Athreya,Morabito}.
 \end{itemize}

 \item Spectral index ($\rm \alpha$) versus projected linear size (D): The sampling of the sources in the D - $\rm \alpha$ plane in Fig.\ \ref{PDZA} (f) shows a weak correlation between the two parameters. The larger the sources, the steeper their spectral index. As the sources grow, they are subjected to radiative losses and thereby steepen the energy distribution. However, it is also reflected in the plot that sources with large projected linear size need not have steep spectra. This result is consistent with the findings of \citet{Blundell}.

 \end{enumerate}

\subsection{Morphology of GRGs}\label{sec:morpdisc}
Using combined information from the VLASS, FIRST, NVSS, and TGSS, we classified the GRGs in our sample into the following types: FR-I, FR-II, HyMoRS, and DDRG.
HyMoRS are RGs with a hybrid morphology \citep{GKhymorph,gkhymors02} that exhibit an FR-I morphology on one side and an FR-II morphology on the other side of the radio core. The earliest example of this class of objects was presented in \citet{saikia96hymorph}. This classification is listed in$^{\rm }$ Col. 15 in Table \ref{tab:maintab}, where HM refers to GRGs that are candidates for HyMoRS. Higher resolution radio maps are needed to confirm the morphology of the four  HyMoRS candidate GRGs. For HyMoRS candidates, the radio power range is P$_{\rm 1400}$ $\sim$\,2.74 $\times$ 10$^{24}$ W Hz$^{-1}$ -  7.78 $\times$ 10$^{25}$ W Hz$^{-1}$.
The most significant result is that about $\text{}$\,92$\%$ of the GRGs in the SGS show an FR-II type (edge-brightened hotspots within radio lobes) radio morphology, whereas only 8 out of 162 the GRGs show an FR-I type radio morphology. The radio power (P$_{\rm 1400}$) for FR-I type GRG ranges from $\sim$\,1.31 $\times$ 10$^{24}$ W Hz$^{-1}$ to 11.1 $\times$ 10$^{25}$ W Hz$^{-1}$, and for FR-II type GRGs. the range extends from $\sim$\,0.5 $\times$ 10$^{24}$ to 8.6 $\times$ 10$^{26}$. Recently, \citet{PDLOTSS} used the LoTSS and also found a similar result: most of the GRGs had an FR-II type morphology.
Moreover, \citet{mingo19} used the LoTSS and reported a significant overlap in radio power of FR-I and FR-II type RGs, along with a new sample of low-luminosity FR-II type RGs. 

\subsection{Environmental analysis of the SGS}\label{sec:sgsenv}
We cross-matched the SGS with one of the largest catalogues of galaxy clusters,  the WHL catalogue \citep{whl12}. We found 18 GRGs from the SGS that were listed as brightest cluster galaxies (BCGs); they are listed in Table~\ref{whl}.
The mass (M$_{200}$) and virial radius (r$_{200}$) of the clusters were obtained from \citet{whl12} and are listed in Table~\ref{whl} for these 18 GRGs. The size is expressed as  r$_{\rm 200}$, which is the radius within which the galaxy cluster mean density is about 200 times the critical density of the Universe, and the mass of the cluster within r$_{\rm 200}$ is denoted by M$_{\rm 200}$. 
We considered the sizes of these 18 GRGs along with 111 non-BCG-GRGs from our sample in the same redshift range (0.063 - 0.369) with median redshifts of 0.174 and 0.205, respectively. We find that the median values for sizes of BCG-GRGs and non-BCG-GRGs are 0.92 Mpc and 1.03 Mpc, respectively. Even though the median values do not vary largely, this indicates that BCG-GRGs have smaller sizes than non-BCGs, or in other words, the immediate environment plays an active role in curtailing their growth.

\section{GRG catalogue: Properties and correlations}
We describe the derived properties of GRGs, consisting of 762 sources with $z < 1$ using the multi-wavelength data. Our multi-wavelength analysis is divided into the following three parts to understand the nature of GRGs, and we investigate the key astrophysical factors governing their growth to megaparsec scales.

We first studied the differences in AGN types of GRGs: Quasars powering giant radio structures (jets and/or lobes) are called GRQs, which constitute less than 20\% of the total known GRG population. The aim is to understand the key differences between giants hosted by quasars and non-quasar AGN. Their properties, such as size, P$_{\rm 1400}$, Q$\rm _{Jet}$,  and $\rm \alpha_{150}^{1400}$ , are compared and discussed in Sec.\ \ref{sec:gq}.
    
Next, we studied the accretion states (LEGRG or HEGRG) of the central nuclei of GRGs: These two sub-classes are investigated  and discussed in Sec.\ \ref{sec:hl} in the context of their properties, such as size, P$_{\rm 1400}$, Q$\rm _{Jet}$, M$\rm _{r}$, $\rm M_{BH}$ , and $\lambdaup_{\rm Edd}$. Finally, we studied the similarities and dissimilarities between RGs and GRGs: We compared their $\rm \alpha_{150}^{1400}$, $\rm M_{BH}$ , and $\lambdaup_{\rm Edd}$ properties, and the findings are discussed in Sec.\ \ref{sec:rggrg}.

To test whether two samples that we compared come from the same distribution, we used the two-sample
Kolmogorov-Smirnov (K-S) test \citep{k1933,s1948,KS-peacock} and the Wilcoxon-Mann-Whitney (WMW) test \citep{Wilcoxon45,mann1947test}. In these we tested the null hypothesis H$_{0}$ : the two samples are drawn from the same distribution, against the alternative hypothesis H$_{1}$: the two samples are not drawn from the same distribution. A lower p-value provides stronger evidence that the null hypothesis is rejected, or in other words, that the two samples are not drawn from the same distribution.

In addition to these comparative studies, we explored the SMBH properties to understand how several aspects such as mass ($\rm M_{BH}$), spin ($a$), Eddington ratio ($\lambdaup_{\rm Edd}$), and jet kinetic power (Q$\rm_{Jet}$) are related in Sec.\ \ref{sec:bhprop}.
Lastly, we discuss in Sec.\ \ref{sec:enva} the environmental properties of GRGs that are found in clusters of galaxies (BCGs), and explore their relationship with M$_{200}$ and other BCG-RGs.

\subsection{Comparison of the properties of GRGs and GRQs}\label{sec:gq}
\begin{figure}
\centering
\includegraphics[scale=0.31]{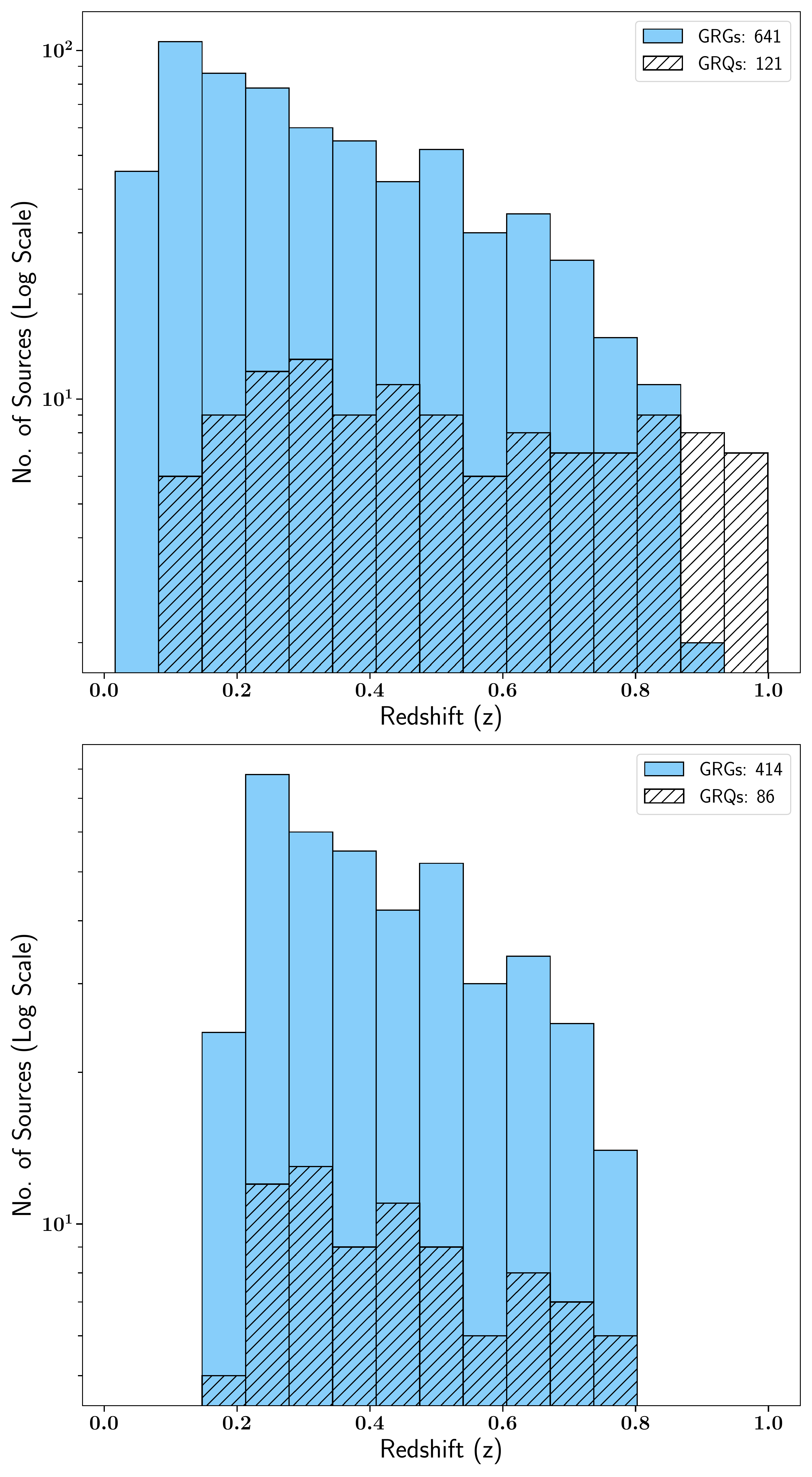}
\caption{The figure shows the redshift ($\rm z$) distributions of GRGs and GRQs for $z <$1, represented in unhatched and hatched bins, respectively for redshift matched (lower plot) and unmatched (upper plot) samples.}
\label{fig:GQ_z}
\end{figure}

As mentioned earlier, we here only considered GRGs with $z$ < 1  for analysis. The redshift distribution (Fig.\ \ref{fig:GQ_z} top panel) of GRGs and GRQs is not similar (the WMW test p-value is 3.4 $\times$ 10$^{-13}$ and the K-S test p-value is 2.9 $\times$ 10$^{-7}$), however, because GRQs extend to higher redshifts than GRGs. The mean and median redshifts of GRGs are 0.331 and 0.284, respectively, for the range of 0.016 to 0.910, and for GRQs, the values are 0.515 and 0.475, respectively, for the range of 0.085 to 0.999.
In order to ensure that our results are not affected by redshift evolution, we considered a redshift-matched (0.2$< z <$0.8) (Fig.\ \ref{fig:GQ_z}  lower panel) sub-sample of GRGs (mean $z$: 0.42 and median $z$: 0.40) and GRQs (mean $z$: 0.44 and median $z$: 0.43) to compare their properties (Fig.\ \ref{fig:GQzmatched}). Here, the redshift distributions of both samples are similar (WMW test p-value is 0.14 and K-S test p-value is 0.54), and our comparison analysis of all properties of GRGs and GRQs gives similar results for the not redshift-matched (Fig.\ \ref{fig:GQ}) and the redshift-matched samples (Fig.\ \ref{fig:GQzmatched}). We therefore present our analysis for the samples of GRGs and GRQs below. The detailed statistics of all properties are presented in Table\ \ref{mean}.

\subsubsection{Distribution of the projected linear size}
The distributions of the projected linear size are shown in Fig.\, \ref{fig:GQ} (a) and Fig.\ \ref{fig:GQzmatched} (a), and the median
values are presented in Table\ \ref{mean}. Although the median value for quasars appears to be marginally higher when all objects with $z <$ 1 in the GRG sample are considered, the difference is reduced and is within 1$\sigma$ when the samples are matched in redshift. However, because the GRG sample has been compiled from multiple heterogeneous sources, we also considered the homogeneous samples of the LoTSS and SGS with matched redshifts of GRGs and GRQs. For these two samples, the median values of sizes for GRGs are 0.89 and 1.20 Mpc, and for GRQs, they are 1.00 and 1.06 Mpc, respectively. The WMW test (p-value for the SGS: 0.06, and for the LoTSS: 0.18) shows that the distributions of the projected linear size of GRQs are not significantly different from those of GRGs.

\subsubsection{Distribution of the radio power (P$_{\rm 1400}$)}
\begin{figure*}
\centering
\includegraphics[scale=0.18]{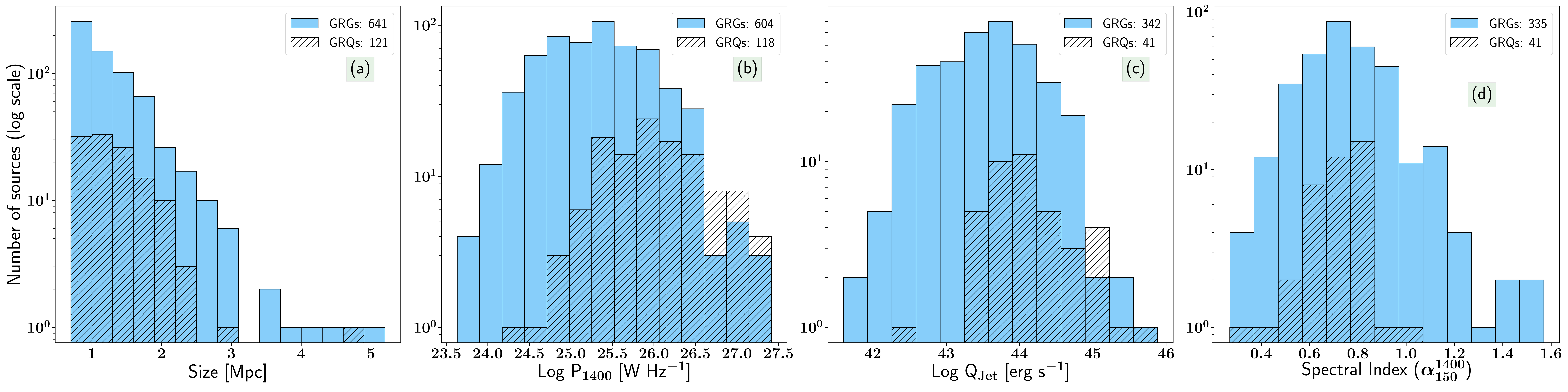}
\caption{Distributions of the unmatched sample of GRGs and GRQs with their different properties, represented in not-hatched and hatched bins, respectively, in the redshift range of 0.01 < $z$ < 1.0. The mean and median values  of the distributions are given in Table\ \ref{mean}. Panel a: Size distribution. Panel b: Distribution of the radio power at 1400 MHz (P$_{\rm 1400}$).  Panel c: Distribution of the jet kinetic power (Q$\rm _{Jet}$).  Panel d: Spectral index ($\rm \alpha_{150}^{1400}$) distribution.}
\label{fig:GQ}
\end{figure*}

\begin{figure*}
\centering
\includegraphics[scale=0.18]{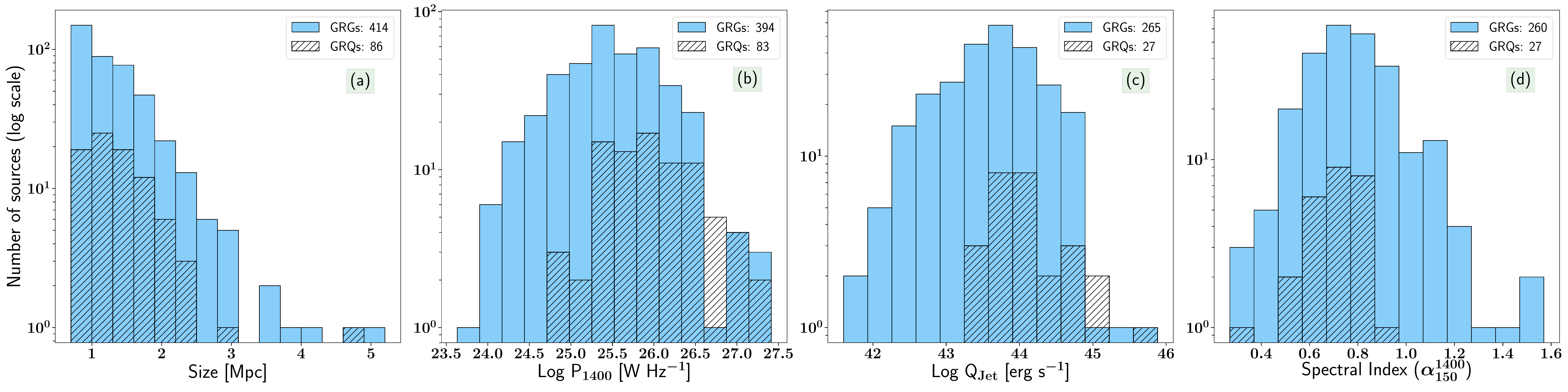}
\caption{Redshift-matched sample of GRG and GRQs in the redshift range of 0.2 < $z$ < 0.8. The description is same as for Fig.\ \ref{fig:GQ}.}
\label{fig:GQzmatched}
\end{figure*}

Based on the availability of data from the NVSS, we were able to estimate P$_{\rm 1400}$ for 604 GRGs and 118 GRQs from the \texttt{GRG catalogue}.
The distributions of P$_{\rm 1400 }$ of GRGs and GRQs are shown in Fig.\ \ref{fig:GQ} (b) and Fig.\ \ref{fig:GQzmatched} (b) for not matched and matched samples, respectively, where we observe that GRQs have a higher radio power than GRGs at 1400 MHz. This is well supported by the K-S test (the p-value for the matched sample is 1.20 $\times$ 10$^{-7}$) and the WMW test (the p-value for the matched sample is 7.88 $\times$ 10$^{-11}$) of the matched redshift samples, as shown in Table\ \ref{mean}, which strongly indicates that the two distributions are significantly different. Clearly, prevailing conditions in the central engines of GRQs  are able to produce more powerful jets, which results in more  radio-luminous sources than for GRGs. GRQs are found to have  a higher jet kinetic power than GRGs (as shown below), and they might host more massive black holes that accrete at a higher Eddington rate. 
Because our sample is restricted to sources with redshift lower than 1, we do not observe even more powerful GRQs, which lie mostly at higher redshifts.

\subsubsection{Distribution of the jet kinetic power ($\rm Q_{Jet}$)} 
The $\rm Q_{Jet}$ of the GRGs and GRQs was estimated using TGSS and the LoTSS, as explained in Sec.\ \ref{sec:jp}. Fig.\ \ref{fig:GQ} (c) and Fig.\ \ref{fig:GQzmatched} (c) show the histograms of $\rm Q_{Jet}$ of GRGs and GRQs. It it is evident that the GRQs have higher values of $\rm Q_{Jet}$, which is also supported by the statistics presented in Table\ \ref{mean}. From the inverse correlation between jet power and dynamical age, that is, Q$_{\rm Jet}$ $\propto$ 1/t$^{-2}_{\rm age}$ \citep{Ito}, it can be inferred that if their sizes are similar, more powerful radio jets of the GRQs would take less time in scaling Mpc distance than the GRGs (if they are placed in an environment of similar ambient density, which is not clear at this stage).
Using a sample of 14 GRGs, \citet{Ursini18xgrg} also reported that $\rm Q_{Jet}$ of GRGs lies in the range of $\sim$\,10$^{42}$ erg s$^{-1}$ to 10$^{44}$ erg s$^{-1}$. It has been observed \citep{mingo14} that some RGs have far higher $\rm Q_{Jet}$ than GRGs, which could be attributed to the severity of the radiative losses suffered by the GRGs over a period of their growth.
 
\citet{Ursini18xgrg}, based on their figure 3, hypothesised that GRGs like RGs at the start of their life have high nuclear luminosities as well as high $\rm Q_{Jet}$, which eventually fades over a period of time. Because their sample of GRGs was hard X-ray selected, it also shows high nuclear luminosities, and other GRG samples (such as our \texttt{GRG-catalogue} sample), which are radio selected, will occupy the lower luminosity part of Fig. 3 in  \citet{Ursini18xgrg}. Our findings based on the \texttt{GRG catalogue} support their hypothesis, as we mostly have radio-selected GRGs with $\rm L_{bol}$ in the range of $\sim$\,10$^{42}$ erg s$^{-1}$ to 10$^{46}$ erg s$^{-1}$, which extends below the $\rm L_{bol}$ range of the \citet{Ursini18xgrg} hard X-ray selected sample of 14 GRGs.

\subsubsection{ Distribution of the spectral index ($\rm \alpha_{150}^{1400}$)}\label{sec:SIdisc}
Fig.\ \ref{fig:GQ} (d) and Fig.\ \ref{fig:GQzmatched} (d) shows the histograms of the spectral indices of GRGs and GRQs. A similar criterion (as mentioned in Sec.\ \ref{sec:SI} and Sec.\ \ref{sec:SISGS}) of considering only those sources that are fully detected in TGSS (or LoTSS) and NVSS was followed for the sources in the \texttt{GRG catalogue}. The mean value, median value, and the p-values of the K-S and WMW tests (Table\ \ref{mean}) confirm that the GRGs and GRQs have the same distribution of the spectral index. This result is also consistent with the recent findings of \citet{PDLOTSS}.

\begin{figure}
\centering
\includegraphics[scale=0.31]{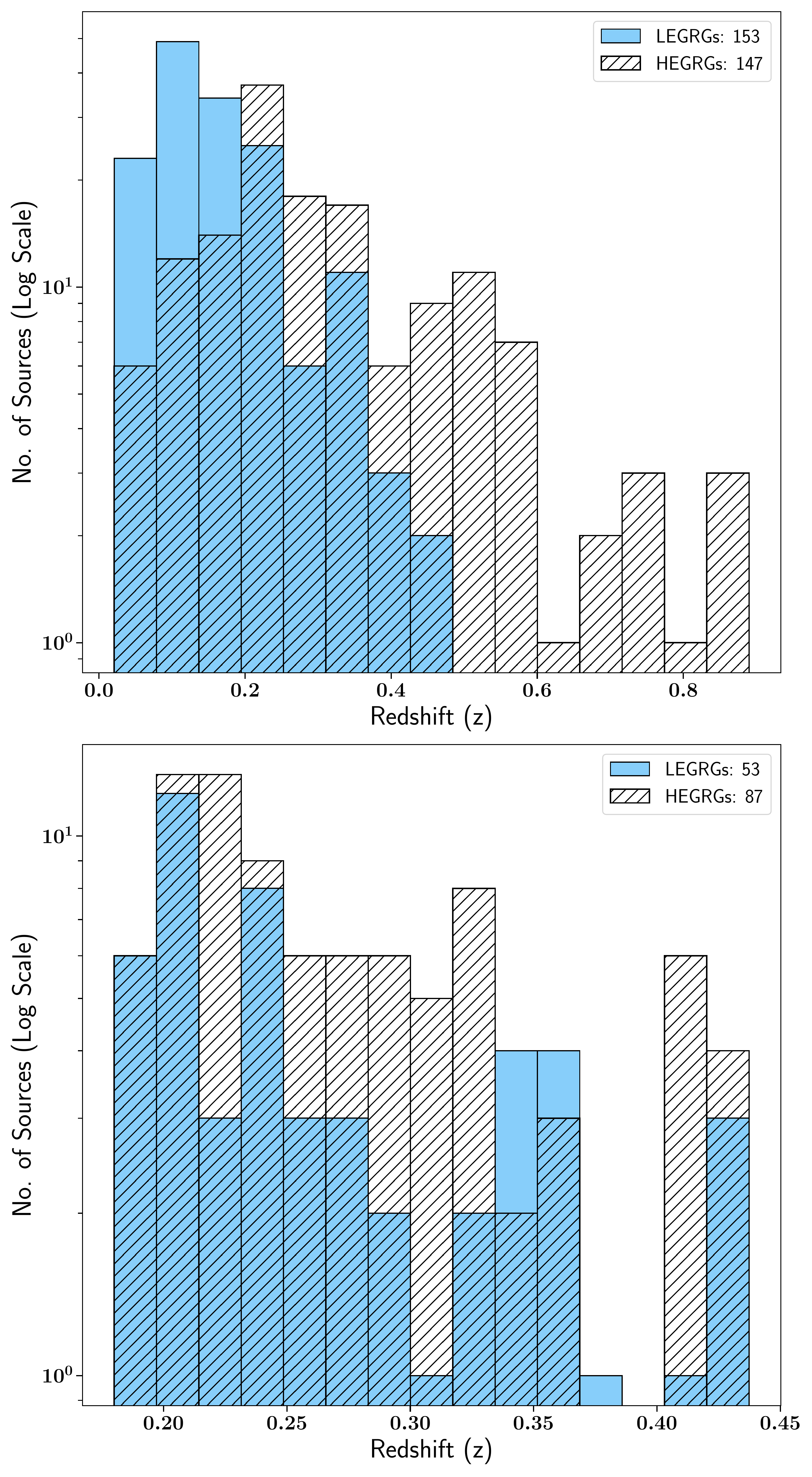}
\caption{The figure shows the redshift ($\rm z$) distributions of LEGRGs and HEGRGs for 0.02 < $z$ < 0.89 (unmatched redshift), represented in not-hatched and hatched bins, respectively in the upper plot. Similarly, in lower plot distributions are shown for matched redshift high-$z$: 0.18 < $z$ < 0.43 sample of LEGRGs and HEGRGs.}
\label{fig:HL_z}
\end{figure}

\subsection{Comparison of the properties of HEGRGs and LEGRGs}\label{sec:hl}
Based on the criteria mentioned in Sec.\ \ref{sec:wisean}, we classified the GRGs into high- and low-excitation, or HEGRGs and LEGRGs, types. When we consider their respective RG counterparts, called LERGs and HERGs, the LERGs are the dominant population as compared to HERGs. About 12$\%$ of the sample of \citet{koziel11} are HERGs, and is similar for \citet{bh12rgs}. As mentioned previously in Sec.\ \ref{sec:wisean}, in order to have a clean sample, we did not consider LERGs or LEGRGs from region III as they are contaminated with possible star formation. Therefore, our LEGRG sample is drastically reduced to 153 because we only considered sources from region II to be LEGRGs. Region I includes 147 HERGs and therefore makes our comparison samples almost equal with each other. In the following subsections, we individually describe the distribution of HEGRGs and LEGRGs in terms of various properties.

The redshift distributions of the HEGRGs and LEGRGs are shown in Fig.\ \ref{fig:HL_z} (upper panel). Their redshifts lie in the range 0.02 < $z$ < 0.89. We note that the redshift distributions are not the same, and we therefore considered two sub-samples matched in redshifts, whose details are described below.\\
Low-$z$: The first sub-sample consists of 27 HEGRGs and 93 LEGRGs in the matched lower redshift range of $\sim$\,0.04 to 0.18. The HEGRGs and LEGRGs have similar mean (LEGRGs: 0.12, and HEGRGs: 0.11) and median (LEGRGs: 0.11, and HEGRGs: 0.10) redshift values as well as the same  distribution (K-S test p-value: 0.52 and WMW test p-value: 0.19). \\
High-$z$: As shown in Fig.\ \ref{fig:HL_z} (lower), the second sub-sample consists of 87 HEGRGs and 53 LEGRGs in the matched higher redshift range $\sim$\,0.18 to 0.43. They have similar mean (LEGRGs: 0.26, and HEGRGs: 0.27) and median (LEGRGs: 0.24, and HEGRGs: 0.25) redshift values as well as the same  distribution (K-S test p-value: 0.47 and WMW test p-value: 0.23).

All three samples of HEGRGs and LEGRGs, that is, the total sample, high-$z$, and low-$z$ , give similar results in the comparison of their different properties. Therefore, below, we present the results of the total sample and of the high-$z$ sample of HEGRGs and LEGRGs.
Fig.\ \ref{fig:HL} and Fig.\ \ref{fig:HLzzmatchhigh} show that for each property, the number of sources vary because it is dependent on the availability of data from public archives such as the SDSS, NED, and Vizier.

\subsubsection{Distribution of the projected linear size}
The projected linear size distributions of LEGRGs and HEGRGs from the \texttt{GRG catalogue} for unmatched and matched redshift samples are shown in Fig.\ \ref{fig:HL} (a) and in Fig.\ \ref{fig:HLzzmatchhigh} (a), respectively.
Both classes seem to follow different distributions, as indicated by the p-values of the K-S and WMW tests, as shown in Table.\ \ref{mean}. Our data suggest that HEGRGs tend to grow to larger sizes than the LEGRGs, as seen from their mean and median sizes presented in Table\ \ref{mean} for both samples. In other words, the radiatively efficient nature of HEGRGs helps GRGs to grow to larger sizes.

Similarly, when we take a sample of RGs ($\sim$\,400 RGs; \citealt{koziel11}) for which we have size estimates along with a high- and low-excitation classification (from \textit{WISE}), we find a similar result: the sizes of high-excitation sources (HERGs) are larger than those of low-excitation sources (LERGs). This particular property is clearly similar for RGs and GRGs and is independent of the overall size of the source.

\subsubsection{Distribution of the radio power (P$_{\rm 1400}$)}
Fig.\ \ref{fig:HL} (b) and Fig.\ \ref{fig:HLzzmatchhigh} (b) show P$_{\rm 1400}$ distributions of LEGRGs and  HEGRGs for unmatched and matched redshift samples, respectively. Based on these, we infer that the HEGRGs spread over a larger range of P$_{\rm 1400}$ than LEGRGs. The distributions are different, as is strongly indicated by the p-values of the K-S and WMW tests. Their distributions also overlap significantly, with HEGRGs having higher P$_{\rm 1400}$ than LEGRGs, as shown in Table.\ \ref{mean}. This result complements the fact that HEGRGs are in a high-accretion state, resulting in the high radio power of GRGs. \citet{koziel11} also found that HERGs, like HEGRGs, which are radiatively efficient, have a higher radio power at 1400 MHz.

\subsubsection{Distribution of the jet kinetic power ($\rm Q_{Jet}$) }
\begin{figure*}
\centering
\includegraphics[scale=0.24]{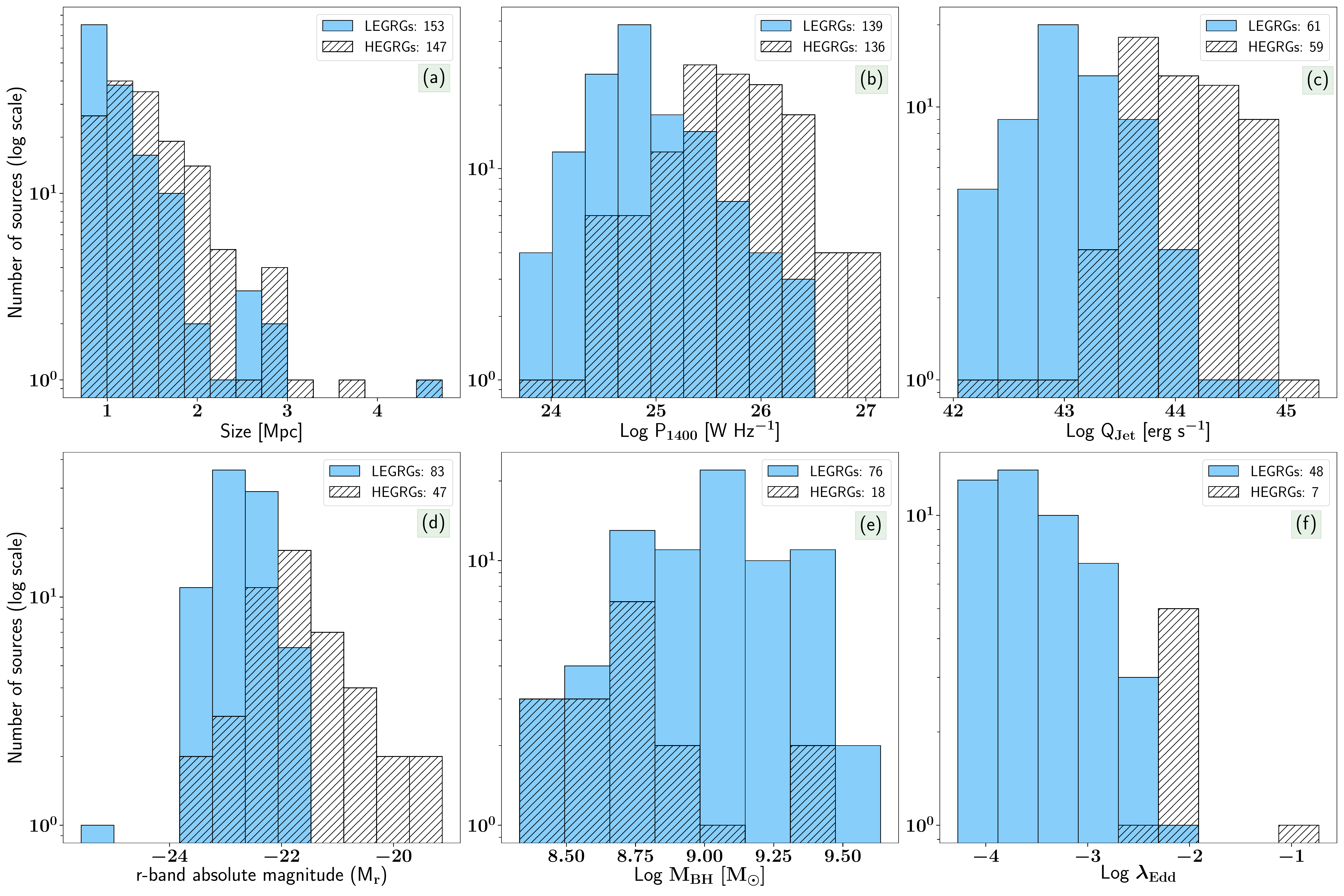}
\caption{ Distributions of LEGRGs and HEGRGs according to their various physical properties. They are represented in not-hatched and hatched bins, respectively, in the redshift range of 0.01 < $z$ < 1.0. The mean and median values of the distributions are given in Table\ \ref{mean}. Panel a: Distribution of the physical size (Mpc). Panel b: Distribution of the radio power at 1400 MHz (P$_{\rm 1400}$). Panel c: Distribution of the jet kinetic power (Q$_{\rm Jet}$). Panel d: Distribution of the absolute r-band magnitude (M$_{\rm r}$). Panel e: Distribution of the black hole mass (M$_{\rm BH}$). Panel f: Distribution of the Eddington ratio ($\lambdaup_{\rm Edd}$).}
\label{fig:HL}
\end{figure*}

\begin{figure*}
\centering
\includegraphics[scale=0.24]{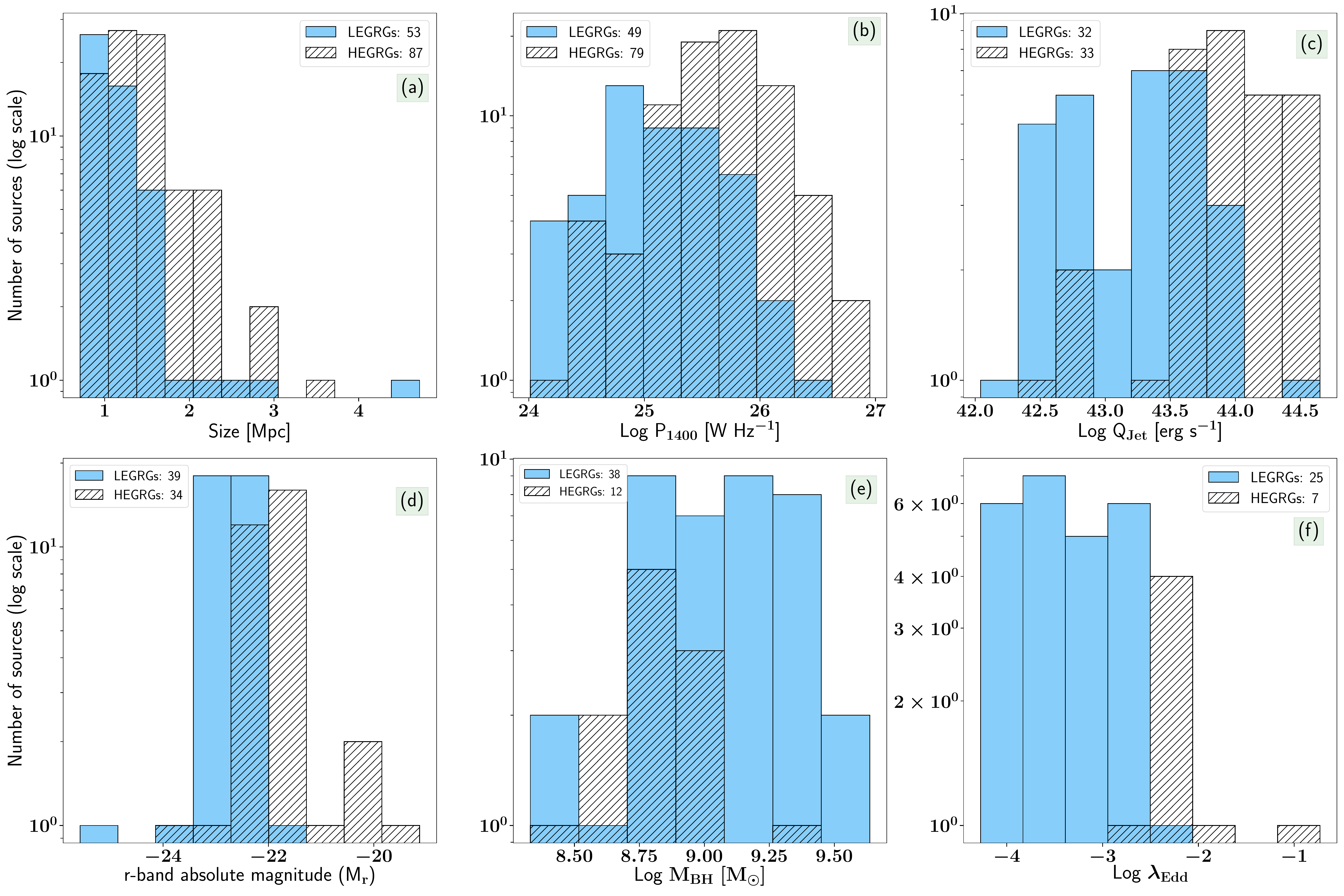}
\caption{Redshift-matched samples of LEGRGs and HEGRGs in the range 0.18$<z<$0.43. The description is same as for Fig.\ \ref{fig:HL}. }
\label{fig:HLzzmatchhigh}
\end{figure*}

The $\rm Q_{Jet}$ of both the populations varies over a wide range, with HEGRGs occupying the higher end of the distributions, as shown in Fig.\ \ref{fig:HL} (c) and Fig.\ \ref{fig:HLzzmatchhigh} (c). The difference in $\rm Q_{Jet}$ is statistically significant; the median $\rm Q_{Jet} \sim 10^{44} erg s^{-1}$ of HEGRGs is about ten times higher than that of LEGRGs (Table\ \ref{mean}).  
The high mean and median values of HEGRGs (as shown in Table.\ \ref{mean}) compared to LEGRGs strongly indicate that the AGNs of GRGs with high-excitation type are responsible for launching higher 
powered jets, resulting in more radio-luminous sources (higher P$_{\rm 1400}$ as above).

 \subsubsection{Distribution of the absolute r-band magnitude (M$_{\rm r}$)}    \label{sec:rbanddis}
The distributions of M$_{\rm r}$ in unmatched and matched redshift bins of LEGRGs and  HEGRGs are presented in Fig.\ \ref{fig:HL} (d) and Fig.\ \ref{fig:HLzzmatchhigh} (d). These figures show two distinct populations, and this is also supported by the low p-value of the K-S and WMW tests (Table\ \ref{mean}). The LEGRGs are found to be optically brighter by $\sim$\,1 magnitude than HEGRGs, indicating that LEGRGs are mostly hosted by bright giant elliptical galaxies comprising an old stellar population, which is quite prominently visible from their absolute r-band  magnitudes. This is consistent with the findings of \citet{hardcastle07}, who have shown this for LERGs.
 
Overall, Fig.\ \ref{fig:HL} (d) and  Fig.\ \ref{fig:HLzzmatchhigh} (d) show the M$_{\rm r}$ distribution of GRGs, which ranges from $\sim$\,$-$19 to $-$25 in brightness. A similar distribution has been observed for host galaxies of normal-sized RGs \citep{govoni00,sadler97,capetiifrii}.

\subsubsection{Distribution of the black hole mass (M$_{\rm BH}$)}
A study of the M$_{\rm BH}$ distribution in LEGRGs and HEGRGs is very crucial for understanding one of the key factors driving the two different accretion modes. For this study, our sample was restricted to 94 sources (unmatched redshift sample) based on the availability of the spectroscopic data from the SDSS. Fig.\ \ref{fig:HL} (e) (unmatched redshift sample) and Fig.\ \ref{fig:HLzzmatchhigh} (e) (matched-redshift sample) show the distributions of M$_{\rm BH}$ in these two classes, in which the mean and median (see Table.\ \ref{mean}) of LEGRGs are found to be greater than HEGRGs. The higher values of mean and median of the M$_{\rm BH}$ of LEGRGS compared to HEGRGs as well as the low p-value of the K-S and WMW tests implies that HEGRGs have lower M$_{\rm BH}$ than LEGRGs.

\subsubsection{ Distribution of the Eddington ratio ($\lambdaup_{\rm Edd}$)}
In this study, the sample was restricted to just 55 (unmatched redshift sample) because of the availability of reliable [OIII] line flux data from the SDSS, which is essential for estimating the L$_{\rm bol}$.
Out of this, a total of 48 GRGs are classified as LEGRG, and 7 are classified as HEGRG in the \texttt{GRG catalogue} based on their mid-IR properties. Nearly 87\% of the sample are LEGRGs, which is consistent with earlier studies \citep{hardcastlenat}, which showed that LERGs are the dominant population of radio galaxies. LERGs or LEGRGs are dominated by the radiatively inefficient accretion flow, or RIAF \citep{ynar14-adaf}. In Fig.\ \ref{fig:HL} (f) and Fig.\ \ref{fig:HLzzmatchhigh} (f) we observe the distribution of the $\lambdaup_{\rm Edd}$ , which ranges from $\sim$\,$ 10^{-4}$ to $10^{-2}$ for LEGRGs and from $10^{-2}$ to $10^{-1}$ for HEGRGs, with a little overlap. The lower $\lambdaup_{\rm Edd}$ values of (Table\ \ref{mean}) of LEGRGs implies a radiatively inefficient state governed by a low accretion rate. HEGRGs, in contrast to LEGRGs, are rarer and have a comparatively higher $\lambdaup_{\rm Edd}$ (Table\ \ref{mean}), indicating their radiatively efficient mode  and higher accretion rate. Our results for GRGs are similar to previous findings for RGs by \citet{koziel11}, \citet{smolic09}, and \citet{bh12rgs}, who have shown that $\lambdaup_{\rm Edd}$ is higher for HERGs than for LERGs. The $\lambdaup_{\rm Edd}$ of LEGRGs peaks around 10$^{-4}$, which is consistent with the findings presented in \citet{ho08}.  

\subsubsection{HEGRG-LEGRG comparison overview}
In the six subsections above we compared HEGRGs to LEGRGs by comparing several properties of LEGRGs and HEGRGs in the total sample as well as redshift-matched ones. The results are similar: (a) The HEGRGs tend to be larger, more luminous, and have a higher jet power than LEGRGs, (b) LEGRGs are optically brighter by $\sim 1$  magnitude  than HEGRGs, (c) LEGRGs have lower Eddington ratios, harbour higher mass black holes, and launch lower power jets (by a factor of 10) than HEGRGs.
This implies  that  higher mass black holes grow in the nucleus of  optically brighter and more massive galaxies, whose central engines are presently  found in  low-excitation, radiatively inefficient, low mass accretion state but are capable of producing sufficiently powerful FR-II jets that result in Mpc-scale GRGs, which  clearly  constitute the  vast majority  ($> 80 \%$) of  all GRGs selected in our study. Although it is beyond the scope of this work, it will be useful in future when the  detailed properties of LERGs are compared with LEGRGs and HERGs with HEGRGs in order to gain  much  deeper insight into the inner workings of the GRG central engine.

\subsection{How similar are GRGs and RGs ?}\label{sec:rggrg}
The following studies were carried out to understand the factors that cause a very small population of RGs to transform into GRGs. We carried out this investigation by studying three of their 
properties:  $\rm \alpha$, M$_{\rm BH}$ , and $\lambdaup_{\rm Edd}$, in this section. Possible differences in black hole spin and their implications are discussed in Sec.\ \ref{sec:spinres}.

\onecolumn
\setlength{\tabcolsep}{3pt}
\begin{landscape}
\begin{longtable}{l c c c c c c c c}
\captionsetup{width=8.8in}
\caption{Mean and median values of the properties of the sources in the \texttt{GRG catalogue} for comparison between their sub-types like GRG-GRQ, HEGRG-LEGRG, and GRG-RG. The p-values correspond to K-S and WMW tests for the respective distributions. The GRG-GRQ comparison of properties is shown in Fig.\ \ref{fig:GQ} for unmatched redshift sample and Fig.\ \ref{fig:GQzmatched} for the  matched-redshift sample. Fig.\ \ref{fig:HL} shows the distribution of HEGRGs and LEGRGs for unmatched redshift sample and Fig.\ \ref{fig:HLzzmatchhigh} shows the distribution for matched redshift sample.} \\
\hline
Property & No of sources &  Mean & Median & No of sources & Mean &  Median &  K-S test  & WMW test  \\ 
 & & & & & & & p-value & p-value \\
\hline
 [0.01 < $z$ < 1 (unmatched)] & \multicolumn{3}{c}{GRG} &   \multicolumn{3}{c}{GRQ}   &  \\ \hline

Size [Mpc]& 641 & 1.26 & 1.09 & 121 &  1.35 & 1.23 & 7.01 $\times$ 10$^{-3}$ & 2.31 $\times$ 10$^{-3}$  \\ 
P$_{\rm 1400}$ [W Hz$^{-1}$] & 604 & 8.53 $\times$ 10$^{25}$ & 2.09 $\times$ 10$^{25}$ &  118  & 26.22 $\times$ 10$^{25}$ & 9.09 $\times$ 10$^{25}$& 1.23 $\times$ 10$^{-16}$ & 8.23 $\times$ 10$^{-22}$  \\ 
Q$_{\rm Jet}$ [erg s$^{-1}$] & 342 & 13.23 $\times$ 10$^{43}$ & 4.03 $\times$ 10$^{43}$ & 41 & 37.85 $\times$ 10$^{43}$ & 12.30 $\times$ 10$^{43}$ & 2.98 $\times$ 10$^{-5}$  & 7.46 $\times$ 10$^{-7}$ \\ 
$\rm \alpha^{\rm 1400}_{\rm 150}$ & 335 & 0.75 & 0.73 & 41 & 0.72 & 0.75 & 0.14  & 0.43  \\ 
\hline

[0.2 < $z$ < 0.8 (matched)] & \multicolumn{3}{c}{GRG} &   \multicolumn{3}{c}{GRQ}   &  \\ \hline
Size [Mpc]& 414 & 1.32 & 1.20 & 86 &  1.40 & 1.29 & 2.28 $\times$ 10$^{-2}$  & 2.14 $\times$ 10$^{-2}$ \\ 
P$_{\rm 1400}$ [W Hz$^{-1}$] & 394 & 11.19 $\times$ 10$^{25}$ & 3.04 $\times$ 10$^{25}$ &  83  & 21.98 $\times$ 10$^{25}$ & 8.72 $\times$ 10$^{25}$& 1.20 $\times$ 10$^{-7}$ & 7.88 $\times$ 10$^{-11}$ \\ 
Q$_{\rm Jet}$ [erg s$^{-1}$] & 265 & 14.38 $\times$ 10$^{43}$ & 4.60 $\times$ 10$^{43}$ & 27 & 36.52 $\times$ 10$^{43}$ & 12.20 $\times$ 10$^{43}$ & 5.18 $\times$ 10$^{-4}$ & 1.02 $\times$ 10$^{-4}$ \\ 
$\rm \alpha^{\rm 1400}_{\rm 150}$ & 260 & 0.77 & 0.75 & 27 & 0.70 & 0.73 & 0.14  & 0.06\\ 
\hline

[0.02 < $z$ < 0.89 (unmatched)] &\multicolumn{3}{c}{HEGRG} &\multicolumn{3}{c}{LEGRG}  &  \\ \hline
Size [Mpc] & 147 & 1.45 & 1.34 & 153 & 1.12 & 0.97 & 5.46 $\times$ 10$^{-11}$ & 1.11 $\times$ 10$^{-11}$ \\ 
P$_{\rm 1400}$ [W Hz$^{-1}$] & 136 & 11.29 $\times$ 10$^{25}$ & 4.94 $\times$ 10$^{25}$ & 139 & 1.68 $\times$ 10$^{25}$ & 0.65 $\times$ 10$^{25}$& 2.56 $\times$ 10$^{-24}$ & 1.50 $\times$ 10$^{-24}$ \\ 
Q$_{\rm Jet}$ [erg s$^{-1}$] & 59 & 20.66 $\times$ 10$^{43}$ & 10.20 $\times$ 10$^{43}$ & 61 & 2.70 $\times$ 10$^{43}$ & 0.95 $\times$ 10$^{43}$ & 3.71 $\times$ 10$^{-15}$ & 4.01 $\times$ 10$^{-14}$\\ 
M$_{\rm r}$ & 47 & -21.67 & -21.84 & 83 & -22.75 & -22.76 & 1.20  $\times$ 10$^{-11}$ & 4.74 $\times$ 10$^{-13}$\\
M$_{\rm BH}$ [M$_{\odot}$] & 18 & 0.72 $\times$ 10$^{9}$ & 0.52 $\times$ 10$^{9}$ & 76 & 1.26 $\times$ 10$^{9}$ & 1.11 $\times$ 10$^{9}$ & 2.24 $\times$ 10$^{-4}$ & 1.20 $\times$ 10$^{-4}$ \\ 
$\lambdaup_{\rm Edd}$ & 7 & 3.25 $\times$ 10$^{-2}$  & 8.15 $\times$ 10$^{-3}$& 48 & 6.75 $\times$ 10$^{-4}$  & 2.86 $\times$ 10$^{-4}$ & 5.72 $\times$ 10$^{-6}$ & 1.82 $\times$ 10$^{-5}$\\ 
 
\hline

[0.18 < $z$ < 0.43 (matched)] &\multicolumn{3}{c}{HEGRG} &\multicolumn{3}{c}{LEGRG}  &  \\ \hline
Size [Mpc] & 87 & 1.42 & 1.35 & 53 & 1.21 & 1.05 & 3.12 $\times$ 10$^{-3}$ & 2.34 $\times$ 10$^{-4}$ \\ 
P$_{\rm 1400}$ [W Hz$^{-1}$] & 79 & 8.26 $\times$ 10$^{25}$ & 4.51 $\times$ 10$^{25}$ & 49 & 2.79 $\times$ 10$^{25}$ & 1.22 $\times$ 10$^{25}$& 6.83 $\times$ 10$^{-6}$  & 6.31 $\times$ 10$^{-7}$\\ 
Q$_{\rm Jet}$ [erg s$^{-1}$] & 33 & 12.37 $\times$ 10$^{43}$ & 8.80 $\times$ 10$^{43}$ & 32 & 3.67 $\times$ 10$^{43}$ & 2.15 $\times$ 10$^{43}$ & 1.71 $\times$ 10$^{-6}$ & 1.19 $\times$ 10$^{-6}$ \\ 
M$_{\rm r}$ & 34 & -21.84 & -21.91 & 39 & -22.71 & -22.73 & 2.91  $\times$ 10$^{-7}$ & 1.52 $\times$ 10$^{-7}$ \\
M$_{\rm BH}$ [M$_{\odot}$] & 12 & 0.74 $\times$ 10$^{9}$ & 0.54 $\times$ 10$^{9}$ & 38 & 1.41 $\times$ 10$^{9}$ & 1.20 $\times$ 10$^{9}$ & 8.98 $\times$ 10$^{-3}$ & 1.57 $\times$ 10$^{-3}$  \\ 
$\lambdaup_{\rm Edd}$ & 7 & 3.25 $\times$ 10$^{-2}$  & 8.15 $\times$ 10$^{-3}$& 25 & 9.85 $\times$ 10$^{-4}$  & 3.11 $\times$ 10$^{-4}$ & 4.85 $\times$ 10$^{-5}$ & 7.73 $\times$ 10$^{-5}$ \\ 
 
\hline

(Matched) & \multicolumn{3}{c}{GRG}  & \multicolumn{3}{c}{RG}   &  \\ \hline
M$_{\rm BH}$ [M$_{\odot}$] & 164 & 1.04 $\times$ 10$^{9}$ & 0.84  $\times$ 10$^{9}$ & 14543 & 1.10 $\times$ 10$^{9}$ & 0.84  $\times$ 10$^{9}$ & 0.86 & 0.46 \\ 
$\lambdaup_{\rm Edd}$ & 99 & 40.5  $\times$ 10$^{-4}$ & 5.41 $\times$ 10$^{-4}$ & 12998 & 60.9 $\times$ 10$^{-4}$ & 25.9 $\times$ 10$^{-4}$ & 2.78 $\times$ 10$^{-14}$ & 1.38 $\times$ 10$^{-14}$ \\ \hline
\label{mean}
\end{longtable}
\end{landscape}
\twocolumn

\subsubsection{Spectral index ($\rm \alpha$)}
Several radio spectral index studies of RGs \citep{kpw69,oort88,gruppioni97,kapahi98,Sirothia09,ishwar10,mahony16} have established the mean spectral index value to be $\sim$\,0.75. Until recently, it was believed that GRGs have steeper spectral index values. However, \citet{PDLOTSS}, using their large sample of GRGs, found that the spectral index of RGs and GRGs is very similar. Now, with our \texttt{GRG catalogue} (which comprises of LoTSS and SGS), we establish this result strongly with a larger (> 250) sample size, as shown in Table\ \ref{mean}: the mean $\rm \alpha$ for GRGs and GRQs is $\sim$\,0.75 and $\sim$\,0.72, respectively, for the unmatched redshift sample. For the matched-redshift sample, the mean $\rm \alpha$ for GRGs and GRQs is 0.77 and 0.70, respectively. These values are similar to those of the RGs and RQs.

\subsubsection{Black hole mass (M$_{\rm BH}$)}
Here, in the first point we describe how we derived the redshift matched RG sample for comparison along with its properties. In the second point we describe our sub-sample of GRGs derived from the \texttt{GRG catalogue}. Lastly, in the third point we present the result of our comparative study.
\begin{enumerate}
    \item RG sample.
    In order to compare the M$_{\rm BH}$ of GRGs and RGs, we used the RG sample from \citet{bh12rgs}. This RG sample was created using the SDSS DR7 \citep{sdssdr7} optical spectroscopic data along  with the FIRST and NVSS radio data. This sample has been filtered from star-forming galaxies (SFGs) and only consists of radio-loud AGNs (RLAGNs) or RGs. Furthermore, we used their stellar velocity dispersion ($\sigma$) information to compute M$_{\rm BH}$ from the SDSS DR14 database \citep{sdssdr14} through CasJobs\footnote{\url{http://skyserver.sdss.org/CasJobs/}}, and retained only galaxies with reliable spectra, that is, a redshift warning flag, ZWarning = 0, from the SDSS DR14 database. Next, only sources with
    $\sigma$ in the range of 80 $<$ $\sigma$ $<$ 420 km $s^{-1}$ were selected based on the criterion provided by \citet{boltonsdss12}.
    Lastly, in order to have a more clean and robust sample, we also imposed an additional filter to the  sources in the sample to have a $\sigma$ error smaller than 30$\%$. 
    The redshift range of the RG sample was restricted to the range of 0.034 to 0.534 to match the redshift range of the GRG sample (from the \texttt{GRG catalogue}) used for this study (see below), and radio-loud quasars were not considered for this study.
    This reduced the original RG sample of \citet{bh12rgs} to 14543. We call this sample as BH12 in the remainder of the paper.

   \item The GRG sample.  
   In our SGS, only 46 GRGs have spectroscopic data from the SDSS, and only these therefore have M$_{\rm BH}$ estimates derived from the M$_{\rm BH}$-$\sigma$ relation. In order to carry out a comparative study of M$_{\rm BH}$ that is statistically significant, we make use of the \texttt{GRG catalogue}. Out of 820 GRGs in the \texttt{GRG catalogue} (including the SGS), 164 have clean, reliable $\sigma$ information from the SDSS. For the purpose of the M$_{\rm BH}$ RG-GRG comparison study, our sample of GRGs was therefore restricted to 164.

   \item Distribution of the black hole mass. Fig.\ \ref{MBH_ERHIST} (a) shows that the distributions of RGs and GRGs are largely similar. 
   The spread of RGs is wider at the lower end of the plot than for GRGs, but both distributions peak at a similar value and have similar mean and median values (GRG: mean = 1.04 $\times$ 10$^{9}$ M$_{\odot}$, and median = 0.84 $\times$ 10$^{9}$ M$_{\odot}$; RG: mean = 1.10 $\times$ 10$^{9}$ M$_{\odot}$, and median = 0.84 $\times$ 10$^{9}$ M$_{\odot}$). This establishes that both RGs and GRGs have the same M$_{\rm BH}$ distribution, and it is also supported by the K-S and WMW tests, where the p-values are 0.86 and 0.46, respectively.

\end{enumerate}

\subsubsection{Eddington ratio ($\lambdaup_{\rm Edd}$)} \label{ER_GRG_RG}
\begin{figure*}[h]
\includegraphics[scale=0.45]{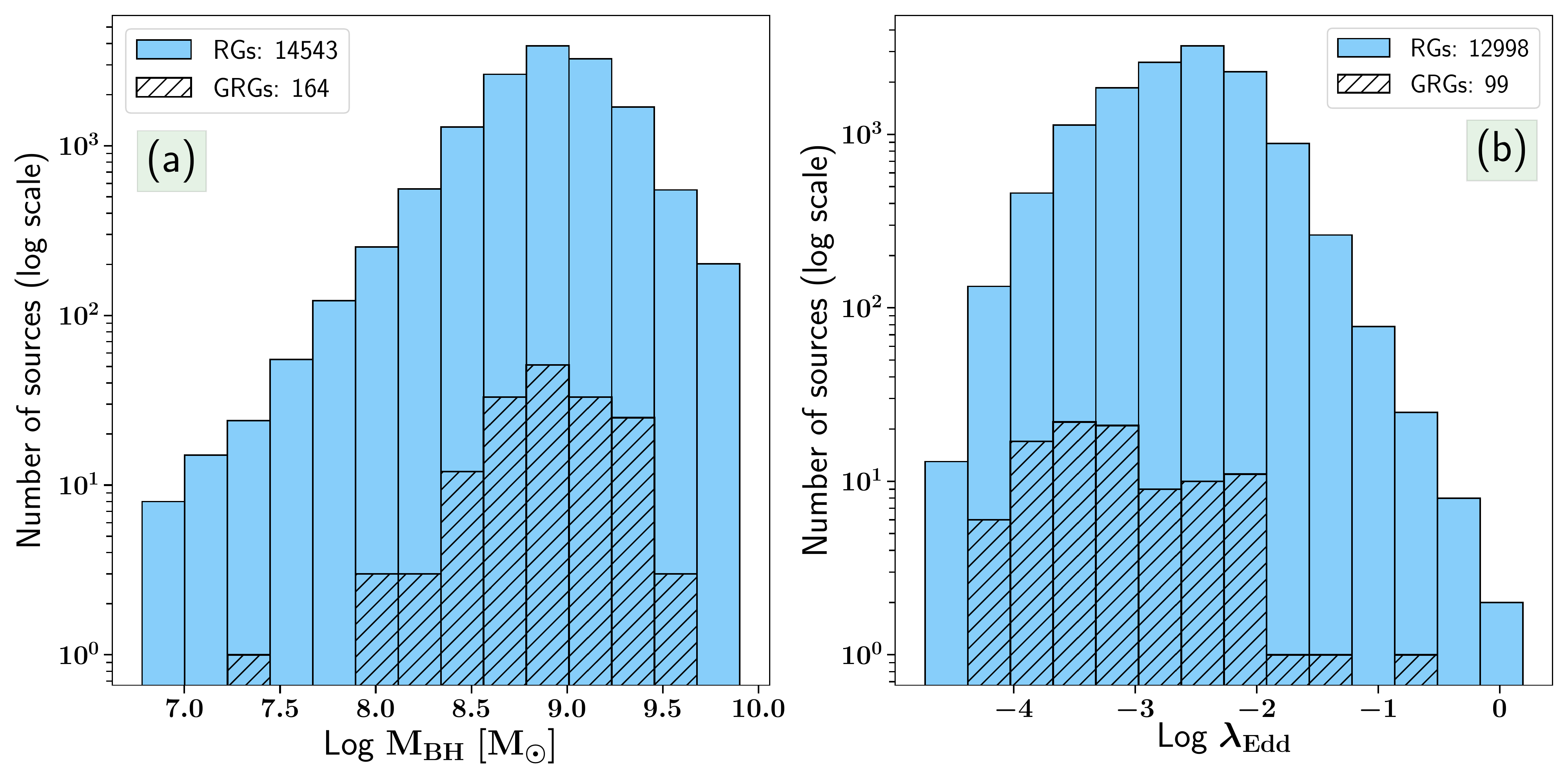} 
\centering
\caption{Distribution of black hole mass (M$_{\rm BH}$) in panel (a) and Eddington ratio ($\lambdaup_{\rm Edd}$) in panel (b) of RGs (not-hatched bins) and GRGs (hatched bins). The RG sample has been derived from \citet{bh12rgs} and the GRG sample is from the \texttt{GRG catalogue} created by us. }
\label{MBH_ERHIST}
\end{figure*}

Based on the availability of the [OIII] line flux information from the SDSS, we have the estimates of $\lambdaup_{\rm Edd}$ for 99 GRGs from the \texttt{GRG catalogue} and 12998 RGs from BH12, whose distribution is shown in Fig.\ \ref{MBH_ERHIST} (b). Based on our analysis of the available data, it appears that the RGs and GRGs have different mean and median values (Table.\ \ref{mean}). The plot shows that the distributions are different, which is statistically evident from the p-value of 2.78 $\times$ 10$^{-14}$ from K-S test and from the p-value of 1.38 $\times$ 10$^{-14}$ from the WMW test. Thus, our data indicate that $\lambdaup_{\rm Edd}$ of RGs is higher than that of GRGs. We can further extrapolate and conjecture that GRGs mostly have RIAF as compared to RGs, as almost all the GRGs have $\lambdaup_{\rm Edd}$ < 0.01 (which is the RIAF regime).

\subsection{Astrophysical processes near accreting black holes in GRGs} \label{sec:bhprop}
In this section we discuss the black hole properties of GRGs based on our estimates of spin, Eddington ratio and jet kinetic power. We also present the possible astrophysical scenarios leading to the observed properties of GRGs.
\subsubsection{Black hole spin ($a$)} \label{sec:spinres}
\begin{figure}[h]
\includegraphics[scale=0.325]{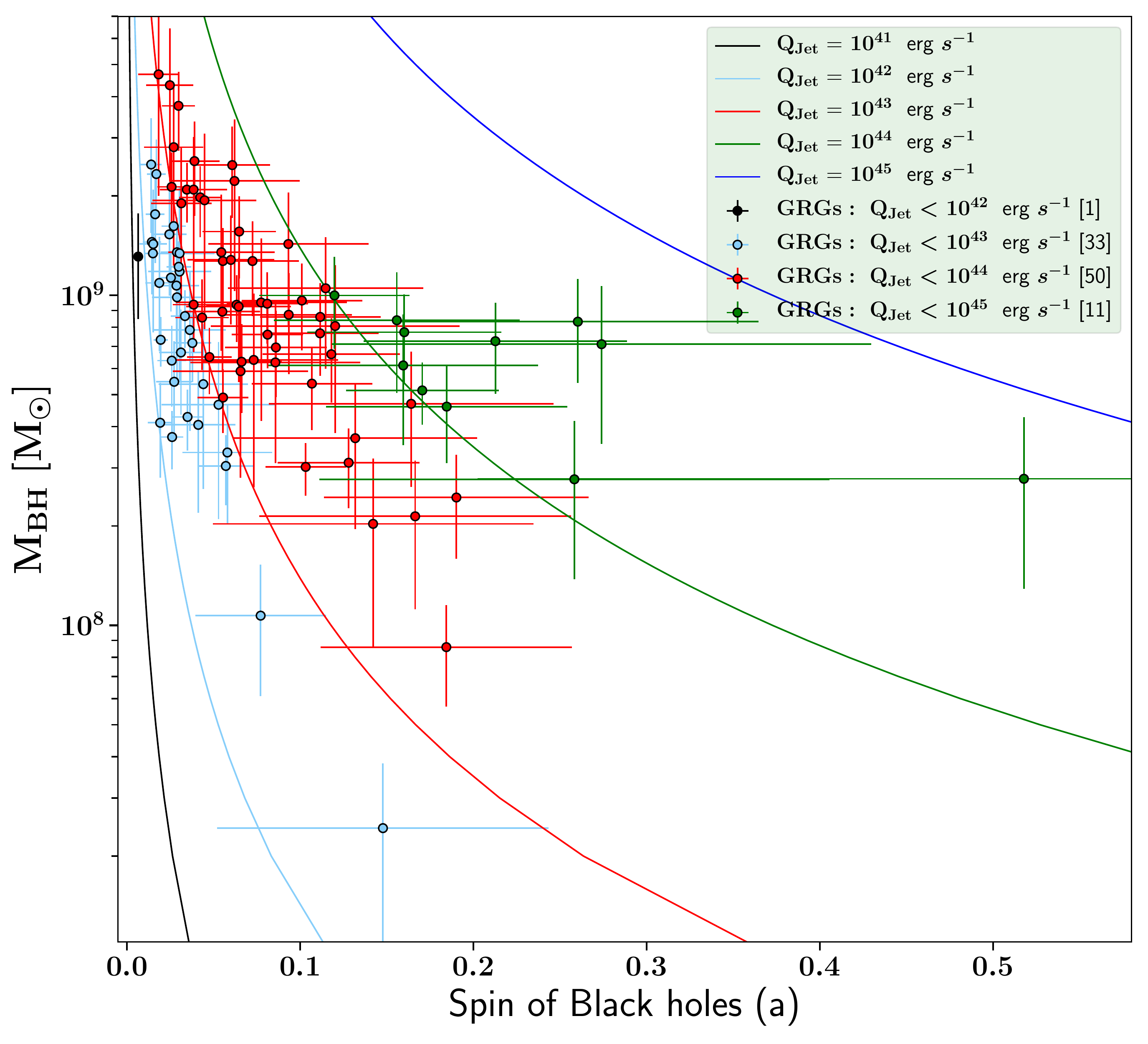} 
\caption{Simulation of black hole spin (using Eq.\ \ref{eq:bhspindisc}) as a function black hole mass over various ranges of jet kinetic power. Coloured points represent estimated values from GRG observations. The solid curved lines with their respective colours indicate the different values of spin attainable by a black hole for a mass range of 10$^{7}$ M$_{\odot}$ to 10$^{9}$ M$_{\odot}$ at that particular jet kinetic power. The solid circles with different colours indicate the GRGs in our sample with their estimated spin values lying in the range of jet kinetic power specified. The total number of GRGs is 95. Figures mentioned in the square brackets in the legend of the plot indicate number of GRGs in respective  kinetic jet power bins. The magnetic field used to compute the black hole spin ($a$) was assumed to be B$_{\rm Edd}$, which is derived from the M$_{\rm BH}$.}
\label{fig:spin}
\end{figure}

The spin of the accreting black hole has been identified as one of the key ingredients for powering the relativistic jets in AGN for a timescale of millions of years or more. Spin is coupled with the strength and geometry of the magnetic field, and by
rearranging equation \ref{eq:bhspin}, we obtain
\begin{equation}\label{eq:bhspindisc}
\rm \textit{a} \propto \frac{\rm \sqrt{\rm Q_{Jet}}} {\rm B \times M_{BH}}
.\end{equation}
The spin ($a$) of the black hole of GRGs was estimated using our M$_{\rm BH}$ and Q$_{\rm jet}$ estimates, as explained in Sec.\ \ref{sec:spininfo} and given in Table.\ \ref{tab:spin}. We emphasise here that these are crude estimates based on certain assumptions.
We rejected sources with an error of 60\% and higher in their M$_{\rm BH}$ estimates, and our sample (\texttt{GRG catalogue}) was therefore restricted to 95 sources for this particular study, as listed in Table \ref{tab:spin}.
Fig.\ \ref{fig:spin} shows the distribution of black hole spin with respect to M$_{\rm BH}$ varying over two orders of magnitude, that is, 10$^{7}$ to 10$^{9}$ M$_{\odot}$. The spin range spans from 0.007 to 0.518, resulting in mean and median values of 0.079 and 0.055, respectively, for GRGs in our sample. We also plot iso-tracks with their respective colours, which indicate the constant jet kinetic powers obtained for various ranges of spin, M$_{\rm BH}$ , and magnetic field strength. 

Fig.\ \ref{fig:spin} shows distribution of spin as a function of black hole mass in various bins of Q$_{\rm Jet}$ from 10$^{41}$ erg s$^{-1}$ to 10$^{45}$ erg s$^{-1}$, which we simulated using equations \ref{eq:bhspin} and \ref{eq:bedd}. With the plausible values of black hole mass, magnetic field, and spin, clearly no GRG is observed to acquire a Q$_{\rm Jet}$ $\geq$ 10$^{45}$ erg s$^{-1}$. For the objects with a low value of Q$_{\rm Jet}$ ($\sim$\,10$^{41}$ erg s$^{-1}$), even a drastic change in the magnetic field or M$_{\rm BH}$ would lead to only a minute change in spin, whereas for the objects with higher Q$_{\rm Jet}$ ($\sim$\,10$^{44}$ erg s$^{-1}$) values, a small change in the magnetic field or M$_{\rm BH}$ would result in a significant change in spin values.
Based on our sample, we observe that the spin values of 99\% of GRGs are constrained within 0.3. For the source with the highest spin value of 0.518, a correct combination of lower mass and high jet power as compared to the rest of the sources served as a favourable condition. AGNs of both types, that is, with and without powerful jets, have SMBHs that spin rapidly, which in turn implies that spin alone is not the driving factor for the production of powerful relativistic jets. 
Our spin estimates are based on assuming the maximum possible magnetic field, that is, B$_{\rm Edd}$, which indicates that these are the lowest possible values. Below we discuss the implications of this result, which is a possible scenario for GRGs.

\citet{kingpringle06} reported that if a black hole has a series of accretion episodes whose principal angular momentum vectors are randomly orientated (a chaotic random-walk process), in which the black hole angular momentum dominates, then it leads to lower black hole spin values. However, the successive accretion flows accumulate the poloidal magnetic flux in the vicinity of the black hole, and this strong magnetic flux threading the black hole accretion disc might lead to the state of a magnetically arrested disc (MAD), which affects the dynamical accretion process onto the black hole itself \citep{Narayan,Tchekhovskoy}. This state disrupts the accretion flow by breaking the flow into small streams. The velocity of the stream decreases with respect to free-fall velocity as the material in the stream has to find its way to the black hole through magnetic interchange and reconnection, which results in lower spin values of the black hole. Similarly, the low spin value of GRGs can be explained when we invoke the state of MAD, or equivalently, a dynamically important magnetic field B$_{\rm dyn}$, which indicates that their accretion history was chaotic rather than a smooth, progressive process. The dynamically important magnetic field field near the black hole has been observationally shown to exist \citep{eatough13nat}.

Another possible scenario might be self-gravity, which fragments the material in the disc into small clouds with randomly oriented angular momentum \citep{Fanidakis}. This leads to randomly oriented numerous accretion episodes that collectively lower the spin of the black hole. The lower spin correlating with higher mass black holes (Fig.\ \ref{eq:bhspindisc}) suggests that SMBHs powering the GRGs were perhaps formed in episodes of quasi-isotropic chaotic accretion events.

Alternatively, when we assume a maximally spinning (Thorne limit: $a$ = 0.998) black hole \citep{Thorne1974} for our GRGs, then using our M$_{\rm BH}$ and Q$_{\rm Jet}$ estimates of a sample 95 GRGs, we can set a lower limit on their magnetic field strength (B or B$_{\rm dyn}$). Thus we obtain the values of B$_{\rm dyn}$ in the range of 1.1 $\times 10^{2} <$ B$_{\rm dyn}$ $<$ 1.8 $\times 10^{4}$  G for our GRGs with the mean value of 2.2 $\times$ 10$^{3}$ G. These limits on B match the limits given using very long baseline interferometry (VLBI) observations of RLAGN, which can probe the parsec to sub-parsec scale region around the black hole \citep{Zamani,vidal15,baczko16,hodgson17,boccardi17}.

\begin{figure*}[h]
\centering
\includegraphics[scale=0.45]{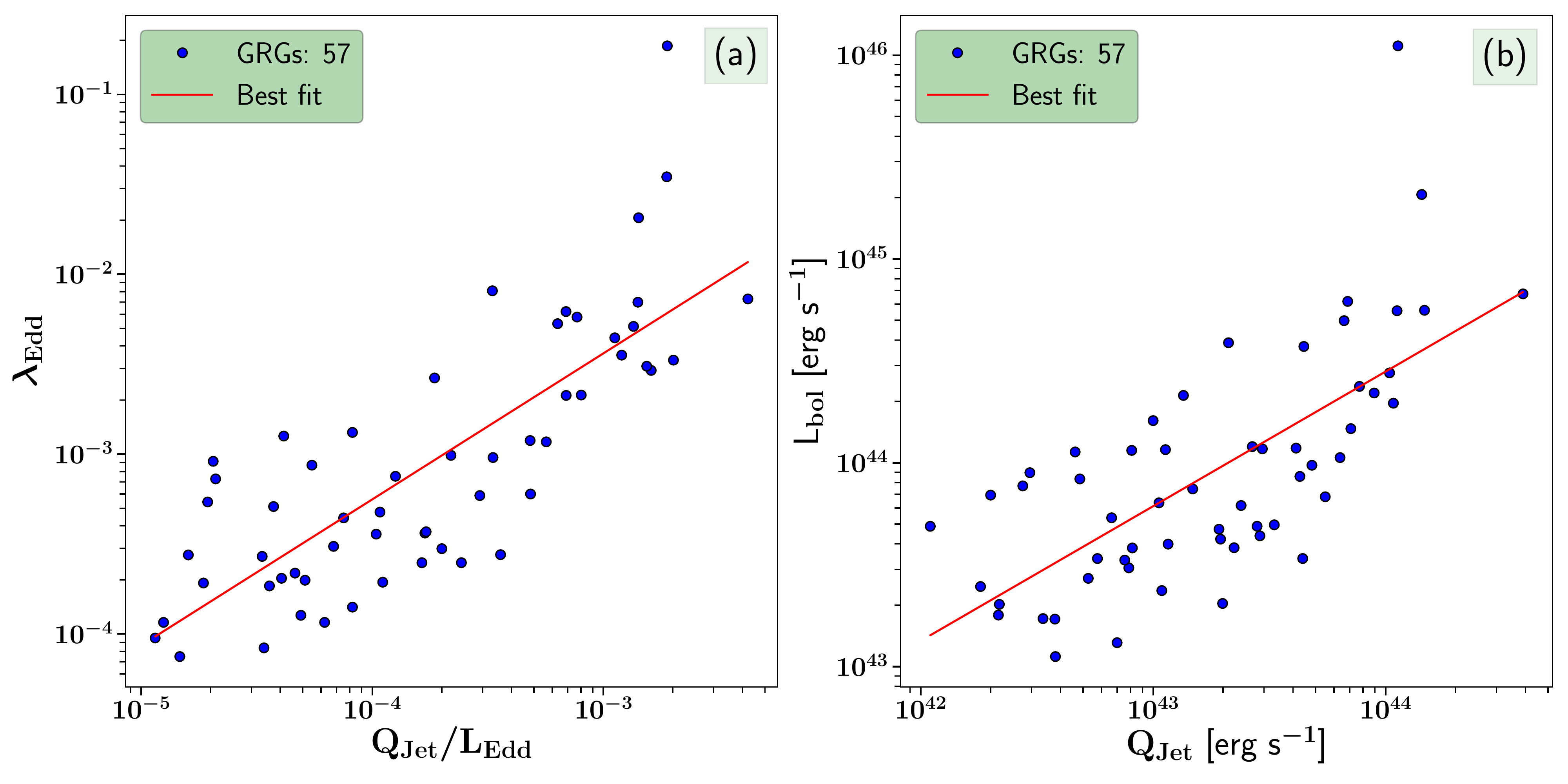} 
\caption{As discussed in Sec.\ \ref{sec:interplay}, in figure
(a) we present a linear correlation for 57 objects between the Eddington ratio ($\lambdaup_{\rm Edd}$) and the ratio of the jet kinetic power (Q$_{\rm Jet}$) to the Eddington luminosity (L$_{\rm Edd}$), which can be represented as log $\lambdaup_{\rm Edd}$ = 0.81 $\times$ log(Q$_{\rm Jet}$/$\rm L_{\rm Edd}$) $-$ 0.005. In figure (b), we show the bolometric luminosity (L$_{\rm bol}$) as a function of jet kinetic power (Q$_{\rm Jet}$), which takes the equation form log L$_{\rm bol}$ =  0.66 $\times$ log(Q$_{\rm Jet}$) + 15.38. The Spearman correlation coefficient is 0.8 and 0.7 for panels (a) and (b), respectively. The red line indicates the best linear fit of the correlations in the log-log plots.}
\label{fig:erall}
\end{figure*}

\subsubsection{Disc-jet coupling: Interplay between Eddington ratio ($\lambdaup_{\rm Edd}$), spin ($a$), and jet kinetic power ($Q_{\rm Jet}$)} \label{sec:interplay}

The role  of the black hole spin  and accretion rate in launching  the observed radio jets is still highly  contentious and an unsolved issue \citep{reynoldspinnat19}. Most often, the jet radio luminosity is taken as  a good indicator of the jet kinetic power, but in reality, it is only a small fraction of the latter. The question then is what the source of  the radio jet  power is that  we observe in black hole systems that  fuels the  extremely large  GRGs. How its  luminosity  scales  with the black hole spin and accretion rate is another open question.

The growth and evolution of an AGN is coupled with the accretion phenomenon and the spin of the central black hole. In order to understand the possible co-dependence of various parameters of the black holes for the purpose of jet launching and propagation, we studied how $\rm \lambdaup_{\rm Edd}$ affects the Q$_{\rm Jet}$. We created a sub-sample of 57 GRGs (based on their data availability) from the \texttt{GRG catalogue} and find a linear correlation between the Eddington ratio ($\lambdaup_{\rm Edd}$) and  Q$_{\rm Jet}$/L$_{\rm Edd}$ (dimensionless jet kinetic power), which is strongly supported by the Spearman correlation coefficient of 0.8, as shown in Fig.\ \ref{fig:erall} (a). The relation between the two parameters takes the following form: log $\lambdaup_{\rm Edd}$ = 0.81 $\times$ log(Q$_{\rm Jet}$/$\rm L_{\rm Edd}$) $-$ 0.005 based on our data. This implies that $\rm L_{\rm bol}$ has a significant effect on the  collimated kinetic jet output (Q$_{\rm Jet}$) of the AGN because $\lambdaup_{\rm Edd} = \frac{\rm L_{\rm bol}}{\rm L_{\rm Edd}}$, and the corresponding equation takes the form log L$_{\rm bol}$ =  0.66 $\times$ log(Q$_{\rm Jet}$) + 15.38, as evident from Fig.\ \ref{fig:erall} (b). The Spearman correlation coefficient for this correlation is 0.7. 

The formation and propagation of highly relativistic magnetised jets is dependent (disc-jet coupling) on the magnetic field strength coupled to magnetic field geometry of the accretion disc \citep{Mckinney2007}. \citet{Zamani} have shown that a strong correlation exists between the jet magnetic field $\rm \phi_{\rm Jet}$ and the accretion disc luminosity ($\rm L_{acc}$ or $\rm L_{bol}$). Furthermore, they also concluded that the launching regions of jets are threaded by dynamically important magnetic fields with the possible effect of the black hole spin. They strongly favoured the MAD model of the accretion disc. Similarly, \citet{sikora-begelman2013} also argued in favour of magnetic flux threading of black hole over spin and Eddington ratio as the governing factor for the launch of collimated powerful radio jets.
\citet{Zamani} reported that the magnetic flux of the black hole $\rm \phi_{\rm BH}$ $\propto$ $\sqrt{\rm L_{bol}}$ $\rm M_{BH}$ , and  in absence of direct measurements of $\rm \phi_{\rm BH}$, we can therefore use $\rm L_{bol}$ as a proxy. Therefore, our L$\rm_{bol}$ - Q$\rm_{Jet}$ correlation (discussed above) is also indicative of a strong effect of $\phi_{\rm BH}$ on $\rm Q_{\rm Jet}$. This strongly indicates a disc-jet coupling phenomenon where the magnetic field in the magnetically arrested disc surrounding the black hole controls the dynamics of accretion flow responsible for the jet launching. A lower magnetic field might therefore also be responsible for the lower Q$\rm_{Jet}$ , as evident from Fig.\ \ref{fig:erall} (b), where $\rm L_{bol}$ acts as a proxy for the magnetic field ($\rm \phi_{\rm BH}$).

In an advection-dominated accretion flow (ADAF, \citealt{Narayan1994}), Q$_{\rm Jet}$ is related to L$_{\rm Edd}$ \citep{Fanidakis} by
\begin{equation}
\rm Q_{\rm Jet} \sim 0.01\textit{a}^{2}L_{\rm Edd}  
,\end{equation}
 where Q$_{\rm Jet}$ and L$_{\rm Edd}$ are in erg s$^{-1}$. This shows that 1\% of L$_{\rm Edd}$ at most can be converted into Q$_{\rm Jet}$ when we consider a maximum spin value of 0.998. Based on our analysis, we infer that $\sim$\,0.06\% of L$_{\rm Edd}$ on an average is converted into Q$_{\rm Jet}$ for GRGs. 
 Therefore it is evident from this result and the discussion that for GRGs the fraction of L$_{\rm Edd}$ that is converted into  Q$_{\rm Jet}$ is quite low. It is therefore expected that the spin values are lower as well. This matches the estimated spin values we presented in Sec.\ \ref{sec:spinres}, where the mean spin value is  $\sim$\,0.079. The spin values for our GRGs were estimated using independently computed Q$_{\rm Jet}$ and 
 M$\rm _{BH}$ values. Here, Q$_{\rm Jet}$ is the lower limit as the Q$_{\rm Jet}$ relation (Eq.\ \ref{jet_kp2}) from \citet{Qjet_Hardcastle} is expected to underestimate the Q$_{\rm Jet}$ for GRGs, and the spin values are therefore also expected to be the lower limits.
From the correlations shown in Fig.\ \ref{fig:erall} (a), we expect a significant correlation between the spin and $\lambdaup_{\rm Edd}$ as well, represented in the relation log $\lambdaup_{\rm Edd}$ = 1.62 $\times$ log(\textit{a}) $-$ 1.02 based on the same data. The Spearman correlation coefficient for the above is 0.8. 

Based on our results, we observe strong evidence of a tight correlation between the spin and $\lambdaup_{\rm Edd}$ at least for the lower regime of spin and $\lambdaup_{\rm Edd}$ values.
Fig.\ \ref{fig:spin} clearly shows that the spin value decreases with increasing black hole mass, which is quite consistent with the findings of \citet{Daly} and \citet{Griffin}, but the magnitude of the spin for GRGs is lower than in their sample studies. 

Observations \citep{reynoldspinnat19} from the X-ray reflection of nearly two dozen SMBHs (prone to bias) have shown that the higher mass black holes ($ \geq 10^{8} \ \rm M_\odot$) tend to have low spin ($a$ $\leq$ 0.7) values, which is quite similar to our results for the GRGs, although our average spins are much lower. It is not ideal to compare properties of an X-ray selected AGN sample ( as mentioned in \citealt{reynoldspinnat19}) and radio-selected AGN (like our GRGs).
 
The two correlations as shown in Fig.\ \ref{fig:erall} (a) and Fig.\ \ref{fig:erall} (b) together may explain the effect of magnetic field and spin on the kinetic output of relativistic jets.
An increment in spin leads to an increment in $\lambdaup_{\rm Edd}$, and this eventually enhances the conversion of L$_{\rm Edd}$ into Q$_{\rm Jet}$. In other words, the higher the Eddington ratio, the higher the accumulation of magnetic flux near the black hole. leading to saturation of the magnetic field (MAD). It drives the magneto-rotational instabilities, which carry away the angular momentum of the in-falling material, and thus control the accretion flow. The magnetic flux in the jetted output scales with the magnetic flux threading the black hole. Furthermore, the spin twists the magnetic field lines around the black hole to transfer energy in the form of Poynting flux through jets. The magneto-hydrodynamics and spin of black holes thus play a crucial role in the formation of the powerful jets which could extend to large distances.


\begin{figure*}[ht]
\centering
\includegraphics[scale=0.45]{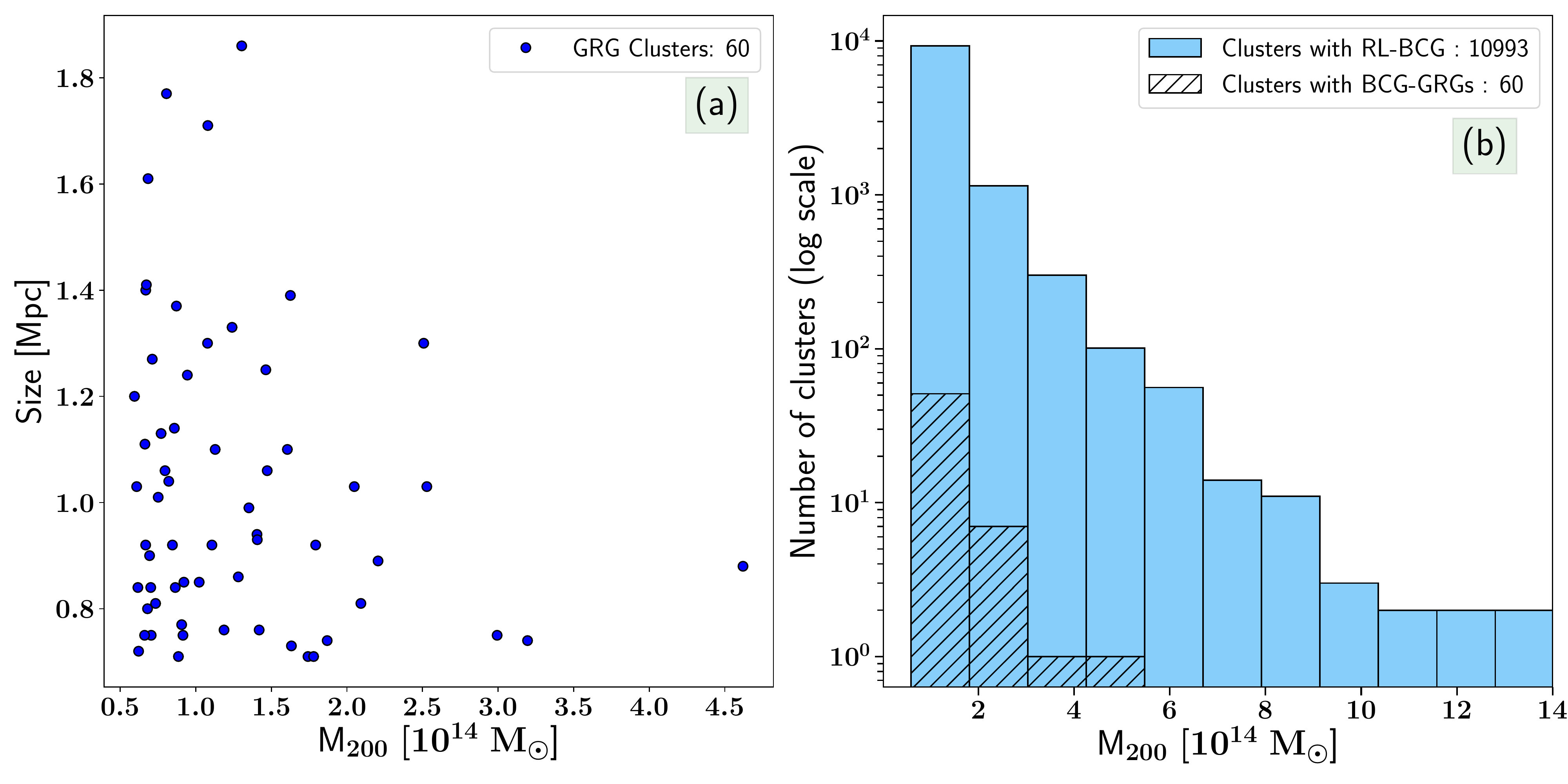} 
\caption{Panel (a): Mass of the galaxy cluster hosting the BCG-GRG (x-axis) and size of the GRG (y-axis). Panel (b): Distribution of M$\rm_{200}$ of clusters with radio-loud BCGs and clusters with GRG as BCGs from the \texttt{GRG catalogue}.}
\label{fig:m200}
\end{figure*}

\subsection{Environment analysis} \label{sec:enva}
We used the \texttt{GRG catalogue} to determine how many GRGs might be hosted by BCGs in galaxy clusters. To do this, we used the WHL galaxy cluster catalogue, as done for SGS in Sec.\ \ref{sec:sgsenv}. We find that a total of 60 GRGs from the \texttt{GRG catalogue} are BCGs, residing in a dense cluster environment. These 60 BCG-GRGs clearly provide strong evidence that sparse environment alone is not responsible for the large size of all GRGs, and powerful AGNs like these can pierce the dense cluster environment to grow to Mpc scales.

In Fig.\ \ref{fig:m200} (a), we explore the relation between the sizes of the GRGs and the M$\rm_{200}$ of the respective galaxy clusters. We observe that GRGs with sizes $\geq$ 1 Mpc are not found in clusters with M$\rm_{200}$ greater than 2.5 $\times \rm 10^{14} M_{\odot}$. 
Fig.\ \ref{fig:m200} (b) shows the distribution of M$\rm_{200}$ of clusters with radio-loud BCGs and BCG-GRGs, where we observe that radio-loud BCGs occur in clusters with a wide range of M$\rm_{200}$. About 95\% of the BCG-GRGs in our sample are located in clusters with M$\rm_{200}$ lower than 2 $\times \rm 10^{14} M_{\odot}$. Below we test whether this result is statistically significant.
Interestingly, no GRG found to be residing in the cluster environment is observed to grow beyond the virial radius (R$\rm_{200}$) of the host cluster. Also, we do not find a correlation between the total P$_{\rm 1400}$ of the GRGs and the mass of the cluster.

The K-S test on the distributions of M$_{200}$ of BCG-GRGs and radio-loud BCGs gives a p-value of 0.83, which indicates that the two distributions are identical. When we take the ratio of the number of BCG-GRGs to the radio-loud BCGs in the first four mass (M$\rm_{200}$) bins, we obtain 0.0055, 0.0061, 0.0033, and 0.0099, respectively, with an average value of 0.0062. If we assume this ratio to be true for all mass bins, then this indicates that we should expect that only 0.62 \% of the radio-loud BCGs in clusters are GRGs. The cosmological halo mass function of evolving structures shows that the abundance of low-mass clusters is much higher than the abundance of high-mass clusters in the local Universe \citep{sheth01}.
Moreover, as the number of high-mass clusters is interestingly very low, we obtain an even lower number of radio-loud BCGs hosting GRGs in the cluster environment. For example, when we take the
mass bin around $7 \times 10^{14} \rm M_{\odot}$ (6$^{\rm th}$ bin), then
there are 14 radio-loud BCGs in this bin. The average value of the ratio, 0.0062, suggests that we needed about 160 clusters of such high masses with a radio-loud BCG to detect at least one GRG. Therefore the probability of finding a radio-loud BCG as a GRG in clusters is indeed very low, and it is very difficult to find them in massive clusters.  

We also observe that the radio morphology of BCG-GRGs is not linear and symmetric, but is rather deformed to a certain degree. This clearly illustrates the effect of the environment on the GRGs, which, despite the resistance of the intra-cluster medium  and cluster weather, are able to grow to megaparsec scales.

In order to study the environment of the GRGs more thoroughly, we need to determine precisely the fraction of the GRG population that resides in clusters (BCG or a cluster member), which can only be achieved with a controlled and complete sample. We will be focusing on this aspect in our future work within project SAGAN.

\section{Discussion}
Based on our study, we further discuss the overall results in the context of GRGs in this section.
Nearly 20-year-old surveys such as the NVSS, when combined with other radio and optical surveys, still have a high discovery potential. This is evident from the large sample (SGS) of 162 GRGs discovered by us, which increases the overall known GRG population by nearly 25\%. 

Our GRG compilation work within project SAGAN, which led to the construction of the \texttt{GRG catalogue,} has revealed that GRGs are not as rare as thought before. Nevertheless, the \texttt{GRG catalogue} consists of only 820 sources, which still is quite small compared to the overall RG population. This large compilation of GRGs has enabled us to reveal new properties of these rare sources, leading to a better understanding.

The central engine powering the GRGs is hosted both by galaxies (GRGs) and quasars (GRQs). We studied the key similarities and dissimilarities in their properties to understand the effect of the central powering engine on their respective properties. Our analysis (Sec.\ \ref{sec:gq}) showed that the projected linear size of the GRQs and GRGs is nearly similar (from the homogeneous sample of LoTSS and SAGAN). The GRQs are also more powerful in terms of their central engine, P$_{1400}$ , and Q$\rm _{Jet}$. This implies that the GRQs might scale megaparsec distances in a shorter time than the GRGs. However, the study of their spectral indices reveals that they have similar distributions and a similar value of the spectral index. It has been shown \citep{thorpe99,Ishwar-saikia00} previously that the hotspots of RQs are steeper than those of RGs, and the same argument can be applicable to GRQs as well. Thus, the flatter core and steeper hotspots of GRQs counterbalance each other, resulting in a similar spectral index to that of GRGs.

AGNs are broadly classified into two categories: the radiatively efficient (RE) mode, and the radiatively inefficient (RI) mode. In the context of radio galaxies, these modes are referred to as HERGs (RE) and LERGs (RI), and similarly for GRGs, as HEGRGs-LEGRGs. In the literature, these modes are also referred to as the radiative mode (RE) and the jet mode (RI). The quasars lie in the RE mode; and in this discussion we primarily focus on non-quasar AGNs below, which does not include the GRQs.

The GRG-AGN activity (HEGRG and LEGRG) can regulate different characteristics of AGN in terms of radiative efficiency and the formation and powering of relativistic jets. \citet{bh12rgs} concluded that HERGs are associated with lower mass black holes with higher accretion rates than LERGs. A similar picture is reflected in the properties of HEGRGs and LEGRGs using the \texttt{GRG catalogue}, where our results (Sec.\ \ref{sec:hl}) also show that the $\lambdaup_{\rm Edd}$ is higher for HEGRGs than LEGRGs. There is a significant overlap in the distribution of $\lambdaup_{\rm Edd}$ of HEGRGs and LEGRGs without a sharp boundary that separates the two populations. This is similar to RGs \citep{bh12rgs,mingo14}, but the observed overlap might also partly be due to the uncertainties in estimating L$_{\rm Bol}$ and M$_{\rm BH}$. The high $\lambdaup_{\rm Edd}$ implies a high accretion rate in HEGRGs, which might be responsible for producing powerful relativistic jets, which in turn might account for the larger sizes of HEGRGs compared to LEGRGs.

One of the most fundamental properties of the black hole is its mass (M$\rm _{BH}$). SMBHs residing in AGNs have masses in the range of $\sim$\,10$^{7}$ to 10$^{10}$ $\rm  M_{\odot}$. The aim of carrying out a comparative study (Sec.\ \ref{sec:rggrg}) of  the AGN properties of RGs and GRGs is to find factors that allow a tiny subset of RGs to grow to GRGs.
Although their M$\rm _{BH}$ distribution is very similar, the $\lambdaup_{\rm Edd}$ distribution varies: GRGs have predominantly lower values than RGs. Our study clearly rules out the possibility that M$\rm _{BH}$ alone is the crucial factor for driving RGs to GRG scales. AGNs with low values ($\lambdaup_{\rm Edd}$ $<$ 0.01) of the Eddington ratio follow the RIAF model, and our study therefore shows that almost all the GRGs (except for GRQs and a few HEGRGs) seem to be fuelled by radiatively inefficient flows at low accretion rates. Objects with RIAFs have a very low Eddington ratio, and quite often, this cause the absence of the so-called big blue bump. This causes the accretion disc to be under-luminous, and accordingly weak in the X-ray regime \citep{ho08,hardcastlenat}. The formation of powerful jets is quite favourable in RIAFs because of the thick-disc structure, which boosts the large-scale poloidal component of the magnetic field, resulting in the launch of collimated jets \citep{rees82,meier99}.
Interestingly, \citet{bagchi14} reached the same conclusion for an extremely unusual GRG hosted by the spiral galaxy 2MASX J23453268-0449256 in an under-dense environment: the launch of powerful FR-II jets in it  is facilitated by an advection-dominated, magnetised accretion flow at low Eddington rate onto a supermassive and  moderately  spinning $a  \sim  0.7$  central black hole. Our previous and present results therefore provide fundamental insight into the accretion disc-relativistic jet coupling process in GRGs.

The RGs, as evident from their distribution of $\lambdaup_{\rm Edd}$, have a higher mass accretion rate as well as a high accretion efficiency, which leads to the formation of powerful relativistic jets that reach hundreds of kiloparsecs. We can further conjecture that GRGs at the start of their lives are RGs with higher $\lambdaup_{\rm Edd}$, which diminishes over the period of their growth to GRGs.
The SMBHs that grow over a period of time through mergers and coherent disc accretion are most likely responsible for fast-spinning black holes (such as quasars, Seyferts, and HERGs). On the other hand, the slowly spinning black holes (such as GRGs) with high-mass black holes may have been formed by more isotropic chaotic accretion \citep{volonteri05,Volonteri2013,reynoldspinnat19}. 

One of the most striking results obtained from the large sample of GRGs available to us is about their radio morphologies, where we find that FR-II population is dominant. The LoTSS and SGS have > 90\% FR-II type GRGs. The FR-I type GRGs prefer low-excitation type AGN and FR-IIs high-excitation type AGN. 

After evaluating all the properties derived by us and their distributions, along with the comparisons with RG samples, it is clear that no one sole factor or parameter, but the correct combination of multiple parameters leads to the growth of GRGs. A possibility to be explored in future studies is estimating and assessing the fueling process of the SMBHs in GRGs in conjunction with their environments.


\section{Summary}\label{sec:sum}
We presented a sample of 162 new GRGs (also called the SAGAN GRG sample, or SGS for short) with a projected linear size above $\sim$\,0.7 Mpc, based on our searches in the NVSS. This investigation used the \citet{proctorGRS} catalogue of probable GRG candidates and employed a visual search for extended sources. The ancillary data from radio surveys such as the FIRST, TGSS, WENSS, and VLASS were very crucial in confirming the morphology of these sources. The identification and classification of the hosts were made using spectroscopic data from the SDSS, Pan-STARRS, and mid-IR data from the \textit{WISE} survey. As a result, 23 out of 162 radio giants are confirmed to be GRQs, and the remainder are GRGs. The mean and median values of several properties of the GRGs are presented in Table\ \ref{mean}.
A brief summary of our results is given below.
\begin{enumerate}
\item GRGs are known to be relatively rare. Considering that only $\sim$\,658 have been known so far, the addition of another 162 new sources increases the known population by $\sim$\,25$\%$. This helps to better study these objects statistically in multiple wavelengths to decipher the reasons for their gigantic nature and rarity.

\item SGS: The projected linear sizes of all sources in SGS lie in the range of $\sim$\,0.71 Mpc to $\sim$\,2.82 Mpc, with P$_{\rm 1400}$ as low as $10^{23}$ W Hz$^{-1}$ at 1400 MHz.

\item SGS: About 92$\%$ of the GRGs in SGS have an FR-II morphology. The remaining sample consists of eight FR-I, four HyMoRS, and one DDRG. It is clear that giants are dominated by the FR-II population.  

\item The study of GRGs of two AGN types (galaxy or quasar) revealed that the radio power and jet kinetic power are higher in GRQs than in GRGs. This clearly shows that GRQs have more powerful central engines than GRGs. However, the projected linear size and spectral index of  the two populations are similar.

\item The study of high- and low-excitation GRGs revealed that those with lower black hole mass, higher accretion efficiency, radio power, and jet kinetic power, that is, the HEGRGs, grow to the maximum extent. 

\item We established using our large database of GRGs, the \texttt{GRG catalogue,} that the radio spectral index (between 150 MHz and 1400 MHz) of RGs and GRGs is similar.

\item  Assuming a maximum magnetic field, we showed that the black holes hosted by GRGs might be spinning slowly, with a maximum spin up to $\sim$\,0.3, which indicates a chaotic accretion history of the GRG host galaxies. 

\item There is a strong correlation between $\lambdaup_{\rm Edd}$ and the amount of Eddington luminosity that is converted into jet kinetic power. It significantly implies that efficient accretion is responsible for the formation of powerful jets, and disc-jet coupling is at play.

\item GRGs tend to avoid a rich cluster environment, and it is rare to find a GRG in a very massive galaxy cluster of M$\rm _{200} >$ 2 $\times 10^{14} M_{\odot}$. We find 60 GRGs hosted by BCGs in dense cluster environments. The analysis of their projected linear sizes indicated that the size of BCG-GRGs is limited with respect to other GRGs in the same redshift range by the dense ambient medium. Our study and the previous results from the literature clearly show that the local environment indeed plays a role in the growth of GRGs, but is not the sole factor that affects their size.

\end{enumerate}

\section*{Acknowledgements}
We thank the anonymous referee for her/his comments which have helped improve the paper significantly. LCH was supported by the National Key R\&D Program of China (2016YFA0400702) and the National Science Foundation of China (11721303, 11991052).  
PD, JB, FC, and BG gratefully acknowledge generous support from the Indo-French Centre for the Promotion of Advanced Research 
(Centre Franco-Indien pour la Promotion de la Recherche Avan\'{c}ee) under programme no. 5204-2. We thank IUCAA 
(especially Radio Physics Lab\footnote{\url{http://www.iucaa.in/~rpl/}}), Pune for providing all the facilities during the period the work was carried out. 
We gratefully acknowledge the use of Edward (Ned) Wright's online Cosmology Calculator. This research has made use of the VizieR catalogue tool, CDS, Strasbourg, France \citep{vizier}

We acknowledge that this work has made use of \textsc{ipython} \citep{per07}, \textsc{astropy} \citep{astropy}, \textsc{aplpy} \citep{apl}, \textsc{matplotlib} \citep{plt} and \textsc{topcat} \citep{top05}.

\bibliographystyle{aa} 
\bibliography{SAGAN.bib}

\begin{appendix}
  
\section{Tables}\label{sec:aptab}

\clearpage
\onecolumn
\setlength{\tabcolsep}{2.4pt}
\begin{tiny}
\begin{landscape}
\begin{longtable}{c c c c c c c c c c c c c c c }
\caption{ In Col. (1), Sr.No with superscript 'i' represents the sources found from an independent manual search.  Col. (2) and Col. (3) represent the right ascension (RA) in HMS and declination (Dec) 
in DMS of the host galaxies of the GRGs. Col. (4) represents the host type, where G stands for galaxy and Q stands for quasar. In Col. (5), m$_{\rm r}$ represents the apparent r-band magnitude of the 
host galaxy from the  SDSS. In Col. (6), z$\dagger$ symbolises spectroscopic redshifts, while the rest are photometric redshift estimates. Col. (7) and Col. (8) list the total  projected linear size of the 
sources in arcmins and megaparsecs, respectively. Col. (9) and Col. (11) list the integrated flux densities ($S_{\rm \nu}$), and Col.(10) and Col. (12) list the total power $P_{\rm \nu}$ of the sources 
at 1400 MHz and 150 MHz, respectively. Col. (13) states the $\rm \alpha_{\rm 150}^{\rm 1400}$ , which is the two-point spectral index between 1400 MHz (NVSS) and 150 MHz (TGSS). Col. (14) contains 
redshift references ($z_{\rm ref}$), given at the end of the table. Column (15) lists the morphological type of the sources:- I represents FR-I type, II represents FR-II type, HM represents 
probable HyMoRS type, and RM represents remnant galaxy.} \\

\hline 
Sr.No & RA & Dec & Type & m$_{\rm r}$ & z & Size  & Size & $\rm S_{1400}$ & $\rm P_{1400}$ & $\rm S_{150}$ & $\rm P_{150}$ & $\rm \alpha^{\rm 1400}_{\rm 150}$ & $z_{\rm ref}$ & Morphology\\
         & (HMS) & (DMS) &  & &  & (\arcmin) & (Mpc) & (mJy) & ($ 10^{25}$W Hz$^{-1}$) & (mJy)& ($ 10^{25}$W Hz$^{-1}$) & \\
(1) &(2)&(3)&(4)&(5)&(6)&(7)&(8)&(9)&(10)&(11)&(12)&(13) & (14) & (15)       \\
\hline
\endfirsthead
\caption{continued.}\\
\hline
Sr.No & RA & Dec & Type & m$_{\rm r}$ & z & Size  & Size & $\rm S_{1400}$ & $\rm P_{1400}$ & $\rm S_{150}$ & $\rm P_{150}$ & $\rm \alpha^{\rm 1400}_{\rm 150}$ & $z_{\rm ref}$ & Morphology \\
         & (HMS) & (DMS) &  & &  & (\arcmin) & (Mpc) & (mJy) & ($ 10^{25}$W Hz$^{-1}$) & (mJy)& ($ 10^{25}$W Hz$^{-1}$) & \\
(1) &(2)&(3)&(4)&(5)&(6)&(7)&(8)&(9)&(10)&(11)&(12)&(13) & (14) & (15)       \\
\hline
\endhead
\hline
\endfoot
1 & 00 04 50.25 & 12 48 40.10 & G & 17.18 $\pm$ 0.01 & 0.14300 $\pm$ 0.00020$^{\dagger}$ & 7.01 & 1.09 & 1983 $\pm$ 60 & 11.10 $\pm$ 0.36 & 10048 $\pm$ 2010 & 56.30 $\pm$ 11.30 & 0.73 $\pm$ 0.21 & 12 & II \\
2 & 00 06 23.24 & 26 35 44.66 & G & 21.29 $\pm$ 0.08 & 0.43600 $\pm$ 0.18820 & 4.88 & 1.71 & 339 $\pm$ 10 & 22.40 $\pm$ 22.90 & 1705 $\pm$ 343 & 113.00 $\pm$ 117.00 & 0.72 $\pm$ 0.21 & 1 & II \\
3 & 00 11 19.35 & 32 17 13.83 & G & 16.28 $\pm$ 0.00 & 0.10714 $\pm$ 0.00002$^{\dagger}$ & 7.80 & 0.95 & 722 $\pm$ 22 & 2.18 $\pm$ 0.07 & 3564 $\pm$ 714 & 10.80 $\pm$ 2.16 & 0.71 $\pm$ 0.21 & 1 & II \\
4 & 00 15 02.43 & 10 10 56.33 & G & 20.98 $\pm$ 0.11 & 0.41900 $\pm$ 0.14980 & 4.67 & 1.60 & 47 $\pm$ 2 & 2.75 $\pm$ 2.32 & 196 $\pm$ 41 & 11.50 $\pm$ 10.00 & 0.64 $\pm$ 0.22 & 1 & II \\
5 & 00 17 41.45 & 08 27 55.72 & Q & 18.82 $\pm$ 0.01 & 0.67841 $\pm$ 0.00007$^{\dagger}$ & 2.75 & 1.20 & 267 $\pm$ 8 & 51.60 $\pm$ 2.90 & 1662 $\pm$ 335 & 321.00 $\pm$ 66.30 & 0.82 $\pm$ 0.21 & 1 & II \\
6 & 00 18 15.21 & 21 41 33.42 & G & 18.51 $\pm$ 0.02 & 0.30256 $\pm$ 0.00005$^{\dagger}$ & 3.60 & 1.00 & 862 $\pm$ 26 & 24.80 $\pm$ 0.95 & 4188 $\pm$ 838 & 121.00 $\pm$ 24.30 & 0.71 $\pm$ 0.21 & 1 & II \\
7 & 00 26 13.63 & 11 43 43.66 & G & 17.82 $\pm$ 0.01 & 0.17500 $\pm$ 0.01620 & 9.72 & 1.78 & 582 $\pm$ 18 & 4.90 $\pm$ 1.02 & 2001 $\pm$ 405 & 16.80 $\pm$ 4.85 & 0.55 $\pm$ 0.21 & 1 & II \\
8 & 00 36 22.99 & $-$02 58 21.55 & G & 18.97 $\pm$ 0.02 & 0.29400 $\pm$ 0.05690 & 4.51 & 1.23 & 188 $\pm$ 6 & 4.96 $\pm$ 2.22 & 745 $\pm$ 155 & 19.70 $\pm$ 9.67 & 0.62 $\pm$ 0.22 & 1 & II \\
9 & 00 38 09.89 & 09 36 01.33 & G & 18.17 $\pm$ 0.02 & 0.29416 $\pm$ 0.00005$^{\dagger}$ & 4.80 & 1.30 & 216 $\pm$ 7 & 5.87 $\pm$ 0.23 & 1089 $\pm$ 219 & 29.60 $\pm$ 5.99 & 0.72 $\pm$ 0.21 & 1 & II \\
10 & 00 45 17.11 & 38 25 03.23  & G & 21.60 $\pm$ 0.13 & 0.89000 $\pm$ 0.09560 & 3.44 & 1.65 & 222 $\pm$ 7 & 86.10 $\pm$ 23.40 & 1673 $\pm$ 336 & 649.00 $\pm$ 219.00 & 0.90 $\pm$ 0.21 & 1 & II \\
11 & 00 46 53.63 & 12 55 41.01 & G & 18.54 $\pm$ 0.02 & 0.26400 $\pm$ 0.06240 & 4.10 & 1.03 & 135 $\pm$ 4 & 2.85 $\pm$ 1.54 & 613 $\pm$ 124 & 12.90 $\pm$ 7.45 & 0.68 $\pm$ 0.21 & 1 & II \\
12$^{i}$ & 00 55 48.79 & $-$22 31 16.50 & G & - & 0.11437 $\pm$ 0.00015$^{\dagger}$ & 6.25 & 0.80 & 179 $\pm$ 6 & 0.61 $\pm$ 0.02 & 555 $\pm$ 115 & 1.89 $\pm$ 0.39 & 0.51 $\pm$ 0.22 & 2 & II \\
13 & 01 00 52.60 & 06 16 41.01 & G & 15.72 $\pm$ 0.00 & 0.11100 $\pm$ 0.00700 & 5.76 & 0.72 & 195 $\pm$ 6 & 0.63 $\pm$ 0.09 & 660 $\pm$ 135 & 2.12 $\pm$ 0.52 & 0.55 $\pm$ 0.21 & 1 & II \\
14 & 01 02 37.34 & 04 21 00.26 & G & 18.44 $\pm$ 0.01 & 0.29200 $\pm$ 0.01000 & 4.20 & 1.14 & 312 $\pm$ 9 & 8.23 $\pm$ 0.72 & 1404 $\pm$ 283 & 37.00 $\pm$ 8.06 & 0.67 $\pm$ 0.21 & 6 & II \\
15 & 01 03 40.05 & 42 39 35.63 & G & 16.26 $\pm$ 0.00 & 0.13500 $\pm$ 0.00720 & 6.90 & 1.02 & 530 $\pm$ 16 & 2.58 $\pm$ 0.31 & 2039 $\pm$ 409 & 9.94 $\pm$ 2.30 & 0.60 $\pm$ 0.21 & 1 & II \\
16 & 01 03 56.32 & 32 06 07.83 & G & 17.05 $\pm$ 0.01 & 0.12600 $\pm$ 0.01290 & 6.20 & 0.87 & 205 $\pm$ 6 & 0.88 $\pm$ 0.20 & - & - & - & 1 & HM? \\
17 & 01 08 03.52 & 27 00 01.99 & G & 19.21 $\pm$ 0.02 & 0.32700 $\pm$ 0.05530 & 4.00 & 1.17 & 696 $\pm$ 21 & 23.60 $\pm$ 9.30 & 3087 $\pm$ 618 & 105.00 $\pm$ 46.20 & 0.67 $\pm$ 0.21 & 1 & II \\
18 & 01 09 36.45 & 25 24 03.76 & G & 22.07 $\pm$ 0.16 & 0.58100 $\pm$ 0.04760 & 3.43 & 1.39 & 202 $\pm$ 6 & 28.90 $\pm$ 5.91 & 1729 $\pm$ 347 & 247.00 $\pm$ 70.50 & 0.96 $\pm$ 0.21 & 1 & II \\
19 & 01 11 04.70 & $-$14 22 32.60 & G & - & 0.10406 $\pm$ 0.00015$^{\dagger}$ & 6.75 & 0.80 & 402 $\pm$ 12 & 1.11 $\pm$ 0.04 & 996 $\pm$ 201 & 2.75 $\pm$ 0.56 & 0.41 $\pm$ 0.21 & 2 & II \\
20$^{i}$ & 01 12 22.34 & $-$28 16 26.60 & Q & - & 0.65312 $\pm$ 0.00015$^{\dagger}$ & 3.61 & 1.55 & 46 $\pm$ 2 & 7.78 $\pm$ 0.38 & - & - & - & 2 & HM? \\
21 & 01 13 41.11 & 01 06 08.52 & G & 18.48 $\pm$ 0.01 & 0.28100 $\pm$ 0.00001 & 4.26 & 1.12 & 383 $\pm$ 12 & 9.36 $\pm$ 0.35 & 1858 $\pm$ 373 & 45.40 $\pm$ 9.15 & 0.71 $\pm$ 0.21 & 1 & II \\
22 & 01 14 30.75 & 05 08 30.64 & G & 17.22 $\pm$ 0.01 & 0.20456 $\pm$ 0.00003$^{\dagger}$ & 6.50 & 1.35 & 229 $\pm$ 7 & 2.73 $\pm$ 0.10 & 902 $\pm$ 181 & 10.70 $\pm$ 2.17 & 0.61 $\pm$ 0.21 & 1 & II \\
23$^{i}$ & 01 14 40.34 & $-$36 16 31.80 & G & - & 0.11207 $\pm$ 0.00011$^{\dagger}$ & 7.04 & 0.89 & 61 $\pm$ 3 & 0.20 $\pm$ 0.01 & - & - & - & 2 & I \\
24 & 01 23 59.92 & 43 12 55.78 & G & 19.84 $\pm$ 0.03 & 0.40900 $\pm$ 0.05650 & 4.90 & 1.65 & 436 $\pm$ 13 & 25.10 $\pm$ 8.27 & 2326 $\pm$ 471 & 134.00 $\pm$ 51.70 & 0.75 $\pm$ 0.21 & 1 & II \\
25 & 01 25 32.16 & 07 03 37.11 & G & 15.91 $\pm$ 0.00 & 0.09800 $\pm$ 0.01010 & 10.40 & 1.17 & 98 $\pm$ 4 & 0.25 $\pm$ 0.06 & - & - & - & 1 & II \\
26 & 01 28 48.89 & 24 31 52.10 & G & 16.12 $\pm$ 0.01 & 0.16455 $\pm$ 0.00002$^{\dagger}$ & 5.45 & 0.95 & 100 $\pm$ 4 & 0.76 $\pm$ 0.03 & - & - & - & 1 & II \\
27 & 01 33 27.24 & $-$08 24 16.52  & G & 15.93 $\pm$ 0.00 & 0.14893 $\pm$ 0.00013$^{\dagger}$ & 5.70 & 0.92 & 234 $\pm$ 7 & 1.40 $\pm$ 0.05 & 878 $\pm$ 178 & 5.27 $\pm$ 1.07 & 0.59 $\pm$ 0.21 & 1 & II \\
28 & 01 35 36.11 & 50 40 38.54 & G & 19.84 $\pm$ 0.04 & 0.47600 $\pm$ 0.01810 & 4.59 & 1.68 & 86 $\pm$ 3 & 6.49 $\pm$ 0.68 & 285 $\pm$ 61 & 21.40 $\pm$ 5.07 & 0.53 $\pm$ 0.23 & 1 & II \\
29 & 01 42 08.56 & $-$06 41 43.4 & G & 15.13 $\pm$ 0.00 & 0.12447 $\pm$ 0.00015$^{\dagger}$ & 6.45 & 0.89 & 138 $\pm$ 5 & 0.56 $\pm$ 0.02 & 431 $\pm$ 89 & 1.75 $\pm$ 0.36 & 0.51 $\pm$ 0.22 & 2 & II \\
30 & 01 48 49.35 & 06 22 43.26 & G & 19.03 $\pm$ 0.03 & 0.32000 $\pm$ 0.03650 & 3.34 & 0.96 & 445 $\pm$ 13 & 15.00 $\pm$ 3.99 & 2724 $\pm$ 545 & 91.60 $\pm$ 30.40 & 0.81 $\pm$ 0.21 & 1 & II \\
31 & 01 52 01.00 & $-$30 33 19.00 & G & - & 0.16999 $\pm$ 0.00021$^{\dagger}$ & 5.22 & 0.94 & 57 $\pm$ 2 & 0.46 $\pm$ 0.02 & - & - & - & 3 & II \\
32 & 01 58 26.09 & 24 51 36.38 & G & 16.59 $\pm$ 0.01 & 0.17660 $\pm$ 0.00002$^{\dagger}$ & 5.00 & 0.92 & 81 $\pm$ 3 & 0.69 $\pm$ 0.03 & 288 $\pm$ 60 & 2.48 $\pm$ 0.52 & 0.57 $\pm$ 0.22 & 1 & II \\
33 & 02 25 27.01 & $-$31 37 37.00 & G & - & 0.11000 $\pm$ 0.00021$^{\dagger}$ & 5.71 & 0.71 & 85 $\pm$ 3 & 0.27 $\pm$ 0.01 & 314 $\pm$ 77 & 0.99 $\pm$ 0.24 & 0.58 $\pm$ 0.26 & 3 & II \\
34 & 03 21 25.91 & 18 06 09.84 & G & 14.12 $\pm$ 0.00 & 0.06155 $\pm$ 0.00014 & 11.85 & 0.87 & 260 $\pm$ 8 & 0.25 $\pm$ 0.01 & - & - & - & 4 & I \\
35 & 03 21 55.75 & 43 46 40.78 & G & 18.21 $\pm$ 0.02 & 0.36530 $\pm$ 0.01000 & 3.28 & 1.03 & 401 $\pm$ 12 & 18.70 $\pm$ 1.43 & 2985 $\pm$ 597 & 139.00 $\pm$ 29.60 & 0.90 $\pm$ 0.21 & 5 & II \\
36 & 03 25 33.95 & $-$39 27 26.80 & G & - & 0.06300 $\pm$ 0.00015$^{\dagger}$ & 12.41 & 0.93 & 209 $\pm$ 7 & 0.21 $\pm$ 0.01 & - & - & - & 18 & II \\
37 & 03 42 53.70 & $-$06 52 24.40 & G & 20.72 $\pm$ 0.07 & 0.55200 $\pm$ 0.03330 & 4.08 & 1.62 & 289 $\pm$ 9 & 31.60 $\pm$ 4.87 & 1180 $\pm$ 238 & 129.00 $\pm$ 32.50 & 0.63 $\pm$ 0.21 & 6 & II \\
38 & 03 49 10.95 & 05 34 26.69 & G & 18.34 $\pm$ 0.02 & 0.33500 $\pm$ 0.04710 & 5.06 & 1.50 & 82 $\pm$ 3 & 3.01 $\pm$ 0.99 & - & - & - & 1 & II \\
39$^{i}$ & 03 50 45.56 & $-$18 18 29.80 & G & - & 0.17474 $\pm$ 0.00015$^{\dagger}$ & 5.50 & 1.01 & 141 $\pm$ 5 & 1.19 $\pm$ 0.04 & 555 $\pm$ 114 & 4.71 $\pm$ 0.96 & 0.62 $\pm$ 0.21 & 2 & II \\
40 & 03 53 39.31 & $-$01 13 19.73  & G & 16.85 $\pm$ 0.01 & 0.19100 $\pm$ 0.00970 & 5.69 & 1.12 & 199 $\pm$ 6 & 2.08 $\pm$ 0.25 & 970 $\pm$ 196 & 10.10 $\pm$ 2.35 & 0.71 $\pm$ 0.21 & 1 & II \\
41 & 04 22 34.14 & $-$26 16 43.20 & G & - & 0.13100 $\pm$ 0.05000$^{\dagger}$ & 5.72 & 0.83 & 1333 $\pm$ 40 & 6.15 $\pm$ 5.08 & 5957 $\pm$ 1191 & 27.50 $\pm$ 23.40 & 0.67 $\pm$ 0.21 & 7 & II \\
42 & 04 39 32.51 & $-$23 31 08.60  & G & - & 0.07457 $\pm$ 0.00015$^{\dagger}$ & 9.95 & 0.87 & 265 $\pm$ 8 & 0.38 $\pm$ 0.01 & - & - & - & 2 & HM? \\
43 & 05 15 36.11 & 15 18 21.31 & G & 19.38 $\pm$ 0.06 & 0.33100 $\pm$ 0.03880 & 5.14 & 1.51 & 66 $\pm$ 2 & 2.38 $\pm$ 0.66 & 382 $\pm$ 78 & 13.70 $\pm$ 4.69 & 0.79 $\pm$ 0.21 & 1 & II \\
44 & 05 43 59.77 & 30 47 48.28 & G & 18.31 $\pm$ 0.07 & 0.30900 $\pm$ 0.04670 & 4.32 & 1.21 & 64 $\pm$ 2 & 1.93 $\pm$ 0.68 & 314 $\pm$ 65 & 9.48 $\pm$ 3.85 & 0.71 $\pm$ 0.22 & 1 & II \\
45 & 06 12 03.50 & $-$32 57 47.00 & G & - & 0.07791 $\pm$ 0.00015$^{\dagger}$ & 10.38 & 0.95 & 341 $\pm$ 11 & 0.52 $\pm$ 0.02 & 1017 $\pm$ 207 & 1.55 $\pm$ 0.32 & 0.49 $\pm$ 0.21 & 8 & II \\
46 & 06 44 08.04 & 10 43 41.40 & G & 14.56 $\pm$ 0.00 & 0.24177 $\pm$ 0.01000 & 11.95 & 2.82 & 250 $\pm$ 8 & 4.41 $\pm$ 0.44 & - & - & - & 5 & II \\
47 & 07 08 44.11 & 30 30 12.73  & G & 21.08 $\pm$ 0.07 & 0.55200 $\pm$ 0.05740 & 3.96 & 1.57 & 103 $\pm$ 4 & 11.60 $\pm$ 2.98 & 495 $\pm$ 102 & 55.80 $\pm$ 18.30 & 0.70 $\pm$ 0.22 & 1 & II \\
48 & 07 25 38.75 & 40 04 12.52 & G & 16.16 $\pm$ 0.00 & 0.16148 $\pm$ 0.00002$^{\dagger}$ & 4.89 & 0.84 & 98 $\pm$ 3 & 0.71 $\pm$ 0.02 & - & - & - & 1 & II \\
49 & 07 28 05.86 & 50 34 45.80 & G & - & 0.35000 $\pm$ 0.05000 & 3.96 & 1.21 & 927 $\pm$ 28 & 38.60 $\pm$ 13.00 & 6131 $\pm$ 1227 & 255.00 $\pm$ 99.50 & 0.85 $\pm$ 0.21 & 15 & II \\
50 & 07 32 05.49 & 15 58 31.06 & G & 18.56 $\pm$ 0.01 & 0.23100 $\pm$ 0.03070 & 6.20 & 1.41 & 326 $\pm$ 10 & 5.56 $\pm$ 1.68 & 3534 $\pm$ 708 & 60.20 $\pm$ 21.80 & 1.07 $\pm$ 0.21 & 1 & II \\
51 & 07 50 21.33 & 16 32 59.35  & G & - & 0.58500 $\pm$ 0.06800 & 4.41 & 1.80 & 185 $\pm$ 6 & 26.10 $\pm$ 7.45 & 1355 $\pm$ 273 & 191.00 $\pm$ 66.40 & 0.89 $\pm$ 0.21 & 1 & II \\
52 & 07 59 31.84 & 08 25 34.59 & G & 17.31 $\pm$ 0.01 & 0.12400 $\pm$ 0.00890 & 5.22 & 0.72 & 103 $\pm$ 4 & 0.42 $\pm$ 0.07 & - & - & - & 1 & RM? \\
53 & 08 29 41.34 & 22 47 58.41  & G & 18.31 $\pm$ 0.01 & 0.25700 $\pm$ 0.02960 & 5.42 & 1.34 & 183 $\pm$ 6 & 3.81 $\pm$ 1.01 & 1307 $\pm$ 263 & 27.20 $\pm$ 9.01 & 0.88 $\pm$ 0.21 & 1 & II \\
54 & 08 47 18.28 & 42 23 41.33 & G & 19.97 $\pm$ 0.04 & 0.47498 $\pm$ 0.00003$^{\dagger}$ & 3.28 & 1.20 & 348 $\pm$ 11 & 29.10 $\pm$ 1.35 & 2181 $\pm$ 437 & 182.00 $\pm$ 37.10 & 0.82 $\pm$ 0.21 & 1 & II \\
55 & 08 47 46.06 & 38 31 39.33 & Q & 16.56 $\pm$ 0.00 & 0.31399 $\pm$ 0.00001$^{\dagger}$ & 5.10 & 1.45 & 86 $\pm$ 3 & 2.72 $\pm$ 0.10 & - & - & - & 1 & II \\
56 & 08 53 49.78 & 14 52 26.04 & G & 14.80 $\pm$ 0.00 & 0.06933 $\pm$ 0.00002$^{\dagger}$ & 9.18 & 0.75 & 127 $\pm$ 4 & 0.15 $\pm$ 0.01 & - & - & - & 1 & II \\
57 & 08 56 32.99 & 59 57 46.89  & Q & 16.72 $\pm$ 0.00 & 0.28347 $\pm$ 0.00003$^{\dagger}$ & 4.00 & 1.06 & 222 $\pm$ 7 & 5.70 $\pm$ 0.22 & 1406 $\pm$ 281 & 36.00 $\pm$ 7.26 & 0.83 $\pm$ 0.21 & 1 & II \\
58 & 09 01 11.78 & 29 43 38.00 & G & 17.31 $\pm$ 0.01 & 0.21979 $\pm$ 0.00003$^{\dagger}$ & 4.30 & 0.94 & 502 $\pm$ 15 & 7.18 $\pm$ 0.22 & - & - & - & 1 & II \\
59 & 09 01 23.31 & 19 14 17.12  & G & 17.87 $\pm$ 0.01 & 0.27649 $\pm$ 0.00004$^{\dagger}$ & 4.36 & 1.13 & 337 $\pm$ 10 & 7.81 $\pm$ 0.29 & 1416 $\pm$ 286 & 32.80 $\pm$ 6.67 & 0.64 $\pm$ 0.21 & 1 & II \\
60 & 09 06 40.80 & 14 25 22.97 & G & 17.01 $\pm$ 0.00 & 0.12600 $\pm$ 0.01180 & 8.33 & 1.16 & 25 $\pm$ 2 & 0.11 $\pm$ 0.02 & - & - & - & 1 & II \\
61 & 09 08 39.13 & 59 45 12.82 & G & 17.04 $\pm$ 0.01 & 0.24004 $\pm$ 0.00003 & 5.55 & 1.30 & 123 $\pm$ 4 & 2.05 $\pm$ 0.08 & 441 $\pm$ 90 & 7.37 $\pm$ 1.51 & 0.57 $\pm$ 0.21 & 1 & II \\
62 & 09 22 56.47 & 24 53 23.68  & Q & 18.35 $\pm$ 0.01 & 0.95360 $\pm$ 0.00024$^{\dagger}$ & 2.94 & 1.44 & 160 $\pm$ 5 & 69.30 $\pm$ 4.89 & 1000 $\pm$ 208 & 433.00 $\pm$ 94.00 & 0.82 $\pm$ 0.22 & 1 & II \\
63 & 09 24 38.24 & 30 28 37.14 & Q & 18.20 $\pm$ 0.01 & 0.27300 $\pm$ 0.00002$^{\dagger}$ & 4.85 & 1.25 & 108 $\pm$ 4 & 2.50 $\pm$ 0.09 & - & - & - & 1 & II \\
64 & 09 25 45.01 & 40 47 39.04  & G & 19.54 $\pm$ 0.03 & 0.45293 $\pm$ 0.00013$^{\dagger}$ & 4.74 & 1.69 & 106 $\pm$ 4 & 7.85 $\pm$ 0.38 & 610 $\pm$ 126 & 45.00 $\pm$ 9.45 & 0.78 $\pm$ 0.22 & 1 & II \\
65 & 09 34 21.58 & 38 23 05.64 & G & 16.41 $\pm$ 0.01 & 0.12299 $\pm$ 0.00004$^{\dagger}$ & 5.99 & 0.82 & 100 $\pm$ 4 & 0.40 $\pm$ 0.02 & - & - & - & 1 & II \\
66$^{i}$ & 09 35 46.29 & $-$22 40 17.30 & G & - & 0.13476 $\pm$ 0.00015$^{\dagger}$ & 5.55 & 0.82 & 26 $\pm$ 2 & 0.13 $\pm$ 0.01 & - & - & - & 2 & I \\
67 & 09 54 55.43 & 40 59 14.11  & Q & 20.21 $\pm$ 0.03 & 0.47798 $\pm$ 0.00005$^{\dagger}$ & 2.78 & 1.02 & 61 $\pm$ 2 & 4.89 $\pm$ 0.29 & 270 $\pm$ 67 & 21.60 $\pm$ 5.46 & 0.66 $\pm$ 0.26 & 1 & II \\
68 & 10 08 25.61 & $-$03 17 34.82 & G & 18.04 $\pm$ 0.01 & 0.24600 $\pm$ 0.02820 & 5.53 & 1.32 & 206 $\pm$ 7 & 3.79 $\pm$ 1.00 & 1115 $\pm$ 235 & 20.50 $\pm$ 6.87 & 0.76 $\pm$ 0.22 & 1 & II \\
69 & 10 08 34.30 & $-$21 39 14.00 & G & - & 0.24600 $\pm$ 0.05000$^{\dagger}$ & 5.13 & 1.23 & 671 $\pm$ 20 & 12.50 $\pm$ 5.77 & 4164 $\pm$ 833 & 77.50 $\pm$ 39.00 & 0.82 $\pm$ 0.21 & 7 & II \\
70 & 10 09 43.50 & 03 37 22.72 & G & 14.78 $\pm$ 0.00 & 0.10513 $\pm$ 0.00002$^{\dagger}$ & 6.31 & 0.75 & 162 $\pm$ 5 & 0.47 $\pm$ 0.01 & - & - & - & 1 & II \\
71 & 10 09 45.37 & 16 25 46.17  & G & 17.88 $\pm$ 0.01 & 0.20800 $\pm$ 0.01490 & 4.10 & 0.86 & 115 $\pm$ 4 & 1.49 $\pm$ 0.24 & 802 $\pm$ 164 & 10.40 $\pm$ 2.70 & 0.87 $\pm$ 0.21 & 1 & II \\
72 & 10 31 29.21 & 16 50 24.53 & G & 18.17 $\pm$ 0.01 & 0.22800 $\pm$ 0.01300 & 5.29 & 1.19 & 91 $\pm$ 3 & 1.39 $\pm$ 0.19 & 404 $\pm$ 85 & 6.16 $\pm$ 1.52 & 0.67 $\pm$ 0.22 & 1 & II \\
73$^{i}$ & 10 46 18.84 & $-$03 26 31.00 & Q & 18.01 $\pm$ 0.01 & 0.83633 $\pm$ 0.00018$^{\dagger}$ & 1.99 & 0.94 & 30 $\pm$ 2 & 8.65 $\pm$ 0.69 & 129 $\pm$ 28 & 36.80 $\pm$ 8.28 & 0.65 $\pm$ 0.23 & 2 & II \\
74 & 10 46 32.22 & 54 35 59.69  & Q & 17.37 $\pm$ 0.01 & 0.14475 $\pm$ 0.00002$^{\dagger}$ & 5.90 & 0.93 & 288 $\pm$ 9 & 1.66 $\pm$ 0.06 & 1554 $\pm$ 312 & 8.97 $\pm$ 1.80 & 0.75 $\pm$ 0.21 & 1 & II \\
75 & 10 52 24.06 & 37 30 04.53  & Q & 18.70 $\pm$ 0.01 & 0.37417 $\pm$ 0.00003$^{\dagger}$ & 4.05 & 1.29 & 369 $\pm$ 11 & 17.00 $\pm$ 0.71 & 1660 $\pm$ 333 & 76.30 $\pm$ 15.50 & 0.67 $\pm$ 0.21 & 1 & II \\
76 & 10 53 09.33 & 26 01 42.13  & G & 18.86 $\pm$ 0.01 & 0.20300 $\pm$ 0.03420 & 5.54 & 1.14 & 315 $\pm$ 9 & 3.77 $\pm$ 1.42 & 1610 $\pm$ 322 & 19.30 $\pm$ 8.22 & 0.73 $\pm$ 0.21 & 1 & II \\
77 & 10 55 39.78 & 14 33 52.23  & G & 18.00 $\pm$ 0.01 & 0.25900 $\pm$ 0.03080 & 4.84 & 1.20 & 362 $\pm$ 11 & 7.37 $\pm$ 2.01 & 1722 $\pm$ 345 & 35.00 $\pm$ 11.80 & 0.70 $\pm$ 0.21 & 1 & II \\
78 & 11 24 22.77 & 15 09 57.90 & G & 16.36 $\pm$ 0.00 & 0.17194 $\pm$ 0.00003$^{\dagger}$ & 5.09 & 0.92 & 52 $\pm$ 3 & 0.44 $\pm$ 0.02 & - & - & - & 1 & II \\
79 & 11 28 00.60 & $-$39 33 16.60 & G & - & 0.10032 $\pm$ 0.00015$^{\dagger}$ & 6.84 & 0.78 & 194 $\pm$ 6 & 0.51 $\pm$ 0.02 & 737 $\pm$ 149 & 1.92 $\pm$ 0.39 & 0.60 $\pm$ 0.21 & 2 & II \\
80$^{i}$ & 11 29 46.01 & $-$01 21 40.55 & Q & 17.53 $\pm$ 0.01 & 0.72638 $\pm$ 0.00003$^{\dagger}$ & 3.60 & 1.61 & 299 $\pm$ 9 & 66.60 $\pm$ 3.86 & 1721 $\pm$ 344 & 383.00 $\pm$ 78.90 & 0.78 $\pm$ 0.21 & 1 & II \\
81 & 11 35 01.03 & 34 44 01.65  & G & 21.79 $\pm$ 0.14 & 0.62100 $\pm$ 0.10070 & 2.75 & 1.15 & 84 $\pm$ 3 & 13.00 $\pm$ 5.18 & 491 $\pm$ 101 & 75.70 $\pm$ 33.80 & 0.79 $\pm$ 0.22 & 1 & II \\
82 & 11 40 53.13 & 25 25 46.53 & Q & 17.74 $\pm$ 0.01 & 0.29878 $\pm$ 0.00002$^{\dagger}$ & 3.97 & 1.09 & 360 $\pm$ 11 & 10.00 $\pm$ 0.39 & 1622 $\pm$ 326 & 45.00 $\pm$ 9.11 & 0.67 $\pm$ 0.21 & 1 & II \\
83 & 11 43 44.42 & 22 29 06.80  & G & 16.56 $\pm$ 0.00 & 0.18079 $\pm$ 0.00003$^{\dagger}$ & 4.23 & 0.80 & 43 $\pm$ 2 & 0.40 $\pm$ 0.02 & - & - & - & 1 & II \\
84 & 11 44 27.19 & 37 08 31.87 & G & 16.24 $\pm$ 0.00 & 0.11482 $\pm$ 0.00002$^{\dagger}$ & 6.64 & 0.86 & 2079 $\pm$ 62 & 7.24 $\pm$ 0.23 & 9212 $\pm$ 1843 & 32.10 $\pm$ 6.43 & 0.67 $\pm$ 0.21 & 1 & II \\
85 & 11 47 38.90 & $-$38 15 42.00 & G & - & 0.08853 $\pm$ 0.00015$^{\dagger}$ & 7.43 & 0.76 & 639 $\pm$ 19 & 1.29 $\pm$ 0.04 & 2893 $\pm$ 580 & 5.84 $\pm$ 1.17 & 0.68 $\pm$ 0.21 & 2 & II \\
86 & 11 49 06.70 & $-$12 04 33.00 & G & - & 0.11700 $\pm$ 0.00100$^{\dagger}$ & 6.93 & 0.91 & 1655 $\pm$ 50 & 6.04 $\pm$ 0.22 & 8270 $\pm$ 1654 & 30.20 $\pm$ 6.07 & 0.72 $\pm$ 0.21 & 11 & II \\
87 & 12 03 43.71 & 23 43 04.72  & G & 16.66 $\pm$ 0.01 & 0.17670 $\pm$ 0.00005$^{\dagger}$ & 5.59 & 1.03 & 409 $\pm$ 12 & 3.59 $\pm$ 0.12 & 1858 $\pm$ 374 & 16.30 $\pm$ 3.29 & 0.68 $\pm$ 0.21 & 1 & II \\
88 & 12 07 05.20 & $-$27 41 47.00 & G & - & 0.02507 $\pm$ 0.00005$^{\dagger}$ & 32.90 & 1.01 & 331 $\pm$ 11 & 0.05 $\pm$ 0.00 & - & - & - & 16 & I \\
89 & 12 08 55.60 & 46 41 13.79 & G & 16.12 $\pm$ 0.00 & 0.10096 $\pm$ 0.00001$^{\dagger}$ & 6.48 & 0.75 & 163 $\pm$ 5 & 0.44 $\pm$ 0.01 & - & - & - & 1 & I \\
90 & 12 16 06.10 & 20 20 54.80 & G & 18.39 $\pm$ 0.01 & 0.26300 $\pm$ 0.01830 & 4.21 & 1.06 & 213 $\pm$ 7 & 4.37 $\pm$ 0.71 & 792 $\pm$ 161 & 16.20 $\pm$ 4.20 & 0.59 $\pm$ 0.21 & 1 & II \\
91 & 12 16 15.21 & 16 24 32.28  & G & 19.53 $\pm$ 0.03 & 0.45908 $\pm$ 0.00003$^{\dagger}$ & 5.09 & 1.83 & 147 $\pm$ 5 & 10.80 $\pm$ 0.51 & 677 $\pm$ 137 & 49.60 $\pm$ 10.20 & 0.68 $\pm$ 0.21 & 1 & II \\
92 & 12 24 44.60 & $-$25 44 42.00 & G & - & 0.16474 $\pm$ 0.00015$^{\dagger}$ & 5.43 & 0.95 & 180 $\pm$ 6 & 1.32 $\pm$ 0.05 & 558 $\pm$ 115 & 4.10 $\pm$ 0.85 & 0.51 $\pm$ 0.22 & 2 & II \\
93 & 12 39 08.61 & 00 18 33.00 & G & 18.90 $\pm$ 0.02 & 0.33005 $\pm$ 0.01000 & 4.70 & 1.38 & 153 $\pm$ 5 & 5.27 $\pm$ 0.43 & 663 $\pm$ 135 & 22.90 $\pm$ 4.97 & 0.66 $\pm$ 0.21 & 5 & II \\
94 & 12 39 44.96 & 19 54 25.41 & Q & 17.52 $\pm$ 0.01 & 0.23936 $\pm$ 0.00006$^{\dagger}$ & 4.38 & 1.03 & 121 $\pm$ 4 & 2.05 $\pm$ 0.08 & 524 $\pm$ 109 & 8.87 $\pm$ 1.86 & 0.66 $\pm$ 0.22 & 1 & II \\
95 & 12 51 54.40 & 35 19 11.97  & G & 17.81 $\pm$ 0.01 & 0.20100 $\pm$ 0.02060 & 5.05 & 1.03 & 179 $\pm$ 6 & 2.15 $\pm$ 0.50 & 1227 $\pm$ 246 & 14.70 $\pm$ 4.49 & 0.86 $\pm$ 0.21 & 1 & II \\
96 & 12 52 04.82 & $-$22 26 45.70 & G & 16.51 $\pm$ 0.00 & 0.13400 $\pm$ 0.00860 & 5.00 & 0.74 & 116 $\pm$ 4 & 0.56 $\pm$ 0.08 & 432 $\pm$ 100 & 2.07 $\pm$ 0.56 & 0.59 $\pm$ 0.24 & 1 & II \\
97 & 12 53 08.78 & $-$01 39 51.73 & G & 18.65 $\pm$ 0.01 & 0.20507 $\pm$ 0.01000 & 5.29 & 1.10 & 169 $\pm$ 5 & 1.99 $\pm$ 0.23 & 541 $\pm$ 118 & 6.37 $\pm$ 1.55 & 0.52 $\pm$ 0.23 & 5 & II \\
98 & 12 58 09.42 & 42 11 09.18 & G & 17.02 $\pm$ 0.01 & 0.24863 $\pm$ 0.00004$^{\dagger}$ & 4.44 & 1.07 & 310 $\pm$ 9 & 5.61 $\pm$ 0.21 & 1133 $\pm$ 229 & 20.50 $\pm$ 4.16 & 0.58 $\pm$ 0.21 & 1 & II \\
99 & 13 01 52.71 & 14 48 43.01 & G & 15.51 $\pm$ 0.00 & 0.13761 $\pm$ 0.00002$^{\dagger}$ & 5.66 & 0.85 & 133 $\pm$ 5 & 0.69 $\pm$ 0.02 & - & - & - & 1 & II \\
100$^{i}$ & 13 08 40.02 & $-$04 56 45.00 & G & - & 0.13520 $\pm$ 0.00021$^{\dagger}$ & 5.03 & 0.75 & 67 $\pm$ 3 & 0.33 $\pm$ 0.01 & - & - & - & 3 & I \\
101 & 13 12 31.35 & 21 15 43.42  & G & 17.22 $\pm$ 0.01 & 0.17086 $\pm$ 0.00001$^{\dagger}$ & 4.78 & 0.86 & 460 $\pm$ 14 & 3.80 $\pm$ 0.13 & 2499 $\pm$ 501 & 20.60 $\pm$ 4.15 & 0.76 $\pm$ 0.21 & 1 & II \\
102 & 13 24 04.20 & 43 34 07.13  & Q & 18.27 $\pm$ 0.01 & 0.33789 $\pm$ 0.00003$^{\dagger}$ & 4.00 & 1.19 & 252 $\pm$ 8 & 9.06 $\pm$ 0.37 & 994 $\pm$ 200 & 35.70 $\pm$ 7.24 & 0.61 $\pm$ 0.21 & 1 & II \\
103 & 13 37 42.35 & 29 42 23.31 & G & 15.16 $\pm$ 0.00 & 0.11547 $\pm$ 0.00002$^{\dagger}$ & 7.12 & 0.92 & 221 $\pm$ 7 & 0.75 $\pm$ 0.02 & 443 $\pm$ 95 & 1.50 $\pm$ 0.32 & 0.31 $\pm$ 0.22 & 1 & II \\
104 & 13 52 28.39 & 09 35 36.02  & G & 17.15 $\pm$ 0.01 & 0.14735 $\pm$ 0.00001$^{\dagger}$ & 5.88 & 0.94 & 219 $\pm$ 7 & 1.30 $\pm$ 0.04 & 1054 $\pm$ 217 & 6.28 $\pm$ 1.29 & 0.70 $\pm$ 0.21 & 1 & I \\
105 & 14 07 00.50 & 10 09 18.77 & G & 16.90 $\pm$ 0.01 & 0.14262 $\pm$ 0.00002$^{\dagger}$ & 5.55 & 0.86 & 407 $\pm$ 13 & 2.28 $\pm$ 0.07 & - & - & - & 1 & II \\
106 & 14 19 47.89 & 08 14 23.39 & G & 18.43 $\pm$ 0.02 & 0.32180 $\pm$ 0.00005$^{\dagger}$ & 3.70 & 1.07 & 298 $\pm$ 9 & 9.74 $\pm$ 0.39 & 1318 $\pm$ 265 & 43.10 $\pm$ 8.73 & 0.67 $\pm$ 0.21 & 1 & II \\
107 & 14 35 48.55 & 20 13 21.38 & Q & 18.55 $\pm$ 0.01 & 0.36800 $\pm$ 0.00001$^{\dagger}$ & 3.20 & 1.01 & 153 $\pm$ 5 & 7.07 $\pm$ 0.30 & 925 $\pm$ 188 & 42.70 $\pm$ 8.76 & 0.80 $\pm$ 0.21 & 17 & II \\
108 & 14 44 08.73 & 26 01 26.43  & G & 18.57 $\pm$ 0.01 & 0.22900 $\pm$ 0.01890 & 4.11 & 0.93 & 258 $\pm$ 8 & 4.08 $\pm$ 0.77 & 1488 $\pm$ 299 & 23.50 $\pm$ 6.44 & 0.78 $\pm$ 0.21 & 1 & II \\
109$^{i}$ & 14 49 28.63 & $-$01 16 17.44 & G & 16.36 $\pm$ 0.00 & 0.20233 $\pm$ 0.00003$^{\dagger}$ & 4.48 & 0.92 & 86 $\pm$ 3 & 1.02 $\pm$ 0.04 & - & - & - & 1 & II \\
110 & 14 53 46.50 & 22 43 14.96 & G & 19.16 $\pm$ 0.02 & 0.33200 $\pm$ 0.03620 & 3.71 & 1.09 & 125 $\pm$ 4 & 4.57 $\pm$ 1.17 & 756 $\pm$ 153 & 27.60 $\pm$ 8.97 & 0.80 $\pm$ 0.21 & 1 & II \\
111 & 14 54 43.40 & 12 25 10.39  & G & 15.81 $\pm$ 0.00 & 0.12199 $\pm$ 0.00002$^{\dagger}$ & 6.90 & 0.94 & 95 $\pm$ 3 & 0.38 $\pm$ 0.01 & - & - & - & 1 & II \\
112 & 15 01 48.34 & 43 46 32.50  & G & 20.29 $\pm$ 0.04 & 0.49979 $\pm$ 0.00004$^{\dagger}$ & 3.78 & 1.42 & 326 $\pm$ 10 & 31.70 $\pm$ 1.53 & 2453 $\pm$ 498 & 239.00 $\pm$ 49.30 & 0.90 $\pm$ 0.21 & 1 & II \\
113 & 15 07 25.32 & 08 29 44.51  & G & 14.20 $\pm$ 0.00 & 0.07857 $\pm$ 0.00001$^{\dagger}$ & 8.32 & 0.77 & 315 $\pm$ 10 & 0.50 $\pm$ 0.02 & - & - & - & 1 & II \\
114 & 15 16 11.83 & $-$38 19 21.50 & G & - & 0.03914 $\pm$ 0.00012$^{\dagger}$ & 17.11 & 0.82 & 447 $\pm$ 14 & 0.17 $\pm$ 0.01 & - & - & - & 4 & II \\
115 & 15 41 31.89 & $-$27 07 36.57 & G & - & 0.23860 $\pm$ 0.01000$^{\dagger}$ & 4.34 & 1.01 & 194 $\pm$ 6 & 3.25 $\pm$ 0.33 & 789 $\pm$ 161 & 13.20 $\pm$ 2.97 & 0.63 $\pm$ 0.21 & 1 & II \\
116 & 15 46 39.13 & 01 24 22.04 & G & 16.27 $\pm$ 0.00 & 0.20836 $\pm$ 0.00004$^{\dagger}$ & 4.40 & 0.93 & 158 $\pm$ 5 & 1.94 $\pm$ 0.07 & 564 $\pm$ 116 & 6.94 $\pm$ 1.43 & 0.57 $\pm$ 0.21 & 1 & II \\
117 & 15 48 17.75 & 07 25 54.85  & G & 18.79 $\pm$ 0.02 & 0.24700 $\pm$ 0.02740 & 6.94 & 1.66 & 765 $\pm$ 23 & 14.10 $\pm$ 3.60 & 4020 $\pm$ 804 & 74.30 $\pm$ 23.90 & 0.74 $\pm$ 0.21 & 1 & II \\
118 & 15 51 40.30 & 10 35 48.66  & G & 18.00 $\pm$ 0.02 & 0.36824 $\pm$ 0.00008$^{\dagger}$ & 3.95 & 1.25 & 517 $\pm$ 16 & 23.20 $\pm$ 0.96 & 2576 $\pm$ 515 & 116.00 $\pm$ 23.40 & 0.72 $\pm$ 0.21 & 1 & II \\
119 & 16 00 27.78 & 08 37 43.04 & Q & 17.64 $\pm$ 0.01 & 0.22687 $\pm$ 0.00003$^{\dagger}$ & 4.58 & 1.03 & 946 $\pm$ 28 & 14.40 $\pm$ 0.51 & 4577 $\pm$ 916 & 69.60 $\pm$ 14.00 & 0.71 $\pm$ 0.21 & 1 & II \\
120 & 16 05 13.74 & 07 11 52.56 & G & 18.43 $\pm$ 0.01 & 0.31117 $\pm$ 0.00004$^{\dagger}$ & 4.66 & 1.32 & 262 $\pm$ 8 & 8.05 $\pm$ 0.32 & 1287 $\pm$ 258 & 39.50 $\pm$ 7.99 & 0.71 $\pm$ 0.21 & 1 & II \\
121 & 16 05 30.66 & $-$09 27 28.99 & G & - & 0.10900 $\pm$ 0.00100$^{\dagger}$ & 6.84 & 0.84 & 3399 $\pm$ 102 & 10.70 $\pm$ 0.40 & 17554 $\pm$ 3511 & 55.20 $\pm$ 11.10 & 0.74 $\pm$ 0.21 & 9 & II \\
122 & 16 11 57.36 & 07 38 56.19 & G & 17.80 $\pm$ 0.01 & 0.19000 $\pm$ 0.02360 & 6.60 & 1.29 & 150 $\pm$ 5 & 1.56 $\pm$ 0.43 & 790 $\pm$ 164 & 8.20 $\pm$ 2.83 & 0.74 $\pm$ 0.22 & 1 & II \\
123 & 16 12 42.06 & 43 13 19.82  & G & 18.14 $\pm$ 0.01 & 0.25069 $\pm$ 0.00003$^{\dagger}$ & 5.70 & 1.38 & 116 $\pm$ 4 & 2.21 $\pm$ 0.09 & 581 $\pm$ 119 & 11.00 $\pm$ 2.28 & 0.72 $\pm$ 0.21 & 1 & II \\
124 & 16 15 34.52 & 09 57 09.96 & Q & 17.18 $\pm$ 0.01 & 0.23423 $\pm$ 0.00001$^{\dagger}$ & 5.18 & 1.19 & 106 $\pm$ 4 & 1.74 $\pm$ 0.06 & - & - & - & 1 & I \\
125 & 17 06 27.63 & 10 24 53.72  & G & 19.26 $\pm$ 0.02 & 0.32000 $\pm$ 0.04470 & 5.33 & 1.54 & 508 $\pm$ 15 & 17.10 $\pm$ 5.58 & 3173 $\pm$ 635 & 107.00 $\pm$ 40.70 & 0.82 $\pm$ 0.21 & 1 & II \\
126 & 17 21 07.89 & 26 24 32.17  & G & 17.68 $\pm$ 0.01 & 0.16963 $\pm$ 0.00002$^{\dagger}$ & 5.38 & 0.96 & 236 $\pm$ 7 & 1.90 $\pm$ 0.07 & 1090 $\pm$ 221 & 8.77 $\pm$ 1.78 & 0.68 $\pm$ 0.21 & 1 & II \\
127$^{i}$ & 17 21 09.49 & 35 42 16.09  & Q & 17.77 $\pm$ 0.01 & 0.28321 $\pm$ 0.00002$^{\dagger}$ & 3.95 & 1.05 & 803 $\pm$ 24 & 17.90 $\pm$ 0.67 & 1473 $\pm$ 297 & 32.80 $\pm$ 6.67 & 0.27 $\pm$ 0.21 & 1 & II \\
128 & 17 31 05.95 & 24 28 51.85 & G & 20.59 $\pm$ 0.05 & 0.55200 $\pm$ 0.03730 & 2.84 & 1.13 & 288 $\pm$ 9 & 33.20 $\pm$ 5.66 & 1542 $\pm$ 310 & 178.00 $\pm$ 46.60 & 0.75 $\pm$ 0.21 & 1 & II \\
129 & 17 42 06.97 & 18 27 20.66 & Q & 16.49 $\pm$ 0.00 & 0.18588 $\pm$ 0.00006$^{\dagger}$ & 5.70 & 1.10 & 1075 $\pm$ 32 & 10.80 $\pm$ 0.37 & 7046 $\pm$ 1411 & 70.90 $\pm$ 14.20 & 0.84 $\pm$ 0.21 & 19 & II \\
130 & 18 08 23.17 & 08 41 03.00 & G & - & 0.04871 $\pm$ 0.00012$^{\dagger}$ & 30.50 & 1.80 & 467 $\pm$ 15 & 0.27 $\pm$ 0.01 & - & - & - & 4 & HM? \\
131 & 18 20 07.41 & 22 51 20.04 & G & 18.68 $\pm$ 0.02 & 0.33400 $\pm$ 0.01920 & 4.89 & 1.45 & 209 $\pm$ 7 & 7.83 $\pm$ 1.09 & 1401 $\pm$ 281 & 52.50 $\pm$ 12.70 & 0.85 $\pm$ 0.21 & 1 & II \\
132 & 18 29 38.62 & 22 30 08.63  & G & 18.05 $\pm$ 0.01 & 0.24000 $\pm$ 0.07050 & 5.07 & 1.19 & 141 $\pm$ 5 & 2.42 $\pm$ 1.61 & 652 $\pm$ 136 & 11.20 $\pm$ 7.77 & 0.69 $\pm$ 0.22 & 1 & II \\
133 & 19 03 04.60 & 36 16 50.92 & G & - & 0.07084 $\pm$ 0.00012$^{\dagger}$ & 10.08 & 0.84 & 473 $\pm$ 14 & 0.59 $\pm$ 0.02 & 1280 $\pm$ 259 & 1.60 $\pm$ 0.32 & 0.45 $\pm$ 0.21 & 4 & II \\
134 & 19 43 49.35 & $-$35 46 46.10  & G & - & 0.09258 $\pm$ 0.00015$^{\dagger}$ & 8.58 & 0.91 & 361 $\pm$ 11 & 0.79 $\pm$ 0.03 & 1373 $\pm$ 297 & 3.02 $\pm$ 0.65 & 0.60 $\pm$ 0.22 & 2 & II \\
135 & 19 58 30.60 & $-$37 38 26.00 & G & - & 0.09458 $\pm$ 0.00015$^{\dagger}$ & 13.01 & 1.41 & 238 $\pm$ 8 & 0.55 $\pm$ 0.02 & - & - & - & 2 & II \\
136 & 20 19 38.37 & 14 17 01.22  & G & 20.59 $\pm$ 0.06 & 0.23800 $\pm$ 0.03980 & 5.36 & 1.25 & 43 $\pm$ 2 & 0.73 $\pm$ 0.28 & - & - & - & 1 & II \\
137 & 20 40 19.60 & $-$06 59 10.19 & G & 18.81 $\pm$ 0.02 & 0.39769 $\pm$ 0.01000 & 4.50 & 1.49 & 192 $\pm$ 6 & 10.00 $\pm$ 0.74 & 827 $\pm$ 167 & 43.40 $\pm$ 9.22 & 0.65 $\pm$ 0.21 & 5 & II \\
138 & 20 59 47.18 & $-$25 06 11.03 & G & - & 0.19953 $\pm$ 0.00015$^{\dagger}$ & 3.85 & 0.78 & 116 $\pm$ 4 & 1.28 $\pm$ 0.05 & 381 $\pm$ 84 & 4.23 $\pm$ 0.93 & 0.53 $\pm$ 0.23 & 2 & II \\
139 & 21 30 39.21 & 07 35 30.28 & G & 19.09 $\pm$ 0.02 & 0.32285 $\pm$ 0.00002$^{\dagger}$ & 3.41 & 0.99 & 105 $\pm$ 4 & 3.41 $\pm$ 0.15 & 411 $\pm$ 89 & 13.30 $\pm$ 2.91 & 0.61 $\pm$ 0.23 & 1 & II \\
140 & 21 37 45.20 & $-$14 32 54.90 & Q & - & 0.20047 $\pm$ 0.00001$^{\dagger}$ & 4.61 & 0.94 & 3796 $\pm$ 114 & 44.20 $\pm$ 1.52 & 19593 $\pm$ 3919 & 228.00 $\pm$ 45.80 & 0.73 $\pm$ 0.21 & 14 & II \\
141 & 21 49 20.70 & 19 40 43.51 & G & 18.77 $\pm$ 0.02 & 0.39800 $\pm$ 0.07250 & 4.44 & 1.47 & 271 $\pm$ 8 & 14.80 $\pm$ 6.40 & 1548 $\pm$ 310 & 84.80 $\pm$ 40.20 & 0.78 $\pm$ 0.21 & 1 & II \\
142 & 22 15 36.84 & 29 02 35.90 & Q & 15.80 $\pm$ 0.00 & 0.22880 $\pm$ 0.00040$^{\dagger}$ & 4.17 & 0.94 & 577 $\pm$ 17 & 8.79 $\pm$ 0.31 & 2351 $\pm$ 471 & 35.80 $\pm$ 7.21 & 0.63 $\pm$ 0.21 & 20 & II \\
143 & 22 18 15.55 & 19 31 43.76 & G & 15.17 $\pm$ 0.00 & 0.10964 $\pm$ 0.00002$^{\dagger}$ & 6.05 & 0.75 & 223 $\pm$ 7 & 0.69 $\pm$ 0.02 & 739 $\pm$ 149 & 2.31 $\pm$ 0.46 & 0.54 $\pm$ 0.21 & 1 & II \\
144 & 22 31 14.32 & 01 00 41.80 & G & 17.71 $\pm$ 0.01 & 0.21246 $\pm$ 0.00003$^{\dagger}$ & 7.33 & 1.57 & 178 $\pm$ 6 & 2.28 $\pm$ 0.09 & 633 $\pm$ 136 & 8.11 $\pm$ 1.75 & 0.57 $\pm$ 0.22 & 1 & II \\
145 & 22 32 49.13 & $-$05 29 58.10 & Q & 17.75 $\pm$ 0.01 & 0.36870 $\pm$ 0.00015$^{\dagger}$ & 2.98 & 0.94 & 205 $\pm$ 6 & 9.35 $\pm$ 0.32 & - & - & - & 2 & II \\
146 & 22 38 36.14 & $-$07 04 57.93 & G & 18.38 $\pm$ 0.01 & 0.31800 $\pm$ 0.01000 & 6.23 & 1.79 & 46 $\pm$ 2 & 1.63 $\pm$ 0.14 & 484 $\pm$ 99 & 17.20 $\pm$ 3.76 & 1.05 $\pm$ 0.22 & 6 & II \\
147$^{i}$ & 22 46 13.00 & $-$16 54 23.10 & G & - & 0.70226 $\pm$ 0.00018$^{\dagger}$ & 3.86 & 1.70 & 39 $\pm$ 2 & 7.83 $\pm$ 0.42 & - & - & - & 2 & II \\
148 & 22 46 21.66 & 31 42 07.94  & G & 16.35 $\pm$ 0.00 & 0.19500 $\pm$ 0.02080 & 4.64 & 0.93 & 327 $\pm$ 10 & 3.43 $\pm$ 0.82 & 951 $\pm$ 192 & 9.97 $\pm$ 3.11 & 0.48 $\pm$ 0.21 & 1 & II \\
149 & 22 51 03.39 & 06 19 25.49 & G & 17.61 $\pm$ 0.01 & 0.17203 $\pm$ 0.00002$^{\dagger}$ & 4.81 & 0.87 & 97 $\pm$ 3 & 0.77 $\pm$ 0.03 & 256 $\pm$ 62 & 2.04 $\pm$ 0.50 & 0.43 $\pm$ 0.25 & 1 & II \\
150 & 22 51 25.53 & 16 07 53.60 & G & 21.38 $\pm$ 0.11 & 0.42800 $\pm$ 0.08770 & 4.30 & 1.49 & 156 $\pm$ 6 & 10.30 $\pm$ 5.00 & 973 $\pm$ 202 & 63.90 $\pm$ 33.80 & 0.82 $\pm$ 0.22 & 1 & II \\
151 & 22 53 21.28 & 16 20 16.77  & G & 19.88 $\pm$ 0.03 & 0.17500 $\pm$ 0.08170 & 9.49 & 1.74 & 304 $\pm$ 10 & 2.66 $\pm$ 2.74 & 1748 $\pm$ 353 & 15.30 $\pm$ 16.00 & 0.78 $\pm$ 0.21 & 1 & II \\
152 & 22 53 56.30 & $-$07 06 38.60 & G & 18.11 $\pm$ 0.01 & 0.23100 $\pm$ 0.11200 & 8.00 & 1.82 & 296 $\pm$ 9 & 4.63 $\pm$ 5.06 & 1266 $\pm$ 255 & 19.80 $\pm$ 22.00 & 0.65 $\pm$ 0.21 & 1 & II \\
153 & 22 59 34.13 & 08 20 40.78  & G & 19.67 $\pm$ 0.04 & 0.40500 $\pm$ 0.04420 & 6.63 & 2.22 & 168 $\pm$ 5 & 9.50 $\pm$ 2.48 & 900 $\pm$ 183 & 50.80 $\pm$ 16.80 & 0.75 $\pm$ 0.21 & 1 & II \\
154 & 23 10 46.70 & $-$21 08 13.90 & G & - & 0.15216 $\pm$ 0.00013$^{\dagger}$ & 4.91 & 0.80 & 439 $\pm$ 13 & 2.75 $\pm$ 0.09 & 1624 $\pm$ 329 & 10.20 $\pm$ 2.07 & 0.59 $\pm$ 0.21 & 10 & II \\
155 & 23 16 22.32 & 22 46 50.28 & G & 19.10 $\pm$ 0.02 & 0.29400 $\pm$ 0.06740 & 8.47 & 2.30 & 91 $\pm$ 4 & 2.49 $\pm$ 1.32 & - & - & - & 1 & II \\
156 & 23 19 56.30 & $-$27 28 12.40 & G & - & 0.17420 $\pm$ 0.00021$^{\dagger}$ & 5.97 & 1.09 & 2826 $\pm$ 85 & 24.10 $\pm$ 0.81 & 13186 $\pm$ 2637 & 112.00 $\pm$ 22.50 & 0.69 $\pm$ 0.21 & 3 & II \\
157 & 23 23 44.52 & 14 57 59.35 & G & 15.11 $\pm$ 0.00 & 0.12600 $\pm$ 0.01000 & 6.50 & 0.91 & 390 $\pm$ 12 & 1.62 $\pm$ 0.28 & 1107 $\pm$ 223 & 4.59 $\pm$ 1.22 & 0.47 $\pm$ 0.21 & 6 & II \\
158 & 23 25 11.80 & $-$32 36 34.60 & Q & - & 0.21600 $\pm$ 0.00015$^{\dagger}$ & 4.10 & 0.89 & 230 $\pm$ 7 & 3.01 $\pm$ 0.11 & 686 $\pm$ 138 & 8.98 $\pm$ 1.81 & 0.49 $\pm$ 0.21 & 13 & II \\
159 & 23 35 12.35 & 17 41 50.38 & G & 18.70 $\pm$ 0.01 & 0.22800 $\pm$ 0.06330 & 6.00 & 1.36 & 433 $\pm$ 13 & 6.68 $\pm$ 4.18 & 2190 $\pm$ 439 & 33.80 $\pm$ 22.20 & 0.73 $\pm$ 0.21 & 1 & II \\
160 & 23 41 37.14 & 08 28 17.25 & G & - & 0.24700 $\pm$ 0.00004$^{\dagger}$ & 7.00 & 1.68 & 164 $\pm$ 5 & 3.00 $\pm$ 0.11 & 765 $\pm$ 156 & 14.00 $\pm$ 2.87 & 0.69 $\pm$ 0.21 & 1 & II \\
161 & 23 48 47.77 & 16 00 10.32 & G & 20.25 $\pm$ 0.04 & 0.40274 $\pm$ 0.00002$^{\dagger}$ & 4.50 & 1.50 & 117 $\pm$ 4 & 7.28 $\pm$ 0.33 & 1276 $\pm$ 257 & 79.20 $\pm$ 16.10 & 1.07 $\pm$ 0.21 & 1 & II \\
162 & 23 59 11.07 & 17 06 10.97  & G & 16.34 $\pm$ 0.00 & 0.09940 $\pm$ 0.00001$^{\dagger}$ & 9.93 & 1.13 & 421 $\pm$ 13 & 1.10 $\pm$ 0.04 & 2683 $\pm$ 539 & 7.01 $\pm$ 1.41 & 0.83 $\pm$ 0.21 & 1 & II \\
\label{tab:maintab}
\end{longtable}
\tablebib{1- SDSS; \citealt{sdssdr14},  2 -6dF; \citealt{6dFJones}, 3 -2dF; \citealt{2dFColless}, 4 -2MASS; \citealt{2MASSHuchra}, 5 - \citealt{5Delli}, 6 - \citealt{6Gao},
 7 - \citealt{7McCarthy}, 8 - \citealt{8Dressler}, 9 - \citealt{9Best}, 10 - \citealt{10Caretta}, 11 - \citealt{11Danziger}, 12 - \citealt{12Drake}, 13 - \citealt{13Grupe}, 14 - \citealt{14Ho}, 15 - \citealt{15Machalski}, 16 - \citealt{16Kaldare}, 17 - \citealt{17McGreer}, 18 - \citealt{18Ratcliffe}, 19 - \citealt{19Veron}, 20 - \citealt{20Wei}}
\end{landscape}
\end{tiny}

\begin{tiny}
\setlength{\tabcolsep}{2.0pt}
\begin{longtable}{c  c c c c c c c c}

\caption{\label{table_2} The table lists the black hole properties, radio core and jet kinetic power of the SGS. Col.(1) : serial number, Col.(2) : GRG name, Col.(3) : galaxy or quasar type host, Col.(4) $\sigma$ : velocity dispersion (km $\rm s^{-1}$) of host galaxies of GRGs, Col. (5) $\rm M_{BH}$ : central black hole mass ($\rm  
M_{\odot} \times 10^{9} $) estimated from the $\rm M_{BH}$-$\sigma$  relation, Col. (6) $\rm  S_{1400 MHz}^{c}$ : core flux density at 1400 MHz (FIRST), Col. (7) $\lambdaup_{\rm Edd}$ : Eddington ratio, and Col.  (8) $\rm 
Q_{Jet}$ is the jet kinetic power.} \\
\hline

Sr.No & Name & Type  & $\sigma$ & $\rm M_{BH}$ & $\rm  S_{1400 MHz}^{c}$  & $\lambdaup_{\rm Edd}$ & $\rm Q_{Jet}$\\
       &  &  &  (km $\rm s^{-1}$) &  ( $ 10^{9} ~ \rm M_{\odot}  $ )& (mJy) & ($10^{-4}$)  & ($ 10^{43}$ erg $\rm s^{-1}$)\\
(1) & (2) & (3) & (4) & (5) & (6) & (7) & (8) \\
\hline
\endfirsthead
\caption{continued.}\\
\hline
Sr.No & Name & Type & $\sigma$ & $\rm M_{BH}$ & $\rm  S_{1400 MHz}^{c}$  & $\lambdaup_{\rm Edd}$ & $\rm Q_{Jet}$\\
       &  &  &  (km $\rm s^{-1}$) &  ( $ 10^{9} ~ \rm M_{\odot}  $ )& (mJy)  & ($10^{-4}$) & ($ 10^{43}$ erg $\rm s^{-1}$)\\
(1) & (2) & (3) & (4) & (5) & (6) & (7) & (8)  \\
\hline
\endhead
\hline
\endfoot
1 & SAGANJ000450.25+124840.10 & G & - & - & -  & - & 18.80 \\
2 & SAGANJ000623.24+263544.66 & G & - & - & -  & - & 37.60 \\
3 & SAGANJ001119.35+321713.83 & G & 245.65 $\pm$ 9.27 & 0.67 $\pm$ 0.14 & -  & - & 3.59 \\
4 & SAGANJ001502.43+101056.33 & G & - & - & -  & - & 3.83 \\
5 & SAGANJ001741.45+082755.72 & Q & - & - & 33.4 $\pm$ 1.0  & - & 107.00 \\
6 & SAGANJ001815.21+214133.42 & G & 250.78 $\pm$ 18.45 & 0.75 $\pm$ 0.31 & -  & - & 40.20 \\
7 & SAGANJ002613.63+114343.66 & G & - & - & 7.5 $\pm$ 0.2  & - & 5.61 \\
8 & SAGANJ003622.99$-$025821.55 & G & - & - & -  & - & 6.56 \\
9 & SAGANJ003809.89+093601.33 & G & 264.24 $\pm$ 24.68 & 1.01 $\pm$ 0.53 & 0.6 $\pm$ 0.2  & - & 9.86 \\
10 & SAGANJ004517.11+382503.23 & G & - & - & -  & - & 216.00 \\
11 & SAGANJ004653.63+125541.01 & G & - & - & -  & - & 4.31 \\
12 & SAGANJ005548.79$-$223116.50 & G & - & - & -  & - & 0.63 \\
13 & SAGANJ010052.60+061641.01 & G & - & - & -  & - & 0.70 \\
14 & SAGANJ010237.34+042100.26 & G & - & 10.31 $\pm$ 0.35 & - & - & - \\
15 & SAGANJ010340.05+423935.63 & G & - & - & -  & - & 3.31 \\
17 & SAGANJ010803.52+270001.99 & G & - & - & -  & - & 34.90 \\
18 & SAGANJ010936.45+252403.76 & G & - & - & -  & - & 82.40 \\
19 & SAGANJ011104.70$-$142232.60 & G & - & - & -  & - & 0.92 \\
21 & SAGANJ011341.11+010608.52 & G & 228.75 $\pm$ 32.77 & 0.45 $\pm$ 0.36 & 1.2 $\pm$ 0.2  & - & 15.10 \\
22 & SAGANJ011430.75+050830.64 & G & 200.27 $\pm$ 11.28 & 0.21 $\pm$ 0.07 & -  & - & 3.58 \\
24 & SAGANJ012359.92+431255.78 & G & - & - & -  & - & 44.70 \\
25 & SAGANJ012532.16+070337.11 & G & - & - & 1.5 $\pm$ 0.2  & - & - \\
26 & SAGANJ012848.89+243152.10 & G & 219.19 $\pm$ 12.93 & 0.35 $\pm$ 0.12 & -  & - & - \\
27 & SAGANJ013327.24$-$082416.52 & G & 309.37 $\pm$ 8.32 & 2.45 $\pm$ 0.37 & 22.6 $\pm$ 0.7  & - & 1.76 \\
28 & SAGANJ013536.11+504038.54 & G & - & - & -  & - & 7.13 \\
29 & SAGANJ014208.56$-$064143.4 & G & - & - & 0.5 $\pm$ 0.1 & -  & 0.58 \\
30 & SAGANJ014849.35+062243.26 & G & - & - & - & - & 30.50 \\
32 & SAGANJ015826.09+245136.38 & G & 291.95 $\pm$ 9.28 & 1.76 $\pm$ 0.32 & -  & - & 0.83 \\
33 & SAGANJ022527.01$-$313737.00 & G & - & - & -  & - & 0.33 \\
35 & SAGANJ032155.75+434640.78 & G & - & - & -  & - & 46.50 \\
37 & SAGANJ034253.70$-$065224.40 & G & - & - & -  & - & 43.00 \\
39 & SAGANJ035045.56$-$181829.80 & G & - & - & -  & - & 1.57 \\
40 & SAGANJ035339.31$-$011319.73 & G & - & - & -  & - & 3.38 \\
41 & SAGANJ042234.14$-$261643.20 & G & - & - & - & - & 9.16 \\
43 & SAGANJ051536.11+151821.31 & G & - & - & -  & - & 4.58 \\
44 & SAGANJ054359.77+304748.28 & G & - & - & -  & - & 3.16 \\
45 & SAGANJ061203.50$-$325747.00 & G & - & - & -  & - & 0.52 \\
47 & SAGANJ070844.11+303012.73 & G & - & - & -  & - & 18.60 \\
48 & SAGANJ072538.75+400412.52 & G & 310.02 $\pm$ 8.49 & 2.48 $\pm$ 0.38 & 39.7 $\pm$ 1.2 & - & - \\
49 & SAGANJ072805.86+503445.80 & G & - & - & - & - & 85.10 \\
50 & SAGANJ073205.49+155831.06 & G & - & - & -  & - & 20.10 \\
51 & SAGANJ075021.33+163259.35 & G & - & - & -  & - & 63.50 \\
53 & SAGANJ082941.34+224758.41 & G & - & - & 2.8 $\pm$ 0.2  & - & 9.08 \\
54 & SAGANJ084718.28+422341.33 & G & 171.43 $\pm$ 31.83 & 0.09 $\pm$ 0.09 & -  & - & 60.80 \\
55 & SAGANJ084746.06+383139.33 & Q & - & - & 6.9 $\pm$ 0.3  & - & - \\
56 & SAGANJ085349.78+145226.04 & G & 282.38 $\pm$ 7.52 & 1.46 $\pm$ 0.22 & 13.1 $\pm$ 0.4  & 0.64 & - \\
57 & SAGANJ085632.99+595746.89 & Q & - & - & 26.3 $\pm$ 0.8  & - & 12.00 \\
58 & SAGANJ090111.78+294338.00 & G & 250.21 $\pm$ 13.82 & 0.74 $\pm$ 0.23 & 15.7 $\pm$ 0.5  & 29.98 & - \\
59 & SAGANJ090123.31+191417.12 & G & 233.86 $\pm$ 21.87 & 0.51 $\pm$ 0.27 & -  & 69.80 & 10.90 \\
60 & SAGANJ090640.80+142522.97 & G & - & - & 2.6 $\pm$ 0.2  & - & - \\
61 & SAGANJ090839.13+594512.82 & G & 305.75 $\pm$ 12.38 & 2.29 $\pm$ 0.52 & 4.3 $\pm$ 0.2  & - & 2.46 \\
62 & SAGANJ092256.47+245323.68 & Q & - & - & 8.9 $\pm$ 0.3  & - & 144.00 \\
63 & SAGANJ092438.24+302837.14 & Q & - & - & 3.9 $\pm$ 0.2  & - & - \\
64 & SAGANJ092545.01+404739.04 & G & 215.24 $\pm$ 39.17 & 0.32 $\pm$ 0.33 & -  & - & 15.00 \\
65 & SAGANJ093421.58+382305.64 & G & 215.21 $\pm$ 17.46 & 0.32 $\pm$ 0.14 & 2.0 $\pm$ 0.2 & 9.69 & - \\
67 & SAGANJ095455.43+405914.11 & Q & - & - & 3.0 $\pm$ 0.2  & - & 7.19 \\
68 & SAGANJ100825.61$-$031734.82 & G & - & - & -  & - & 6.83 \\
69 & SAGANJ100834.30$-$213914.00 & G & - & - & -  & - & 25.80 \\
70 & SAGANJ100943.50+033722.72 & G & 274.69 $\pm$ 7.15 & 1.25 $\pm$ 0.18 & 13.3 $\pm$ 0.4  & - & - \\
71 & SAGANJ100945.37+162546.17 & G & - & - & - & - & 3.47 \\
72 & SAGANJ103129.21+165024.53 & G & - & - & - & - & 2.05 \\
73 & SAGANJ104618.84$-$032631.00 & Q & - & - & 8.4 $\pm$ 0.3  & - & 12.30 \\
74 & SAGANJ104632.22+543559.69 & Q & - & - & 7.8 $\pm$ 0.3  & - & 2.99 \\
75 & SAGANJ105224.06+373004.53 & Q & - & - & 2.0 $\pm$ 0.3  & - & 25.40 \\
76 & SAGANJ105309.33+260142.13 & G & - & - & 1.0 $\pm$ 0.2  & - & 6.43 \\
77 & SAGANJ105539.78+143352.23 & G & - & - & -  & - & 11.70 \\
78 & SAGANJ112422.77+150957.90 & G & 269.40 $\pm$ 10.48 & 1.12 $\pm$ 0.25 & 2.3 $\pm$ 0.3 & 3.49 & - \\
79 & SAGANJ112800.60$-$393316.60 & G & - & - & -  & - & 0.64 \\
80 & SAGANJ112946.01$-$012140.55 & Q & - & - & 7.4 $\pm$ 0.2  & - & 128.00 \\
81 & SAGANJ113501.03+344401.65 & G & - & - & -  & - & 25.20 \\
82 & SAGANJ114053.13+252546.53 & Q & - & - & -  & - & 15.00 \\
83 & SAGANJ114344.42+222906.80 & G & 323.02 $\pm$ 12.69 & 3.12 $\pm$ 0.69 & 1.5 $\pm$ 0.4  & 0.53 & - \\
84 & SAGANJ114427.19+370831.87 & G & 224.78 $\pm$ 9.13 & 0.40 $\pm$ 0.09 & 4.7 $\pm$ 0.4  & 29.19 & 10.70 \\
85 & SAGANJ114738.90$-$381542.00 & G & - & - & -  & - & 1.95 \\
86 & SAGANJ114906.70$-$120433.00 & G & - & - & -  & - & 10.10 \\
87 & SAGANJ120343.71+234304.72 & G & 253.57 $\pm$ 18.06 & 0.80 $\pm$ 0.32  & - & 5.99 & 5.43 \\
89 & SAGANJ120855.60+464113.79 & G & 209.64 $\pm$ 6.31 & 0.27 $\pm$ 0.05 & 63.7 $\pm$ 2.0  & 10.18 & - \\
90 & SAGANJ121606.10+202054.80 & G & - & - & -  & - & 5.41 \\
91 & SAGANJ121615.21+162432.28 & G & 216.58 $\pm$ 38.15 & 0.33 $\pm$ 0.33 & -  & - & 16.50 \\
92 & SAGANJ122444.60$-$254442.00 & G & - & - & -  & - & 1.37 \\
93 & SAGANJ123908.61+001833.00 & G & - & - & 10.0 $\pm$ 0.3  & - & 7.62 \\
94 & SAGANJ123944.96+195425.41 & Q & - & - & 2.4 $\pm$ 0.2  & - & 2.96 \\
95 & SAGANJ125154.40+351911.97 & G & - & - & 3.0 $\pm$ 0.2  & - & 4.91 \\
96 & SAGANJ125204.82$-$222645.70 & G  & - & - & -  & - & 0.69 \\
97 & SAGANJ125308.78$-$013951.73 & G & - & - & -  & - & 2.12 \\
98 & SAGANJ125809.42+421109.18 & G  & 259.38 $\pm$ 15.52 & 0.91 $\pm$ 0.30 & -  & 11.69 & 6.83 \\
99 & SAGANJ130152.71+144843.01 & G & 278.55 $\pm$ 9.42 & 1.35 $\pm$ 0.26 & 4.2 $\pm$ 0.2  & - & - \\
100 & SAGANJ130840.02$-$045645.00 & G & - & - & 24.1 $\pm$ 0.8  & - & - \\
101 & SAGANJ131231.35+211543.42 & G & 246.19 $\pm$ 12.53 & 0.68 $\pm$ 0.19 & 12.2 $\pm$ 0.4  & 61.96 & 6.88 \\
102 & SAGANJ132404.20+433407.13 & Q  & - & - & 3.6 $\pm$ 0.2  & - & 11.90 \\
103 & SAGANJ133742.35+294223.31 & G  & 268.86 $\pm$ 9.55 & 1.11 $\pm$ 0.22 & 3.1 $\pm$ 0.2 & 1.85 & 0.50 \\
104 & SAGANJ135228.39+093536.02 & G & 149.29 $\pm$ 10.56 & 0.04 $\pm$ 0.02 & 39.8 $\pm$ 1.3  & 347.24 & 2.09 \\
105 & SAGANJ140700.50+100918.77 & G & 186.41 $\pm$ 11.70 & 0.14 $\pm$ 0.05 & -  & 75.74 & - \\
106 & SAGANJ141947.89+081423.39 & G  & 251.32 $\pm$ 21.57 & 0.76 $\pm$ 0.37 & - & 51.20 & 14.40 \\
107 & SAGANJ143548.55+201321.38 & Q  & - & - & 7.5 $\pm$ 0.3  & - & 14.20 \\
108 & SAGANJ144408.73+260126.43 & G  & - & - & 1.6 $\pm$ 0.3  & - & 7.82 \\
109 & SAGANJ144928.63$-$011617.44 & G & 325.03 $\pm$ 12.69 & 3.23 $\pm$ 0.71 & 40.2 $\pm$ 1.2  & 1.09 & - \\
110 & SAGANJ145346.50+224314.96 & G  & - & - & 2.8 $\pm$ 0.2  & - & 9.19 \\
111 & SAGANJ145443.40+122510.39 & G & 226.73 $\pm$ 9.89 & 0.42 $\pm$ 0.10 & 16.2 $\pm$ 0.6  & - & - \\
112 & SAGANJ150148.34+434632.50 & G  & 226.67 $\pm$ 45.40 & 0.42 $\pm$ 0.48 & 1.0 $\pm$ 0.2 & - & 79.60 \\
113 & SAGANJ150725.32+082944.51 & G  & 322.86 $\pm$ 7.17 & 3.11 $\pm$ 0.39 & 3.2 $\pm$ 0.1  & 0.83 & - \\
115 & SAGANJ154131.89$-$270736.57 & G  & - & - & - - & - & 4.39 \\
116 & SAGANJ154639.13+012422.04 & G  & 353.67 $\pm$ 14.72 & 5.20 $\pm$ 1.22 & 2.3 $\pm$ 0.2  & 1.26 & 2.31 \\
117 & SAGANJ154817.75+072554.85 & G & - & - & 2.3 $\pm$ 0.2  & - & 24.80 \\
118 & SAGANJ155140.30+103548.66 & G  & 241.95 $\pm$ 26.76 & 0.61 $\pm$ 0.38 & 4.3 $\pm$ 0.2  & 71.83 & 38.60 \\
119 & SAGANJ160027.78+083743.04 & Q  & - & - & -  & - & 23.20 \\
120 & SAGANJ160513.74+071152.56 & G & 194.99 $\pm$ 21.78 & 0.18 $\pm$ 0.11 & -  & - & 13.20 \\
121 & SAGANJ160530.66$-$092728.99 & G  & - & - & - & - & 18.40 \\
122 & SAGANJ161157.36+073856.19 & G  & - & - & 12.2 $\pm$ 0.4  & - & 2.73 \\
123 & SAGANJ161242.06+431319.82 & G  & 240.66 $\pm$ 14.56 & 0.59 $\pm$ 0.20 & -  & - & 3.68 \\
124 & SAGANJ161534.52+095709.96 & Q  & - & - & 7.1 $\pm$ 0.3  & - & - \\
125 & SAGANJ170627.63+102453.72 & G  & - & - & - & - & 35.60 \\
126 & SAGANJ172107.89+262432.17 & G & 181.63 $\pm$ 23.37 & 0.12 $\pm$ 0.09 & 2.6 $\pm$ 0.2  & 44.30 & 2.92 \\
127 & SAGANJ172109.49+354216.09 & Q  & - & - & 393.0 $\pm$ 11.8  & - & 10.90 \\
128 & SAGANJ173105.95+242851.85 & G  & - & - & -  & - & 59.20 \\
129 & SAGANJ174206.97+182720.66 & Q  & - & - & -  & - & 23.60 \\
131 & SAGANJ182007.41+225120.04 & G  & - & - & -  & - & 17.50 \\
132 & SAGANJ182938.62+223008.63 & G  & - & - & -  & - & 3.72 \\
133 & SAGANJ190304.60+361650.92 & G  & - & - & -  & - & 0.53 \\
134 & SAGANJ194349.35$-$354646.10 & G  & - & -  & - & - & 1.01 \\
137 & SAGANJ204019.60$-$065910.19 & G  & - & - & -  & - & 14.50 \\
138 & SAGANJ205947.18$-$250611.03 & G  & - & - & -  & - & 1.41 \\
139 & SAGANJ213039.21+073530.28 & G & 208.24 $\pm$ 21.71 & 0.26 $\pm$ 0.15 & -  & - & 4.45 \\
140 & SAGANJ213745.20$-$143254.90 & Q & - & - & -  & - & 76.10 \\
141 & SAGANJ214920.70+194043.51 & G  & - & - & - & - & 28.30 \\
142 & SAGANJ221536.84+290235.90 & Q  & - & - & -  & - & 11.90 \\
143 & SAGANJ221815.55+193143.76 & G  & 273.87 $\pm$ 7.91 & 1.23 $\pm$ 0.20 & -  & - & 0.77 \\
144 & SAGANJ223114.32+010041.80 & G  & 276.10 $\pm$ 13.64 & 1.29 $\pm$ 0.36 & 6.1 $\pm$ 0.2  & - & 2.70 \\
145 & SAGANJ223249.13$-$052958.10 & Q  & - & - & 21.8 $\pm$ 0.7  & - & - \\
146 & SAGANJ223836.14$-$070457.93 & G  & - & -  & - & - & 5.72 \\
148 & SAGANJ224621.66+314207.94 & G  & - & - & -  & - & 3.32 \\
149 & SAGANJ225103.39+061925.49 & G & 199.24 $\pm$ 9.68 & 0.20 $\pm$ 0.06 & 3.2 $\pm$ 0.2  & - & 0.68 \\
150 & SAGANJ225125.53+160753.60 & G  & - & - & -  & - & 21.30 \\
151 & SAGANJ225321.28+162016.77 & G & - & - & -  & - & 5.09 \\
152 & SAGANJ225356.30$-$070638.60 & G  & - & - & -  & - & 6.60 \\
153 & SAGANJ225934.13+082040.78 & G  & - & - & 1.5 $\pm$ 0.2 & - & 16.90 \\
154 & SAGANJ231046.70$-$210813.90 & G  & - & - & -  & - & 3.40 \\
156 & SAGANJ231956.30$-$272812.40 & G  & - & - & -  & - & 37.50 \\
157 & SAGANJ232344.52+145759.35 & G  & - & - & -  & - & 1.53 \\
158 & SAGANJ232511.80$-$323634.60 & Q  & - & - & - & - & 2.99 \\
159 & SAGANJ233512.35+174150.38 & G  & - & - & - & - & 11.30 \\
160 & SAGANJ234137.14+082817.25 & G & 289.43 $\pm$ 15.33 & 1.68 $\pm$ 0.50 & 1.3 $\pm$ 0.2  & - & 4.66 \\
161 & SAGANJ234847.77+160010.32 & G  & 243.11 $\pm$ 15.53 & 0.63 $\pm$ 0.23 & - & - & 26.40 \\
162 & SAGANJ235911.07+170610.97 & G  & 198.95 $\pm$ 6.42 & 0.20 $\pm$ 0.04 & -  & - & 2.34 \\

\label{tab:table_2}
\end{longtable}


\begin{center}
\setlength{\tabcolsep}{4.5pt}
\begin{longtable}{c c c c c c c}
\caption{ The table lists the jet kinetic power and black hole properties of GRGs from \texttt{GRG catalogue} as discussed in Sec.\ \ref{sec:bhprop}. Col. (1) indicates the sample from which the GRGs originate, where `Others' refers to GRGs from the GRG catalogue that are not part of LoTSS or SGS. Col. (2) and Col. (3) represent the right ascension (RA) in HMS and declination (Dec) in DMS of the GRG host galaxies. Col. (4) consists of the jet kinetic power (Q$_{\rm Jet}$). Col. (5) lists $\rm M_{BH}$ : the central black hole mass ($ 10^{9}$ $ \rm M_{\odot} $) estimated from the $\rm M_{BH}$-$\sigma$  relation. Col. (6) includes the B$_{\rm Edd}$: Eddington magnetic field ($ 10^{4} $ G), and Col. (7) lists the estimated GRG spin.}\\
\hline
Sample & RA & Dec & Q$_{\rm Jet}$ & M$_{\rm BH}$ & B$_{\rm Edd}$ & Spin \\ 
 & (HMS) & (DMS)  & ($10^{43}$ erg s$^{-1}$) & ($10^{9}$ M$_{\odot}$)& ($10^{4}$ G)  &   \\
(1) & (2) & (3) & (4) & (5) & (6) & (7) \\
\hline
\endfirsthead
\caption{continued.}\\
\hline
Sample & RA & Dec & Q$_{\rm Jet}$ & M$_{\rm BH}$ & B$_{\rm Edd}$ & Spin \\ 
& (HMS) & (DMS)  & ($10^{43}$ erg s$^{-1}$) & ($10^{9}$ M$_{\odot}$)& ($10^{4}$ G)  &   \\
(1) & (2) & (3) & (4) & (5) & (6) & (7) \\
\hline
\endhead
\hline
\endfoot

SGS & 00 11 19.35 & 32 17 13.83 & 3.60 $\pm$ 0.72 & 0.76 $\pm$ 0.16 & 2.18 $\pm$ 0.23 & 0.08 $\pm$ 0.02 \\ 
SGS & 00 18 15.21 & 21 41 33.42 & 40.60 $\pm$ 8.13 & 0.83 $\pm$ 0.29 & 2.08 $\pm$ 0.36 & 0.26 $\pm$ 0.10 \\ 
SGS & 00 38 09.89 & 09 36 01.33 & 9.91 $\pm$ 1.98 & 1.05 $\pm$ 0.45 & 1.86 $\pm$ 0.40 & 0.11 $\pm$ 0.06 \\ 
SGS & 01 14 30.75 & 05 08 30.64 & 3.67 $\pm$ 0.73 & 0.31 $\pm$ 0.08 & 3.40 $\pm$ 0.46 & 0.13 $\pm$ 0.04 \\ 
SGS & 01 33 27.24 & $-$08 24 16.52 & 1.79 $\pm$ 0.36 & 2.09 $\pm$ 0.43 & 1.31 $\pm$ 0.14 & 0.03 $\pm$ 0.01 \\ 
SGS & 01 58 26.09 & 24 51 36.38 & 0.85 $\pm$ 0.17 & 1.62 $\pm$ 0.34 & 1.49 $\pm$ 0.16 & 0.03 $\pm$ 0.01 \\ 
Others & 07 46 33.68 & 17 08 09.63 & 0.81 $\pm$ 0.17 & 0.33 $\pm$ 0.13 & 3.28 $\pm$ 0.64 & 0.06 $\pm$ 0.03 \\ 
Others & 07 51 08.93 & 42 31 23.6 & 2.70 $\pm$ 0.54 & 0.94 $\pm$ 0.21 & 1.96 $\pm$ 0.22 & 0.06 $\pm$ 0.02 \\ 
Others & 08 57 01.76 & 01 31 30.93 & 2.89 $\pm$ 0.58 & 1.35 $\pm$ 0.67 & 1.63 $\pm$ 0.41 & 0.05 $\pm$ 0.03 \\ 
SGS & 09 01 23.31 & 19 14 17.12 & 11.20 $\pm$ 2.25 & 0.61 $\pm$ 0.26 & 2.42 $\pm$ 0.52 & 0.16 $\pm$ 0.08 \\ 
Others & 09 01 36.70 & 21 46 33.80 & 1.12 $\pm$ 0.25 & 4.67 $\pm$ 2.67 & 0.88 $\pm$ 0.25 & 0.02 $\pm$ 0.01 \\ 
SGS & 09 08 39.13 & 59 45 12.82 & 2.55 $\pm$ 0.51 & 1.98 $\pm$ 0.48 & 1.35 $\pm$ 0.16 & 0.04 $\pm$ 0.01 \\ 
Others & 09 11 54.57 & 08 12 31.13 & 6.64 $\pm$ 1.33 & 0.66 $\pm$ 0.19 & 2.33 $\pm$ 0.33 & 0.12 $\pm$ 0.04 \\ 
Others & 09 32 38.30 & 16 11 57.22 & 14.30 $\pm$ 2.85 & 0.77 $\pm$ 0.23 & 2.16 $\pm$ 0.32 & 0.16 $\pm$ 0.06 \\ 
Others & 10 21 31.84 & 05 19 02.90 & 1.00 $\pm$ 0.02 & 0.94 $\pm$ 0.26 & 1.96 $\pm$ 0.28 & 0.04 $\pm$ 0.01 \\ 
Others & 10 32 58.88 & 56 44 53.27 & 0.22 $\pm$ 0.04 & 1.34 $\pm$ 0.23 & 1.64 $\pm$ 0.14 & 0.01 $\pm$ 0.00 \\ 
Others & 10 34 03.90 & 18 40 49.00 & 4.43 $\pm$ 0.89 & 0.54 $\pm$ 0.15 & 2.58 $\pm$ 0.36 & 0.11 $\pm$ 0.03 \\ 
Others & 10 36 36.30 & 38 35 07.50 & 0.81 $\pm$ 0.16 & 1.35 $\pm$ 0.37 & 1.63 $\pm$ 0.22 & 0.03 $\pm$ 0.01 \\ 
Others & 10 48 43.30 & 11 08 00.30 & 4.45 $\pm$ 0.89 & 0.94 $\pm$ 0.23 & 1.95 $\pm$ 0.24 & 0.08 $\pm$ 0.02 \\ 
LoTSS & 10 57 09.25 & 48 40 41.03 & 8.40 $\pm$ 1.68 & 0.81 $\pm$ 0.42 & 2.11 $\pm$ 0.56 & 0.12 $\pm$ 0.07 \\ 
LoTSS & 11 04 33.11 & 46 42 25.76 & 0.46 $\pm$ 0.09 & 0.11 $\pm$ 0.05 & 5.79 $\pm$ 1.24 & 0.08 $\pm$ 0.04 \\ 
LoTSS & 11 05 15.27 & 54 41 09.30 & 0.75 $\pm$ 0.15 & 0.79 $\pm$ 0.40 & 2.14 $\pm$ 0.54 & 0.04 $\pm$ 0.02 \\ 
LoTSS & 11 18 57.28 & 55 06 56.96 & 9.09 $\pm$ 1.82 & 0.47 $\pm$ 0.21 & 2.77 $\pm$ 0.61 & 0.16 $\pm$ 0.08 \\ 
LoTSS & 11 21 26.44 & 53 44 56.71 & 0.38 $\pm$ 0.08 & 0.43 $\pm$ 0.09 & 2.90 $\pm$ 0.30 & 0.04 $\pm$ 0.01 \\ 
Others & 11 21 45.00 & 17 24 25.30 & 0.74 $\pm$ 0.15 & 0.72 $\pm$ 0.17 & 2.24 $\pm$ 0.27 & 0.04 $\pm$ 0.01 \\ 
LoTSS & 11 24 35.86 & 49 03 25.92 & 3.31 $\pm$ 0.67 & 0.63 $\pm$ 0.32 & 2.40 $\pm$ 0.60 & 0.09 $\pm$ 0.05 \\ 
LoTSS & 11 27 13.18 & 51 13 26.35 & 0.78 $\pm$ 0.16 & 1.18 $\pm$ 0.62 & 1.75 $\pm$ 0.46 & 0.03 $\pm$ 0.02 \\ 
LoTSS & 11 32 02.31 & 47 28 24.14 & 0.28 $\pm$ 0.06 & 1.09 $\pm$ 0.41 & 1.81 $\pm$ 0.34 & 0.02 $\pm$ 0.01 \\ 
LoTSS & 11 32 50.67 & 50 57 04.68 & 0.89 $\pm$ 0.18 & 1.34 $\pm$ 0.79 & 1.64 $\pm$ 0.48 & 0.03 $\pm$ 0.02 \\ 
LoTSS & 11 35 03.20 & 48 26 12.12 & 0.66 $\pm$ 0.13 & 1.53 $\pm$ 0.53 & 1.53 $\pm$ 0.27 & 0.02 $\pm$ 0.01 \\ 
Others & 11 35 35.23 & 39 01 45.77 & 0.38 $\pm$ 0.08 & 0.02 $\pm$ 0.01 & 12.20 $\pm$ 3.48 & 0.15 $\pm$ 0.10 \\ 
SGS & 11 44 27.19 & 37 08 31.87 & 10.80 $\pm$ 2.16 & 0.52 $\pm$ 0.11 & 2.64 $\pm$ 0.28 & 0.17 $\pm$ 0.04 \\ 
LoTSS & 11 48 14.98 & 54 57 16.49 & 0.11 $\pm$ 0.02 & 0.41 $\pm$ 0.13 & 2.96 $\pm$ 0.47 & 0.02 $\pm$ 0.01 \\ 
SGS & 12 03 43.71 & 23 43 04.72 & 5.49 $\pm$ 1.10 & 0.87 $\pm$ 0.30 & 2.03 $\pm$ 0.34 & 0.09 $\pm$ 0.04 \\ 
LoTSS & 12 18 49.88 & 50 26 17.59 & 11.30 $\pm$ 2.26 & 0.46 $\pm$ 0.15 & 2.80 $\pm$ 0.46 & 0.18 $\pm$ 0.07 \\ 
LoTSS & 12 20 28.13 & 52 51 44.89 & 0.75 $\pm$ 0.15 & 0.54 $\pm$ 0.28 & 2.59 $\pm$ 0.67 & 0.04 $\pm$ 0.03 \\ 
LoTSS & 12 22 55.24 & 49 26 42.32 & 0.34 $\pm$ 0.07 & 1.76 $\pm$ 0.51 & 1.43 $\pm$ 0.21 & 0.02 $\pm$ 0.01 \\ 
LoTSS & 12 25 31.36 & 49 46 43.95 & 0.24 $\pm$ 0.05 & 1.43 $\pm$ 0.66 & 1.59 $\pm$ 0.36 & 0.02 $\pm$ 0.01 \\ 
LoTSS & 12 29 36.25 & 50 13 04.65 & 0.94 $\pm$ 0.19 & 0.47 $\pm$ 0.26 & 2.78 $\pm$ 0.76 & 0.05 $\pm$ 0.03 \\ 
LoTSS & 12 32 04.95 & 53 06 27.31 & 0.49 $\pm$ 0.10 & 0.41 $\pm$ 0.19 & 2.98 $\pm$ 0.69 & 0.04 $\pm$ 0.02 \\ 
LoTSS & 12 41 42.34 & 51 35 14.32 & 6.17 $\pm$ 1.24 & 2.22 $\pm$ 1.19 & 1.27 $\pm$ 0.34 & 0.06 $\pm$ 0.04 \\ 
Others & 12 47 33.31 & 67 23 16.34 & 0.20 $\pm$ 0.04 & 0.73 $\pm$ 0.13 & 2.22 $\pm$ 0.19 & 0.02 $\pm$ 0.00 \\ 
Others & 12 53 03.20 & 45 00 44.80 & 0.70 $\pm$ 0.14 & 0.86 $\pm$ 0.17 & 2.04 $\pm$ 0.20 & 0.03 $\pm$ 0.01 \\ 
Others & 12 55 50.13 & 58 18 41.72 & 1.48 $\pm$ 0.31 & 2.81 $\pm$ 1.58 & 1.13 $\pm$ 0.32 & 0.03 $\pm$ 0.02 \\ 
SGS & 12 58 09.42 & 42 11 09.18 & 7.08 $\pm$ 1.42 & 0.96 $\pm$ 0.28 & 1.93 $\pm$ 0.28 & 0.10 $\pm$ 0.03 \\ 
LoTSS & 13 03 31.08 & 53 59 48.65 & 0.29 $\pm$ 0.06 & 0.55 $\pm$ 0.18 & 2.56 $\pm$ 0.42 & 0.03 $\pm$ 0.01 \\ 
LoTSS & 13 03 32.18 & 52 20 02.02 & 0.63 $\pm$ 0.13 & 1.07 $\pm$ 0.42 & 1.83 $\pm$ 0.36 & 0.03 $\pm$ 0.01 \\ 
SGS & 13 12 31.35 & 21 15 43.42 & 6.87 $\pm$ 1.37 & 0.77 $\pm$ 0.20 & 2.17 $\pm$ 0.28 & 0.11 $\pm$ 0.03 \\ 
Others & 13 13 57.70 & 64 25 55.00 & 1.95 $\pm$ 0.40 & 0.89 $\pm$ 0.30 & 2.01 $\pm$ 0.34 & 0.06 $\pm$ 0.02 \\ 
LoTSS & 13 14 04.60 & 54 39 37.88 & 8.91 $\pm$ 1.79 & 1.43 $\pm$ 0.62 & 1.59 $\pm$ 0.35 & 0.09 $\pm$ 0.05 \\ 
LoTSS & 13 22 29.07 & 50 48 44.53 & 0.30 $\pm$ 0.06 & 0.63 $\pm$ 0.17 & 2.38 $\pm$ 0.33 & 0.03 $\pm$ 0.01 \\ 
LoTSS & 13 24 35.19 & 50 41 02.31 & 4.27 $\pm$ 0.86 & 0.21 $\pm$ 0.10 & 4.10 $\pm$ 0.97 & 0.17 $\pm$ 0.09 \\ 
LoTSS & 13 31 35.25 & 45 59 55.53 & 1.91 $\pm$ 0.38 & 4.33 $\pm$ 2.10 & 0.91 $\pm$ 0.22 & 0.02 $\pm$ 0.01 \\ 
LoTSS & 13 32 58.28 & 53 53 55.60 & 2.23 $\pm$ 0.45 & 2.09 $\pm$ 0.93 & 1.31 $\pm$ 0.29 & 0.04 $\pm$ 0.02 \\ 
LoTSS & 13 36 18.75 & 53 39 52.12 & 0.47 $\pm$ 0.09 & 0.67 $\pm$ 0.24 & 2.32 $\pm$ 0.41 & 0.03 $\pm$ 0.01 \\ 
SGS & 13 37 42.35 & 29 42 23.31 & 0.53 $\pm$ 0.10 & 1.13 $\pm$ 0.24 & 1.79 $\pm$ 0.19 & 0.03 $\pm$ 0.01 \\ 
LoTSS & 13 42 06.98 & 47 25 53.04 & 1.02 $\pm$ 0.20 & 2.13 $\pm$ 0.58 & 1.30 $\pm$ 0.18 & 0.03 $\pm$ 0.01 \\ 
Others & 13 45 03.60 & 39 52 31.00 & 1.99 $\pm$ 0.40 & 0.63 $\pm$ 0.19 & 2.39 $\pm$ 0.36 & 0.07 $\pm$ 0.02 \\ 
LoTSS & 13 45 57.55 & 54 03 16.62 & 6.36 $\pm$ 1.27 & 0.24 $\pm$ 0.08 & 3.84 $\pm$ 0.67 & 0.19 $\pm$ 0.08 \\ 
LoTSS & 13 49 27.92 & 46 20 15.11 & 2.46 $\pm$ 0.49 & 0.64 $\pm$ 0.38 & 2.38 $\pm$ 0.70 & 0.07 $\pm$ 0.05 \\ 
Others & 13 50 00.70 & 29 47 21.40 & 0.22 $\pm$ 0.06 & 1.45 $\pm$ 0.44 & 1.58 $\pm$ 0.24 & 0.01 $\pm$ 0.01 \\ 
SGS & 13 52 28.39 & 09 35 36.02 & 2.10 $\pm$ 0.42 & 0.09 $\pm$ 0.03 & 6.47 $\pm$ 1.10 & 0.18 $\pm$ 0.07 \\ 
LoTSS & 13 56 28.50 & 52 42 19.23 & 1.35 $\pm$ 0.27 & 1.90 $\pm$ 0.92 & 1.38 $\pm$ 0.34 & 0.03 $\pm$ 0.02 \\ 
LoTSS & 13 59 51.16 & 47 03 21.03 & 0.59 $\pm$ 0.12 & 0.99 $\pm$ 0.43 & 1.91 $\pm$ 0.42 & 0.03 $\pm$ 0.01 \\ 
Others & 14 00 43.40 & 30 19 19.00 & 7.71 $\pm$ 1.54 & 0.86 $\pm$ 0.23 & 2.05 $\pm$ 0.27 & 0.11 $\pm$ 0.03 \\ 
LoTSS & 14 07 18.48 & 51 32 04.63 & 53.70 $\pm$ 10.80 & 0.28 $\pm$ 0.15 & 3.60 $\pm$ 0.97 & 0.52 $\pm$ 0.32 \\ 
LoTSS & 14 16 25.89 & 54 25 45.85 & 4.81 $\pm$ 0.96 & 1.27 $\pm$ 0.41 & 1.68 $\pm$ 0.27 & 0.07 $\pm$ 0.03 \\ 
Others & 14 18 37.75 & 37 46 23.00 & 1.06 $\pm$ 0.21 & 0.65 $\pm$ 0.15 & 2.35 $\pm$ 0.27 & 0.05 $\pm$ 0.01 \\ 
SGS & 14 19 47.89 & 08 14 23.39 & 14.70 $\pm$ 2.94 & 0.84 $\pm$ 0.33 & 2.07 $\pm$ 0.41 & 0.16 $\pm$ 0.07 \\ 
LoTSS & 14 28 57.66 & 54 36 27.81 & 4.11 $\pm$ 0.83 & 0.95 $\pm$ 0.54 & 1.95 $\pm$ 0.55 & 0.08 $\pm$ 0.05 \\ 
LoTSS & 14 29 33.45 & 54 43 35.29 & 1.09 $\pm$ 0.22 & 0.49 $\pm$ 0.11 & 2.71 $\pm$ 0.30 & 0.06 $\pm$ 0.01 \\ 
Others & 14 31 03.40 & 33 45 41.60 & 0.48 $\pm$ 0.11 & 2.33 $\pm$ 0.64 & 1.24 $\pm$ 0.17 & 0.02 $\pm$ 0.01 \\ 
LoTSS & 14 31 36.99 & 52 27 24.90 & 3.32 $\pm$ 0.67 & 1.28 $\pm$ 0.47 & 1.68 $\pm$ 0.30 & 0.06 $\pm$ 0.03 \\ 
LoTSS & 14 44 10.50 & 55 47 45.64 & 6.60 $\pm$ 1.32 & 2.48 $\pm$ 0.76 & 1.20 $\pm$ 0.18 & 0.06 $\pm$ 0.02 \\ 
LoTSS & 14 45 20.87 & 54 03 29.62 & 2.80 $\pm$ 0.56 & 1.94 $\pm$ 1.15 & 1.36 $\pm$ 0.40 & 0.04 $\pm$ 0.03 \\ 
Others & 14 45 27.40 & 09 32 18.00 & 0.18 $\pm$ 0.04 & 0.37 $\pm$ 0.07 & 3.11 $\pm$ 0.31 & 0.03 $\pm$ 0.01 \\ 
LoTSS & 14 46 07.20 & 48 41 37.79 & 2.76 $\pm$ 0.56 & 0.92 $\pm$ 0.38 & 1.97 $\pm$ 0.40 & 0.06 $\pm$ 0.03 \\ 
Others & 14 53 02.93 & 33 08 40.80 & 10.30 $\pm$ 2.07 & 1.00 $\pm$ 0.31 & 1.90 $\pm$ 0.29 & 0.12 $\pm$ 0.04 \\ 
LoTSS & 14 57 02.81 & 48 06 46.64 & 0.04 $\pm$ 0.01 & 1.31 $\pm$ 0.46 & 1.66 $\pm$ 0.29 & 0.01 $\pm$ 0.00 \\ 
LoTSS & 15 01 32.11 & 50 34 55.14 & 0.35 $\pm$ 0.07 & 2.49 $\pm$ 0.95 & 1.20 $\pm$ 0.23 & 0.01 $\pm$ 0.01 \\ 
LoTSS & 15 06 24.10 & 53 55 02.61 & 1.16 $\pm$ 0.23 & 0.86 $\pm$ 0.26 & 2.05 $\pm$ 0.32 & 0.04 $\pm$ 0.02 \\ 
Others & 15 24 44.60 & 19 59 57.08 & 1.83 $\pm$ 0.40 & 0.59 $\pm$ 0.31 & 2.47 $\pm$ 0.65 & 0.07 $\pm$ 0.04 \\ 
SGS & 15 46 39.13 & 01 24 22.04 & 2.39 $\pm$ 0.48 & 3.75 $\pm$ 1.01 & 0.98 $\pm$ 0.13 & 0.03 $\pm$ 0.01 \\ 
SGS & 15 51 40.30 & 10 35 48.66 & 38.40 $\pm$ 7.71 & 0.71 $\pm$ 0.36 & 2.25 $\pm$ 0.56 & 0.27 $\pm$ 0.16 \\ 
SGS & 16 05 13.74 & 07 11 52.56 & 13.30 $\pm$ 2.66 & 0.28 $\pm$ 0.14 & 3.61 $\pm$ 0.91 & 0.26 $\pm$ 0.15 \\ 
SGS & 16 12 42.06 & 43 13 19.82 & 3.70 $\pm$ 0.74 & 0.69 $\pm$ 0.20 & 2.28 $\pm$ 0.33 & 0.09 $\pm$ 0.03 \\ 
SGS & 17 21 07.89 & 26 24 32.17 & 2.95 $\pm$ 0.59 & 0.20 $\pm$ 0.12 & 4.21 $\pm$ 1.21 & 0.14 $\pm$ 0.09 \\ 
SGS & 21 30 39.21 & 07 35 30.28 & 4.61 $\pm$ 0.92 & 0.37 $\pm$ 0.17 & 3.12 $\pm$ 0.73 & 0.13 $\pm$ 0.07 \\ 
Others & 21 45 04.53 & $-$06 59 07.76 & 2.79 $\pm$ 0.58 & 2.55 $\pm$ 0.81 & 1.19 $\pm$ 0.19 & 0.04 $\pm$ 0.01 \\ 
SGS & 22 18 15.55 & 19 31 43.76 & 0.78 $\pm$ 0.16 & 1.22 $\pm$ 0.23 & 1.72 $\pm$ 0.17 & 0.03 $\pm$ 0.01 \\ 
SGS & 22 31 14.32 & 01 00 41.80 & 2.80 $\pm$ 0.56 & 1.27 $\pm$ 0.33 & 1.69 $\pm$ 0.22 & 0.06 $\pm$ 0.02 \\ 
SGS & 22 51 03.39 & 06 19 25.49 & 0.71 $\pm$ 0.14 & 0.30 $\pm$ 0.07 & 3.44 $\pm$ 0.41 & 0.06 $\pm$ 0.02 \\ 
SGS & 23 41 37.14 & 08 28 17.25 & 4.72 $\pm$ 0.94 & 1.56 $\pm$ 0.43 & 1.52 $\pm$ 0.21 & 0.06 $\pm$ 0.02 \\ 
SGS & 23 48 47.77 & 16 00 10.32 & 23.60 $\pm$ 4.74 & 0.73 $\pm$ 0.22 & 2.23 $\pm$ 0.34 & 0.21 $\pm$ 0.08 \\ 
SGS & 23 59 11.07 & 17 06 10.97 & 2.32 $\pm$ 0.46 & 0.30 $\pm$ 0.05 & 3.45 $\pm$ 0.31 & 0.10 $\pm$ 0.02 \\

\label{tab:spin}
\end{longtable}
\end{center}
\end{tiny}

\clearpage
\setlength{\tabcolsep}{4pt}
\begin{table}
\centering
\caption{RA and  Dec represent the right ascension and declination of the GRG host galaxies in HMS and DMS, respectively. Col. 5 shows the m$_{\rm r}$ of the BCGs. Parameters such as  $\rm 
r_{200}$ and $\rm R_{L*}$ have been obtained from the WHL galaxy cluster catalogue, and  $\rm M_{200}$ has been estimated using equation 2 of \citet{whl12}. $r_{200}$ is the virial radius (Mpc) of the 
galaxy cluster,
$\rm M_{200}$ is the mass (10$^{14} \rm M_{\odot}$ ) of the cluster within $\rm r_{200}$, and $\rm N_{200}$ is the number of galaxies present within $\rm r_{200}$. Column 11 shows the comoving number density of galaxies within $r_{200}$. Col. 12 includes the ratio of half of the linear extent of the source to the virial radius.}

\begin{tabular}{c c c c c c c c c c c c}
\hline

\multicolumn{1}{l}{No} & WHL Cluster &  RA &  Dec & m$_{\rm r}$ & $z$ & \multicolumn{1}{l}{$\rm r_{200}$} & \multicolumn{1}{l}{$\rm R_{L*}$} & \multicolumn{1}{l}{\rm $\rm M_{200}$} & 
\multicolumn{1}{l}{$\rm N_{200}$} & \multicolumn{1}{l}{Density}  & \multicolumn{1}{l}{Ratio}\\
  & & (HMS) & (DMS) & & & (Mpc) & & (10$^{14}  \rm M_{\odot}$) & & Mpc$^{-3}$ & \\ 
(1) & (2) & (3) & (4) & (5) & (6) & (7) & (8) & (9) & (10) & (11) & (12) \\
\hline
1 & J013327.2$-$082417 & 01 33 27.24 & $-$08 24 16.52  & 15.90 & 0.14893 & 0.87 & 16.26 & 0.85 & 15 & 5.44 & 0.53\\
2 & J014208.5$-$064143 & 01 42 08.56 & $-$06 41 43.40 & 15.12 & 0.12447 & 1.30 & 36.91 & 2.21 & 22 & 2.39 & 0.34 \\
3 & J072538.8+400413 & 07 25 38.75 & 40 04 12.52 & 16.14 & 0.16148 & 0.85 & 12.42 & 0.62 & 9 & 3.50 & 0.49\\
4 & J085349.8+145226 & 08 53 49.78 & 14 52 26.04 & 14.79 & 0.06933 & 0.81 & 13.18 & 0.66 & 9 & 4.04 & 0.46\\
5 & J090123.3+191417 & 09 01 23.31 & 19 14 17.12  & 17.87 & 0.27649 & 0.90 & 15.02 & 0.77 & 10 & 3.27 & 0.63\\
6 & J090839.1+594513 & 09 08 39.13 & 59 45 12.82 & 17.01 & 0.24004 & 1.36 & 41.19 & 2.51 & 37 & 3.51 & 0.48 \\
7 & J100943.5+033723 & 10 09 43.50 & 03 37 22.72 & 14.77 & 0.10513 & 0.95 & 17.40 & 0.92 & 8 & 2.23 & 0.39\\
8 & J112422.8+150958 & 11 24 22.77 & 15 09 57.90 & 16.34 & 0.17194 & 1.17 & 30.92 & 1.79 & 22 & 3.28 & 0.39 \\
9 & J114344.4+222907 & 11 43 44.42 & 22 29 06.80  & 16.55 & 0.18079 & 0.76 & 13.51 & 0.68 & 14 & 7.61 & 0.53 \\
10 & J120343.7+234305 & 12 03 43.71 & 23 43 04.72  & 16.64 & 0.17670 & 1.36 & 41.48 & 2.53 & 40 & 3.80 & 0.38 \\
11 & J121606.1+202055 & 12 16 06.10 & 20 20 54.80 & 18.39 & 0.26300 & 0.89 & 15.45 & 0.80 & 13 & 4.40 & 0.60\\
12 & J125154.4+351912 & 12 51 54.40 & 35 19 11.97  & 17.81 & 0.20100 & 0.83 & 12.28 & 0.61 & 11 & 4.59 & 0.62\\
13 & J130152.7+144843 & 13 01 52.71 & 14 48 43.01 & 15.51 & 0.13761 & 0.98 & 17.50 & 0.92 & 12 & 3.04 & 0.43 \\
14 & J133742.3+294223 & 13 37 42.35 & 29 42 23.31 & 15.16 & 0.11547 & 0.93 & 13.30 & 0.67 & 9 & 2.67 & 0.49\\
15 & J144928.6$-$011617 & 14 49 28.63 & $-$01 16 17.44 & 16.32 & 0.20233 & 1.04 & 20.47 & 1.11 & 15 & 3.18 & 0.44\\
16 & J154639.1+012422 & 15 46 39.13 & 01 24 22.04 & 16.22 & 0.20836 & 1.04 & 25.13 & 1.41 & 12 & 2.55 & 0.45 \\
17 & J155140.3+103549 & 15 51 40.30 & 10 35 48.66  & 17.98 & 0.36824 & 1.13 & 25.99 & 1.46 & 23 & 3.81 & 0.55\\
18 & J221815.6+193144 & 22 18 15.55 & 19 31 43.76 & 15.14 & 0.10964 & 1.35 & 47.92 & 2.99 & 28 & 2.72 & 0.28\\
\hline
\end{tabular}
\label{whl}
\end{table}

\section{Notes on individual sources from SGS}
Here we present our findings and important notes related to some interesting GRGs from our sample.
\begin{itemize}
\item  SAGANJ000450.25+124840.10 and SAGANJ011341.11+010608.52. These two sources show two symmetric winged back-flows that emanate from the two hotspots. They can therefore be referred to as X-shaped radio galaxies. For X-shaped radio galaxies, three models exist: i) the twin AGN model, ii) the rapid-jet reorientation models, and iii) the back-flow diversion model. They have been proposed to explain this phenomenon in RGs. After inspecting high-resolution maps of FIRST and VLASS, we found no evidence of the presence of twin AGN at the core. 
However, our observations support the model that the pair of wings arises from the diversion of synchrotron plasma from the hotspots due to ambient pressure gradient.

\item SAGANJ075931.84+082534.59.
The radio core of the source is only detected at 3000 MHz in the high-resolution survey VLASS (2$\arcsec$). It coincides with a galaxy at a redshift of 0.124 with an r-band magnitude of 17.31. The diffuse plasma, spread along the jet axis on either side of the core, is seen properly in the NVSS, but only partially well in the TGSS, as shown in Fig.\ \ref{montage3}. Thus, the overall intricacies of this object make it a good candidate for a remnant radio-loud AGN \citep{Parma2007, Mahatma2018} with a projected linear size of $\sim$\,0.72 Mpc. This is likely to be a young active source with a fading plasma from the earlier activity of the source. The integrated two-point spectral index of $\sim$\,0.68 also supports this argument. This source is important in order to understand the last phase of the duty cycle of AGN activity after the jets have switched off. 

\item SAGANJ105309.33+260142.13.  Contamination is observed near the western side of the core, which is quite clear in the contours of FIRST  in the montage in Fig.\ \ref{montage4}. The measured flux density at 1400 MHz (NVSS) is 336.2 mJy, which includes the flux density of the contaminating source (henceforth source A). For source A, using FIRST, where it is sufficiently resolved, we estimate a flux density of 21.6 mJy, and when this value is subtracted from the total measured flux density, we obtain the corrected flux density value of 314.6 mJy. A similar method is followed for the flux density correction at 150 MHz (TGSS). Because source A is not sufficiently well resolved in TGSS map, the 3$\sigma$ contours of the source from the FIRST map are overlaid on the TGSS map, and the flux density of the corresponding region (3$\sigma$) is considered to be the flux density of source A in TGSS. The respective value (77.9  mJy) is then subtracted from the measured flux density (1688.3 mJy) of the source, and therefore the corrected flux density is 1610.4 mJy, as given in Table \ref{tab:maintab}. The two corrected flux densities at the respective frequencies are used for further calculations of radio power.

\item SAGANJ112422.77+150957.90.
This is the only double-double radio galaxy (DDRG) identified in our sample of 162 GRGs based on the available radio maps. The DDRG clearly displays aligned radio components, consisting of radio core and two inner and two outer lobes. The double-double morphology is prominently seen in the montage
Fig.\ \ref{montage4}, where images from the NVSS (45$\arcsec$), TGSS (25$\arcsec$), and FIRST (5$\arcsec$) are overlaid. The outer lobes are well detected in the NVSS (1400 MHz), but the inner structure is unresolved there. In the TGSS (150 MHz), there is a hint of detection of the outer and inner components of the source, but they are not bright enough to confirm the DDRG nature. However, the high-resolution FIRST map confirms the inner structure, showing that the two inner lobes have an edge-brightened FR-II morphology, and VLASS (3000 MHz) with its higher resolution of 2$\arcsec$ clearly reveals the radio core. Because of the relatively low surface brightness sensitivity, the FIRST (1400 MHz) and VLASS (3000 MHz) surveys have resolved out the outer components of the source. The inner doubles are very compact compared to the outer ones. The two outer hotspots are quite prominent in the NVSS map, and a winged flow is observed to be emitted from the northern hotspot. The angular size of the inner double is 0.82$\arcmin$  , projecting a linear size of $\sim$\,0.15 Mpc, whereas the outer double spans up to $\sim$\,0.92 Mpc. The radio core of the DDRG coincides with an optical galaxy with an r-band magnitude of 16.36. This source has been classified as a DDRG by \citet{koziel19}.

\item SAGANJ114427.19+370831.87. A winged back-flow is observed to emanate from the northern hotspot, but no such feature is seen near the southern hotspot. However, this source has been classified as an X-shaped radio galaxy by \citet{koziel19}.
\item SAGANJ225321.28+162016.77.
The source was mentioned in \citet{Solovyov14} as a candidate GRG.
\end{itemize}

\section{Radio maps of all GRGs from the SGS.}
\label{sec:apfig}
Below we present the maps of GRGs from the SGS. Blue represents the radio emission detected in the NVSS, and black and yellow contours are from the TGSS and the FIRST, respectively. Absence of  contours from any survey means either that the survey did not cover the part of the sky or that the survey did not detect the source.
The red crosses represent the location of the host galaxy, which is also given in the name of the source at the top of each image. The lower right corner of each image shows the angular scale for reference. Numbers in the lower left corner represent the serial numbers from Table\ \ref{tab:maintab}. 

\begin{figure*}
\centering
\includegraphics[scale=0.83]{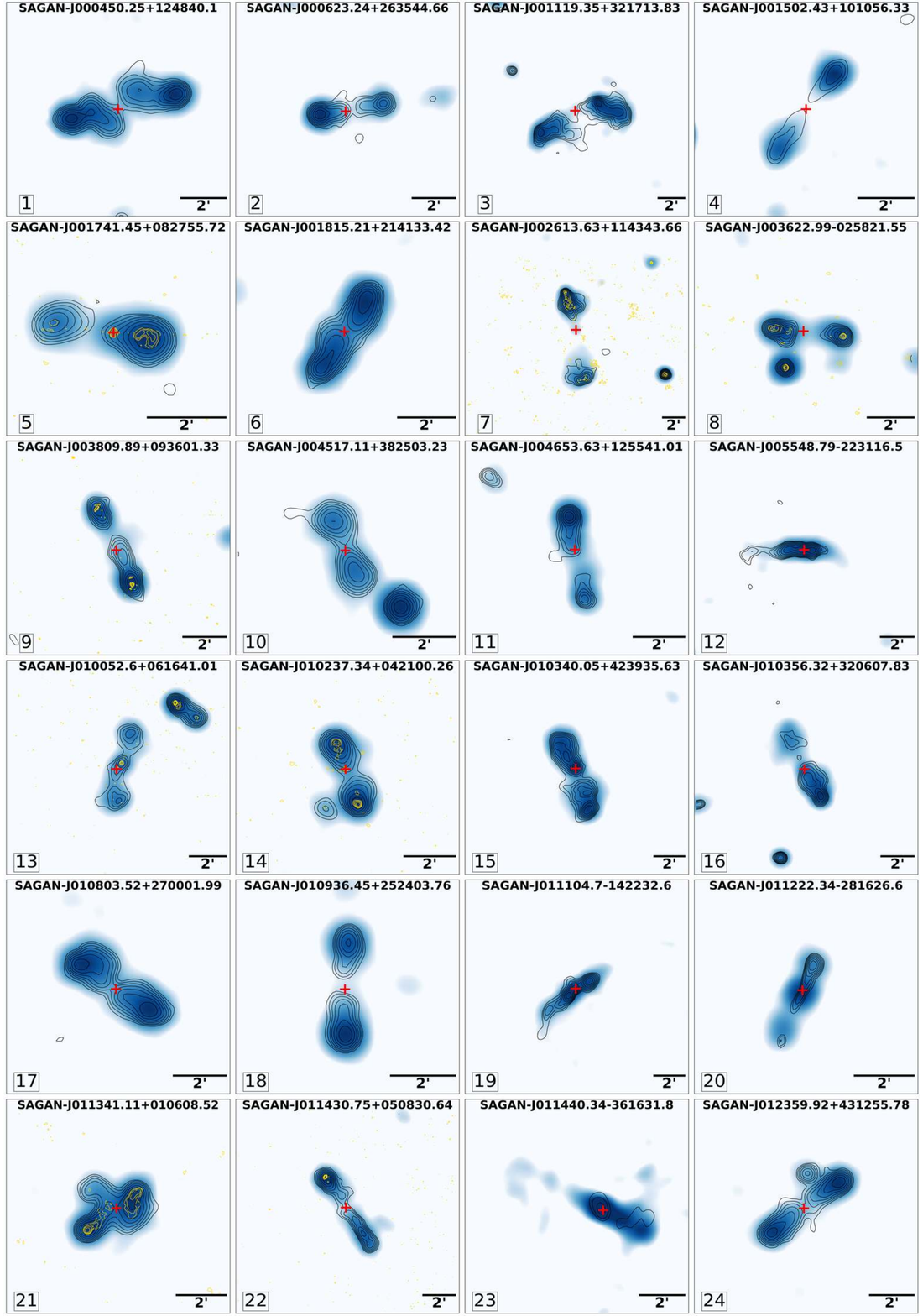}
\caption{Radio maps of 1 to 24 GRGs from the SGS as described in Sec.\ \ref{sec:apfig}.}
\label{montage1}
\end{figure*}

\begin{figure*}
\centering
\includegraphics[scale=0.83]{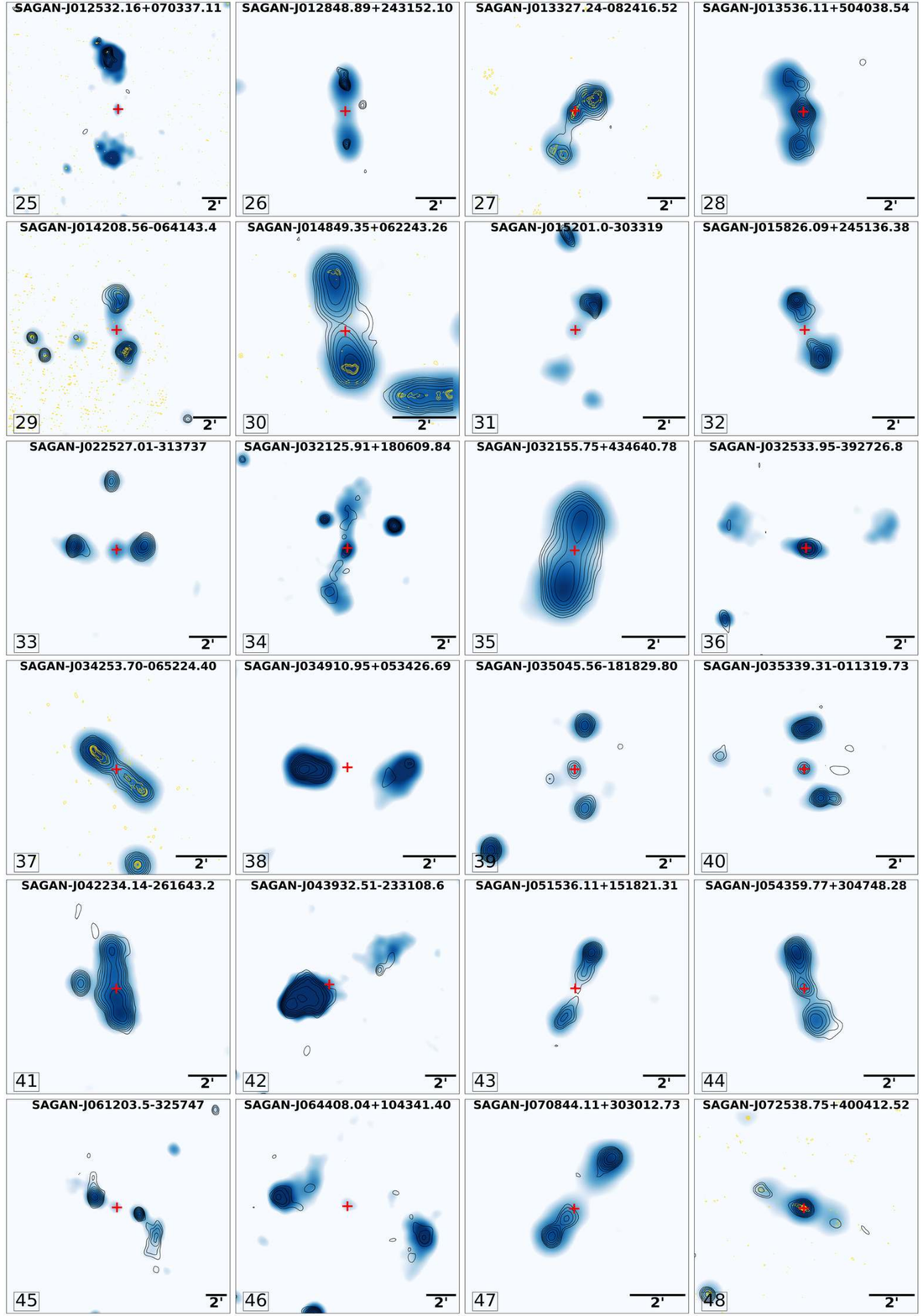} 
\caption{Radio maps of 24 to 48 GRGs from the SGS as described in Sec.\ \ref{sec:apfig}.}
\label{montage2}
\end{figure*}

\begin{figure*}
\centering
\includegraphics[scale=0.83]{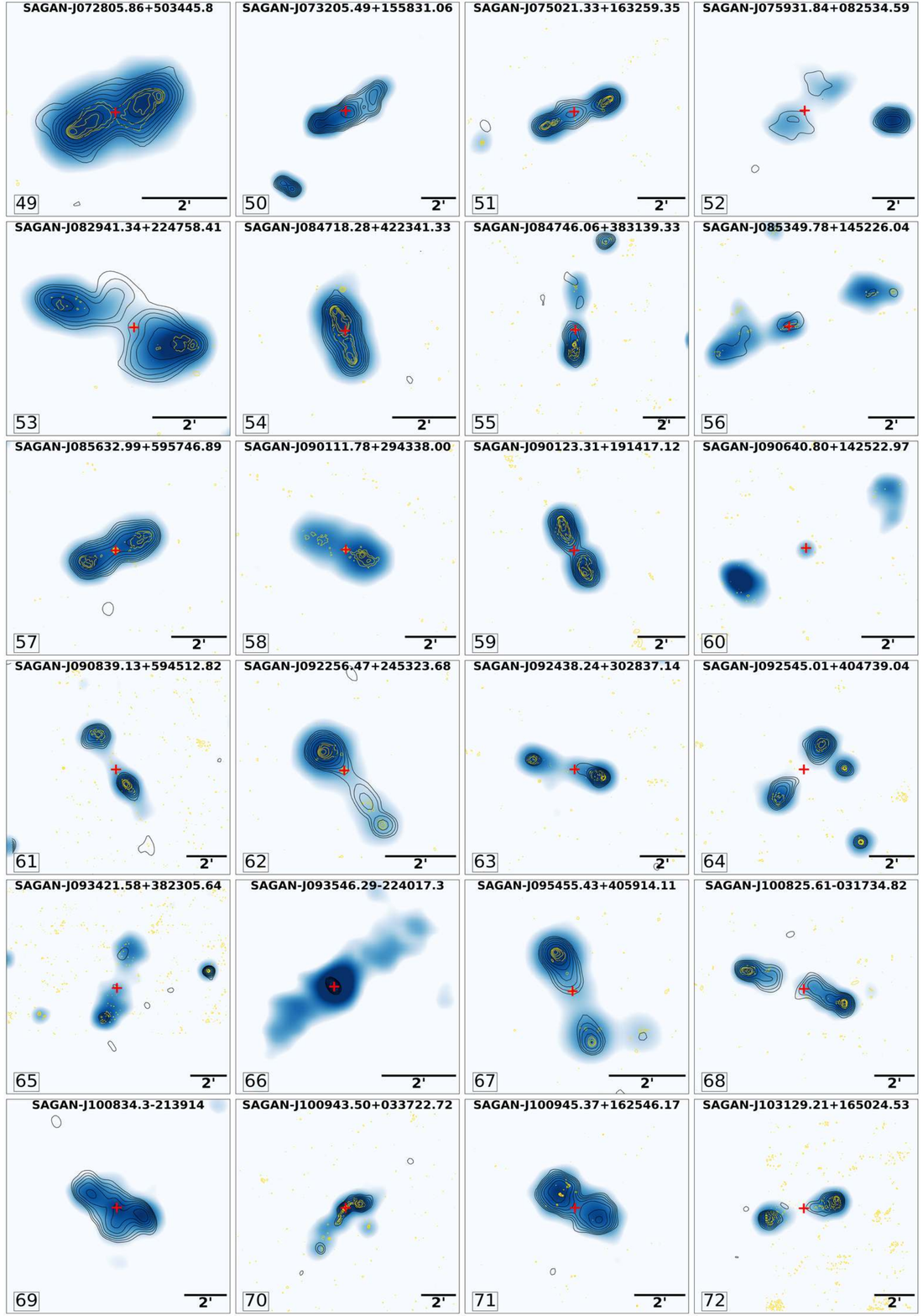} 
\caption{Radio maps of 49 to 72 GRGs from the SGS as described in Sec.\ \ref{sec:apfig}.}
\label{montage3}
\end{figure*}

\begin{figure*}
\centering
\includegraphics[scale=0.83]{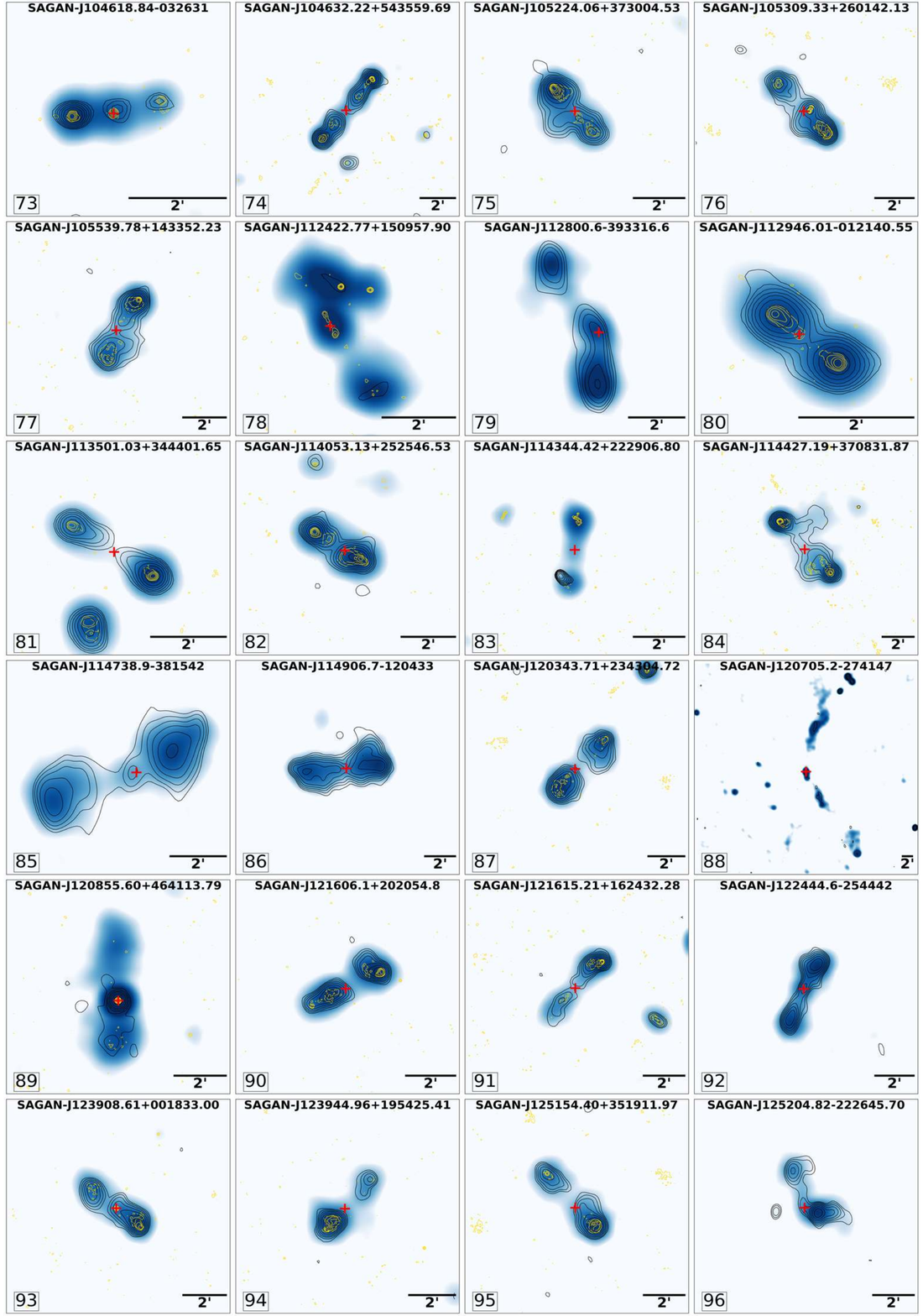} 
\caption{Radio maps of 73 to 96 GRGs from the SGS as described in Sec.\ \ref{sec:apfig}.}
\label{montage4}
\end{figure*}

\begin{figure*}
\centering
\includegraphics[scale=0.83]{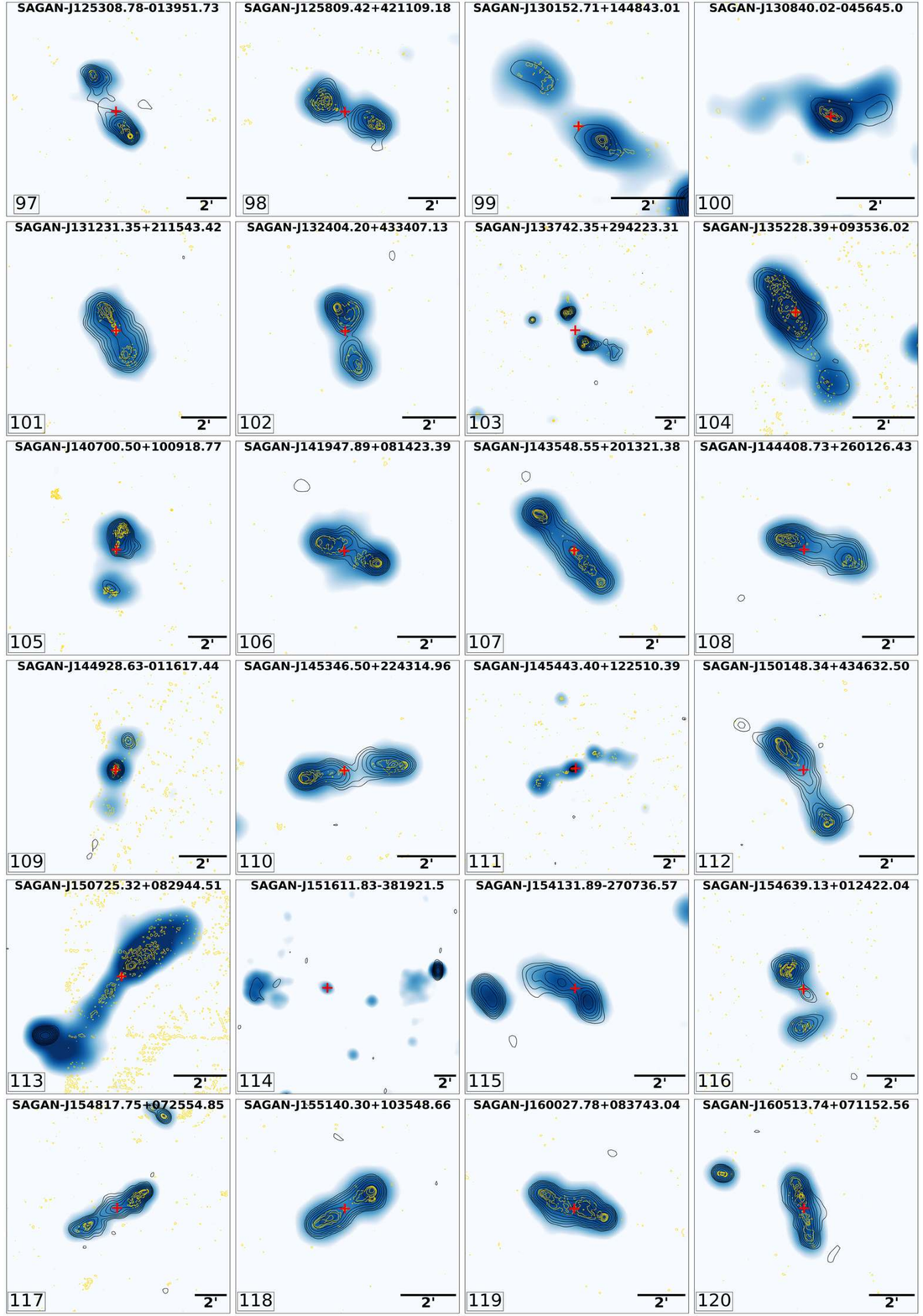} 
\caption{Radio maps of 97 to 120 GRGs from the SGS as described in Sec.\ \ref{sec:apfig}.}
\label{montage5}
\end{figure*}

\begin{figure*}
\centering
\includegraphics[scale=0.83]{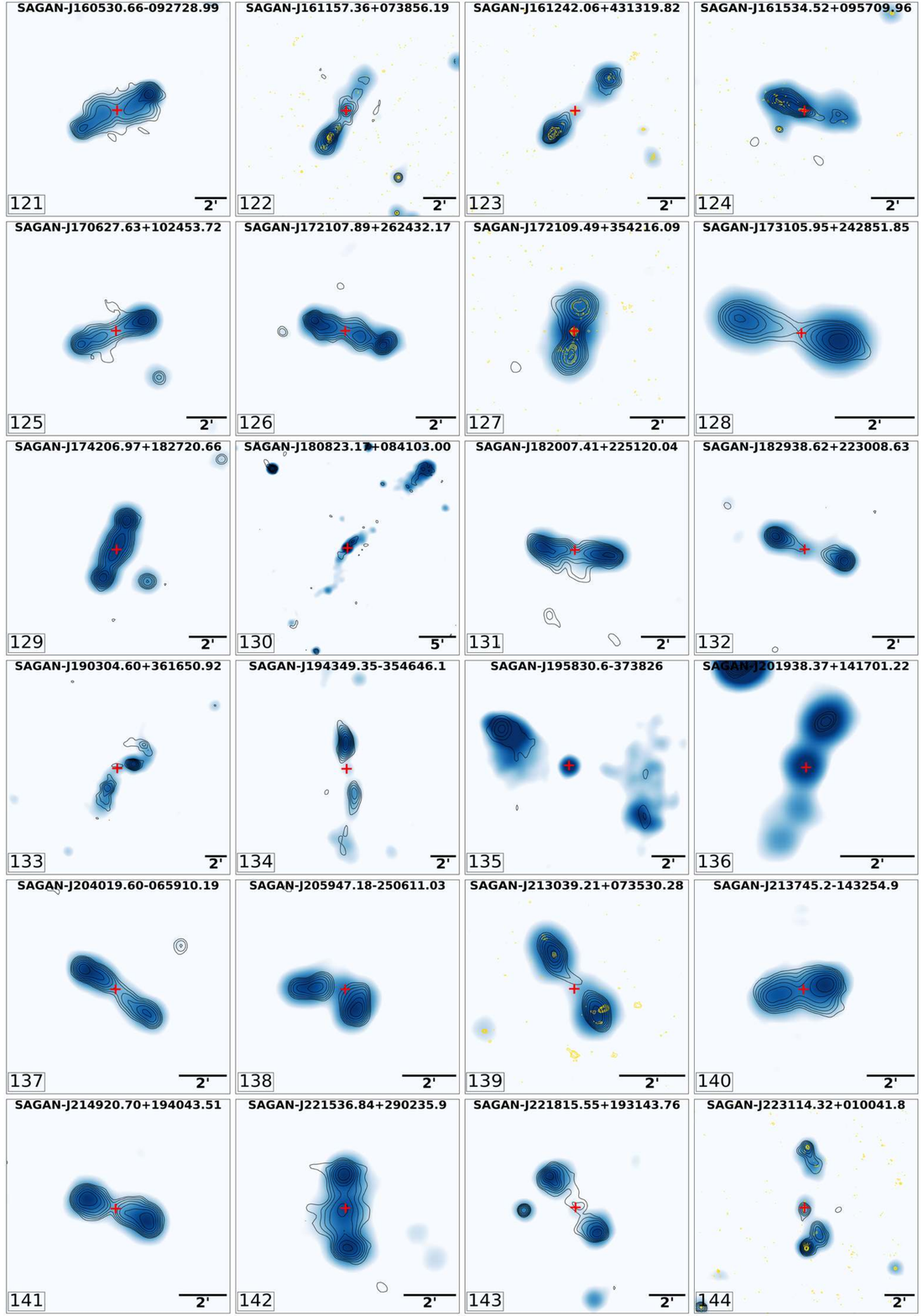} 
\caption{Radio maps of 121 to 144 GRGs from the SGS as described in Sec.\ \ref{sec:apfig}.}
\label{montage6}
\end{figure*}

\begin{figure*}
\centering
\includegraphics[scale=0.83]{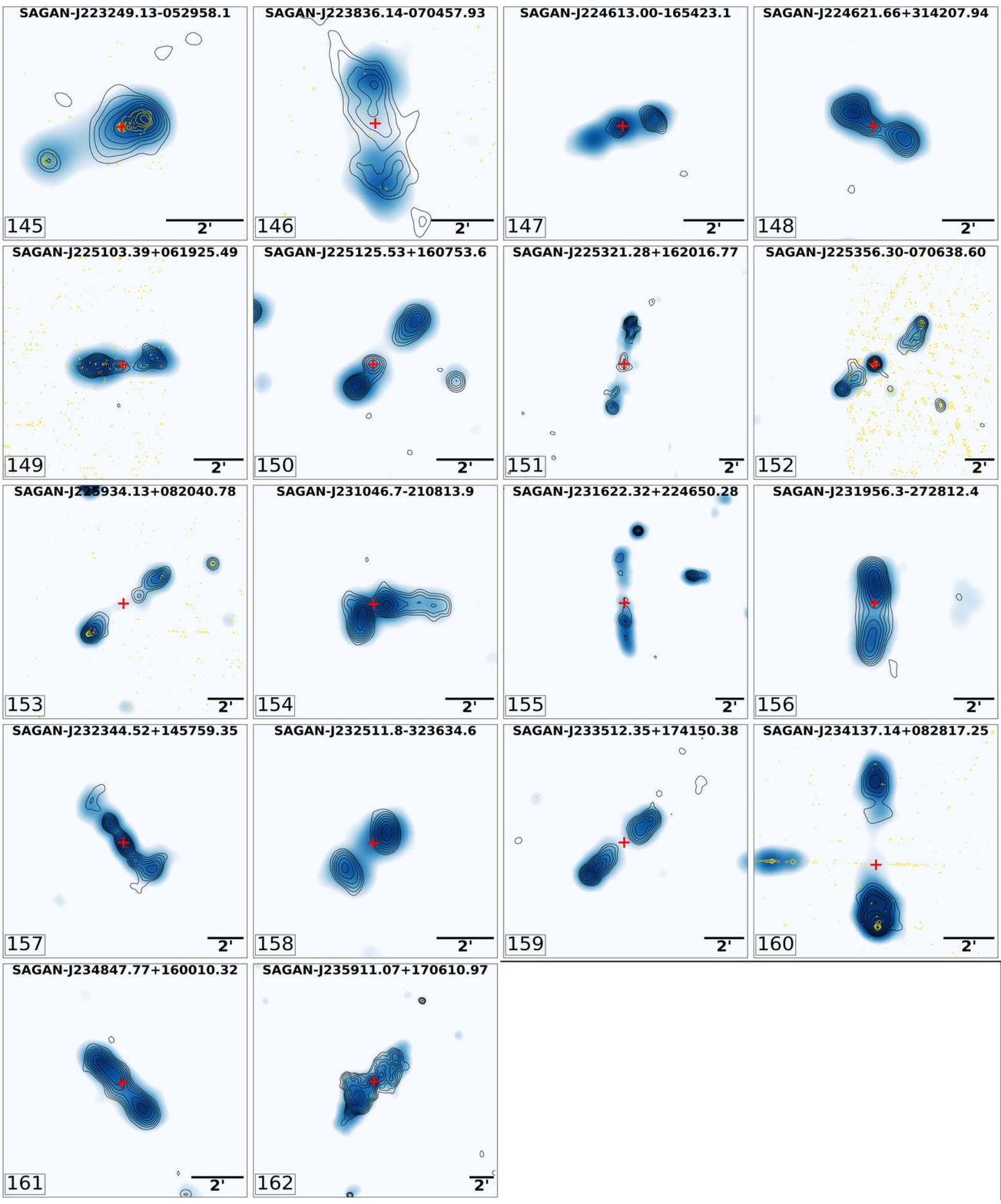} 
\caption{Radio maps of 145 to 162 GRGs from the SGS as described in Sec.\ \ref{sec:apfig}.}
\label{montage7}
\end{figure*}
\end{appendix}


\end{document}